\documentclass[prd,aps,twocolumn,superscriptaddress,showpacs,preprintnumbers,
amsmath,amssymb,floatfix]{revtex4}

\def\gtwid{\mathrel{\raise.3ex\hbox{$>$\kern-.75em\lower1ex\hbox{$\sim$}}}}
\def\ltwid{\mathrel{\raise.3ex\hbox{$<$\kern-.75em\lower1ex\hbox{$\sim$}}}}

\usepackage{graphicx}
\usepackage{dcolumn}
\usepackage{bm}

\def\~GeV{~GeV/c$^2$}
\def\eg{{{\em e.g.}}}

\long\def\symbolfootnote[#1]#2{\begingroup%
\def\thefootnote{\fnsymbol{footnote}}\footnote[#1]{#2}\endgroup}

\begin{document}

\preprint{CDMS \today}

\title{Exclusion Limits on the WIMP-Nucleon Cross-Section from the First Run of the Cryogenic Dark Matter Search in the Soudan Underground Lab}

\author{D.S.~Akerib} \affiliation{Department of Physics, Case Western Reserve University, Cleveland, OH  44106, USA}
\author{M.S.~Armel-Funkhouser} \affiliation{Department of Physics, University of California, Berkeley, CA 94720, USA}
\author{M.J.~Attisha} \affiliation{Department of Physics, Brown University, Providence, RI 02912, USA}
\author{C.N.~Bailey} \affiliation{Department of Physics, Case Western Reserve University, Cleveland, OH  44106, USA}
\author{L.~Baudis} \affiliation{Department of Physics, University of Florida, Gainesville, FL 32611, USA}
\author{D.A.~Bauer} \affiliation{Fermi National Accelerator Laboratory, Batavia, IL 60510, USA}
\author{P.L.~Brink} \affiliation{Department of Physics, Stanford University, Stanford, CA 94305, USA}
\author{R.~Bunker} \affiliation{Department of Physics, University of California, Santa Barbara, CA 93106, USA}
\author{B.~Cabrera} \affiliation{Department of Physics, Stanford University, Stanford, CA 94305, USA}
\author{D.O.~Caldwell} \affiliation{Department of Physics, University of California, Santa Barbara, CA 93106, USA}
\author{C.L.~Chang} \affiliation{Department of Physics, Stanford University, Stanford, CA 94305, USA}
\author{M.B.~Crisler} \affiliation{Fermi National Accelerator Laboratory, Batavia, IL 60510, USA}
\author{P.~Cushman} \affiliation{School of Physics \& Astronomy, University of Minnesota, Minneapolis, MN 55455, USA}
\author{M.~Daal} \affiliation{Department of Physics, University of California, Berkeley, CA 94720, USA}
\author{R.~Dixon} \affiliation{Fermi National Accelerator Laboratory, Batavia, IL 60510, USA}
\author{M.R.~Dragowsky} \affiliation{Department of Physics, Case Western Reserve University, Cleveland, OH  44106, USA}
\author{D.D.~Driscoll} \affiliation{Department of Physics, Case Western Reserve University, Cleveland, OH  44106, USA}
\author{L.~Duong} \affiliation{School of Physics \& Astronomy, University of Minnesota, Minneapolis, MN 55455, USA}
\author{R.~Ferril} \affiliation{Department of Physics, University of California, Santa Barbara, CA 93106, USA}
\author{J.~Filippini} \affiliation{Department of Physics, University of California, Berkeley, CA 94720, USA}
\author{R.J.~Gaitskell} \affiliation{Department of Physics, Brown University, Providence, RI 02912, USA}
\author{R.~Hennings-Yeomans} \affiliation{Department of Physics, Case Western Reserve University, Cleveland, OH  44106, USA}
\author{D.~Holmgren} \affiliation{Fermi National Accelerator Laboratory, Batavia, IL 60510, USA}
\author{M.E.~Huber} \affiliation{Department of Physics, University of Colorado at Denver and Health Sciences Center, Denver, CO 80217, USA}
\author{S.~Kamat} \affiliation{Department of Physics, Case Western Reserve University, Cleveland, OH  44106, USA}
\author{A.~Lu} \affiliation{Department of Physics, University of California, Berkeley, CA 94720, USA}
\author{R.~Mahapatra} \affiliation{Department of Physics, University of California, Santa Barbara, CA 93106, USA}
\author{V.~Mandic} \affiliation{Department of Physics, University of California, Berkeley, CA 94720, USA}
\author{J.M.~Martinis} \affiliation{National Institute of Standards and Technology, Boulder, CO 80303, USA}
\author{P.~Meunier} \affiliation{Department of Physics, University of California, Berkeley, CA 94720, USA}
\author{N.~Mirabolfathi} \affiliation{Department of Physics, University of California, Berkeley, CA 94720, USA}
\author{H.~Nelson} \affiliation{Department of Physics, University of California, Santa Barbara, CA 93106, USA}
\author{R.~Nelson} \affiliation{Department of Physics, University of California, Santa Barbara, CA 93106, USA}
\author{R.W.~Ogburn} \affiliation{Department of Physics, Stanford University, Stanford, CA 94305, USA}
\author{T.A.~Perera} \affiliation{Department of Physics, Case Western Reserve University, Cleveland, OH  44106, USA}
\author{M.C.~Perillo~Issac} \affiliation{Department of Physics, University of California, Berkeley, CA 94720, USA}
\author{E.~Ramberg} \affiliation{Fermi National Accelerator Laboratory, Batavia, IL 60510, USA}
\author{W.~Rau} \affiliation{Department of Physics, University of California, Berkeley, CA 94720, USA}
\author{A.~Reisetter} \affiliation{School of Physics \& Astronomy, University of Minnesota, Minneapolis, MN 55455, USA}
\author{R.R.~Ross} \thanks{Deceased}\affiliation{Department of Physics, University of California, Berkeley, CA 94720, USA} \affiliation{Lawrence Berkeley National Laboratory, Berkeley, CA 94720, USA}
\author{T.~Saab} \affiliation{Department of Physics, Stanford University, Stanford, CA 94305, USA}
\author{B.~Sadoulet} \affiliation{Department of Physics, University of California, Berkeley, CA 94720, USA} \affiliation{Lawrence Berkeley National Laboratory, Berkeley, CA 94720, USA}
\author{J.~Sander} \affiliation{Department of Physics, University of California, Santa Barbara, CA 93106, USA}
\author{C.~Savage} \affiliation{Department of Physics, University of California, Santa Barbara, CA 93106, USA}
\author{R.W.~Schnee} \affiliation{Department of Physics, Case Western Reserve University, Cleveland, OH  44106, USA}
\author{D.N.~Seitz} \affiliation{Department of Physics, University of California, Berkeley, CA 94720, USA}
\author{B.~Serfass} \affiliation{Department of Physics, University of California, Berkeley, CA 94720, USA}
\author{K.M.~Sundqvist} \affiliation{Department of Physics, University of California, Berkeley, CA 94720, USA}
\author{J-P.F.~Thompson} \affiliation{Department of Physics, Brown University, Providence, RI 02912, USA}
\author{G.~Wang} \affiliation{Department of Physics, Case Western Reserve University, Cleveland, OH  44106, USA}
\author{S.~Yellin} \affiliation{Department of Physics, Stanford University, Stanford, CA 94305, USA} \affiliation{Department of Physics, University of California, Santa Barbara, CA 93106, USA}
\author{B.A.~Young} \affiliation{Department of Physics, Santa Clara University, Santa Clara, CA 95053, USA}

\collaboration{CDMS Collaboration}
\date{\today}

\begin{abstract}
The Cryogenic Dark Matter Search (CDMS-II) employs low-temperature Ge and Si detectors to seek Weakly Interacting Massive Particles (WIMPs) via their elastic scattering interactions with nuclei. Simultaneous measurements of both ionization and phonon energy provide discrimination against interactions of background particles. For recoil energies above 10~keV, events due to background photons are rejected with $>$~99.99\% efficiency. Electromagnetic events very near the detector surface can mimic nuclear recoils because of reduced charge collection, but these surface events are rejected with $>$~96\% efficiency by using additional information from the phonon pulse shape. Efficient use of active and passive shielding, combined with the the 2090 m.w.e. overburden at the experimental site in the Soudan mine, makes the background from neutrons negligible for this first exposure. All cuts are determined in a blind manner from {\it in situ} calibrations with external radioactive sources without any prior knowledge of the event distribution in the signal region. Resulting efficiencies are known to $\sim$10\%. A single event with a recoil of 64~keV passes all of the cuts and is consistent with the expected misidentification rate of surface-electron recoils. Under the assumptions for a standard dark matter halo, these data exclude previously unexplored parameter space for both spin-independent and spin-dependent WIMP-nucleon elastic scattering. The resulting limit on the spin-independent WIMP-nucleon elastic-scattering cross-section has a minimum of  $4\times10^{-43}$ cm$^{2}$ at a WIMP mass of 60~GeV~c$^{-2}$. The minimum of the limit for the spin-dependent WIMP-neutron elastic-scattering cross-section is $2\times10^{-37}$ cm$^{2}$ at a WIMP mass of 50~GeV~c$^{-2}$. 
\end{abstract}


\maketitle

\section{Introduction}
This paper reports new results from the Cryogenic Dark Matter Search (CDMS-II) experiment, a search for non-luminous, non-baryonic Weakly Interacting Massive Particles (WIMPs)~\cite{lee} that could form the majority of the matter in the universe~\cite{bergstrom,gaitskell04}. The scientific case for WIMPs continues to grow stronger, most recently combining the WMAP results~\cite{WMAP} with studies of large-scale clustering, the Sloan Digital Sky Survey~\cite{SDSS,tegmark}, and supernova redshift data~\cite{paper:perlmutter,paper:riess}.
 
One attractive WIMP candidate is the lightest supersymmetric particle (LSP), which arises naturally in many models of supersymmetry~\cite{jkg,Baer:2003jb} and is stable if R-parity is conserved~\cite{ellis}. Some of these recent models of supersymmetry favor a LSP mass in the range 50--500~GeV~c$^{-2}$.

A reasonable model for the distribution of WIMPs in our own galaxy that is consistent with measurements of spiral galaxy rotation curves~\cite{salucci} is a roughly isothermal spherical halo around our galaxy with a mean velocity of $\sim230~$km~s$^{-1}$.  With this mass range and velocity, the kinetic energy imparted to a nucleus in an elastic WIMP-nucleon scattering event~\cite{goodman,primack} would range from a few keV to tens of keV~\cite{lewin}. The expected event rate ($<$1~event~keV$^{-1}$~kg$^{-1}$~day$^{-1}$) is estimated from previous experimental exclusion limits, inferences from the annihilation cross-section consistent with the present WIMP relic abundance estimate, and the latest supersymmetric models.

The small recoil energy, coupled with the expected low event rate, means that it is vital to suppress backgrounds effectively. Active and passive shielding are used to reduce backgrounds produced outside of the experimental apparatus, leaving decays of radioactive contaminants inside of the shielding as the dominant natural background. The products of these decays all interact electromagnetically, so that it is particularly useful to discriminate between electron-recoil events (most backgrounds) and nuclear-recoil events (WIMPs and background neutrons).

The CDMS-II experiment discriminates nuclear recoils from electron recoils by measuring both the ionization and phonon energies of interactions within Ge and Si detectors. The simultaneous measurements provide a primary discrimination of recoil type, as the ionization signal for nuclear recoils is suppressed relative to electron recoils. The simultaneous measurements also allow an accurate determination of the recoil energy for both nuclear and electron recoils. The pulse shape of the phonon signal gives further discrimination against events that could potentially be misidentified as nuclear recoils because of incomplete charge collection near the detector surface~\cite{tomltd9,shuttthesis}. These events are primarily due to low-energy electrons.
 
The remaining background comes from neutrons, which produce nuclear-recoil events identical to WIMPs and so must be distinguished by other means. First, neutrons often scatter several times within the detector array, while a single WIMP will not. Second, while Ge and Si have similar scattering rates per nucleon for neutrons, the WIMP-nucleon scattering rate is expected to be 5-7 times greater in Ge than in Si for all but the lowest-mass WIMPs. Third, the kinematics of neutron elastic scattering give a recoil energy spectrum scaled in energy by a factor of $\sim2$ in Si compared to Ge, whereas the factor would be $\sim$1 or less for WIMP elastic scattering. The multiple-scatter consideration can be used to identify a fraction of the WIMP candidate events as being uniquely due to neutrons. All three of these methods can be used, in conjunction with Monte Carlo simulations, to statistically subtract any neutron background. 
 
The CDMS collaboration is operating the CDMS-II experiment in the Soudan Underground Laboratory (Minnesota, U.S.A.)~\cite{SoudanMine}. The 780~m (2090~meters water equivalent) of rock overburden reduces the surface muon flux by a factor of $5\times10^4$. In the shallow underground site at Stanford University~\cite{R19prd,R21}, hereafter referred to as SUF, neutrons produced by cosmic-ray muons had become the limiting background of the CDMS-I and -II experiments. With this background reduced by a factor of $\sim300$ at Soudan, the existing CDMS detectors have a greatly improved sensitivity to rare WIMP-scattering events. 

This paper presents results from the first dark matter run taken in the Soudan mine, from October 2003 through January 2004, with the same six detectors previously run at the SUF shallow site~\cite{R21,saab,driscoll}. For the first dark matter run of these six detectors at the Soudan mine, we will describe in this paper the methods and results of the two analyses used. The first analysis~\cite{R18prl} is referred to here as the ``initial" analysis and the second is referred to as the ``current" analysis. Both analyses use identical cuts which were set without examining the WIMP-search data. This blind method of setting cuts guarantees that the cuts are determined without bias from details of the WIMP-search  data. The two analyses of the WIMP-search data differ slightly in the algorithm used to estimate the ionization pulse height (see Sec.~\ref{sec:fitting}). In the initial analysis, the ionization pulses of approximately half of the data were inadvertently analyzed using a time-domain algorithm with poorer resolution in one of the ionization channels (see Sec.~\ref{sec:fitting}). The loss of resolution caused one of the cuts to be unintentionally overly severe (see Sec.~\ref{sec:qinner}), leading to a slight reduction in our exposure (see Sec.~\ref{sec:totaleff}). This initial analysis found no WIMP candidates in the WIMP-search data. The current analysis uses the intended algorithm for analyzing all the ionization pulses. Since the current analysis was conducted after the WIMP-search data had been studied using the initial analysis, we do not consider the results from the current analysis to be blind in a formal sense. We emphasize, however, that since the cuts for the current analysis were set {\it before} examining the WIMP-search data with either analyses described, this later analysis still remains unbiased. The current WIMP search analysis finds one WIMP candidate event, but this is also consistent with the expected background. For either analysis, the lack of a signal rules out a significant new range of WIMP models.

Section~\ref{sec:installation} of this paper describes the CDMS-II experimental apparatus at Soudan, including the cryogenics, shielding, veto, ZIP detectors, readout electronics, and data acquisition systems. Section~\ref{sec:overview} presents an overview of the experimental run, triggering, and calibrations with external sources. Section~\ref{sec:Backgrounds} reports on the neutron, photon, beta and alpha backgrounds observed during the WIMP-search run and comparisons with expectations from Monte Carlo simulations. Section~\ref{sec:analysis} presents the analysis of our data. In that section, we describe our pulse-shape analysis, event reconstruction, WIMP selection cut definitions and efficiencies, estimates of our exposure and backgrounds, and estimates of our systematic uncertainties. Section~\ref{sec:results} presents the results and discusses the importance of our new WIMP-nucleon cross-section exclusion limits. 

\section{The Soudan Installation}
\label{sec:installation}

\subsection{The Soudan Underground Laboratory}
The CDMS-II experiment is housed in one of the two excavated caverns of the Soudan Underground Laboratory, operated by the University of Minnesota. The laboratory is located in the Soudan Underground Mine State Park, operated by the Department of Natural Resources (DNR) of the State of Minnesota. Access to the mine is provided by DNR personnel, and the laboratory technical and management staff is employed by the University of Minnesota. The CDMS infrastructure in the Soudan Underground Laboratory was designed and constructed jointly by Fermilab and the University of Minnesota. The laboratory is situated at a depth of 780~m, and muon flux measurements indicate a water equivalent depth of about 2090~m.

\subsection{Cryogenics}
The CDMS-II detectors operate at a temperature of 50~mK inside a specially made cold volume cooled by an Oxford Instruments Kelvinox 400-S dilution refrigerator. The cryogenic system has followed the same basic design used in the original CDMS low-background facility at Stanford University~\cite{R21,saab,driscoll}, but with many additions to allow automatic and remote control.  These improvements make it possible to maintain a stable and robust detector environment even when physical access to the laboratory is limited or unavailable.

The dilution refrigerator, attached cold volume, and surrounding shielding for CDMS-II are housed in an RF-shielded enclosure at the Soudan Underground Laboratory. The enclosure, known as the RF room, has been measured to be a class-10,000 clean room during working hours, and better than a class-1000 clean room when unoccupied. The pumps, cryogen supplies and cryogenic control systems for the dilution refrigerator are situated outside of the RF room in the ``cryo-pad" area. The detector front-end electronics are in crates inside the RF room, adjacent to an electrical break-out box attached to the detector cold volume. All other readout electronics for the detectors and veto system are housed outside the RF room in an electronics room.

The detector cold volume is the innermost of six nested, cylindrical cans that together make up the CDMS cryostat or ``icebox"~\cite{icebox,pdbthesis}. These cans are made from OFHC, low-radioactivity copper and are individually thermally coupled to the various temperature stages of the dilution refrigerator through a horizontal configuration of five nested copper tubes and one solid cold finger, collectively called the ``cold stem.'' The innermost can, connected to the mixing chamber of the refrigerator by the solid copper finger at the center of the cold stem, is 30~cm in diameter and 30~cm high, and can accommodate up to seven stacks, or ``towers,'' of six ZIP detectors each at $\lesssim$~50~mK. The detector volume is cooled efficiently by conduction through the copper stems but kept away from all cryogenic liquids and any radioactivity in the commercial refrigerator itself. A separate horizontal set of copper stems, called the ``electronics stem'' contains striplines connecting individual detector stacks to the room temperature Front-end Electronics Boards (FEBs) in the RF room.

Only radioactively-screened, low-background materials were used in the fabrication of the icebox. The icebox is surrounded by a 2-mm-thick mu-metal shield at room temperature which screens the detectors and cold electronics from ambient magnetic fields.

A Moore APACS industrial control unit controls much of the cryogenic system. It continuously monitors temperatures, pressures, and flows in the cryogenic system, all of which are stored and displayed via commercial software (Intellution) running on a PC and are also accessible via a web interface. The APACS unit has been programmed via the Intellution software to automatically perform liquid helium and nitrogen transfers, and to respond to certain situations that could otherwise require operator intervention. Whenever any of the measured temperatures, pressures, liquid levels, or flow rates fall outside normal bounds, the operator is contacted by phone and can remotely resolve most problems using a secure computer interface to APACS. A separate system, the Oxford Instruments ``Intelligent Gas Handling'' (IGH) unit, monitors and directs the circulation of the $^{3}$He/$^{4}$He mixture. This system, which uses LabVIEW software,  also supports remote monitoring and operation. Uninterruptible Power Supplies (UPSs) provide stability when electrical power is interrupted, as happens a few times a year at Soudan during summer thunderstorms.

Except for a few brief cryogenic excursions (described in~\cite{mandicthesis}), the six detectors described here were kept at base temperature ($\sim50$~mK) from July of 2003 until August, 2004. During this entire run, the dilution refrigerator suffered from a small leak from the liquid helium reservoir into the insulating (outer) vacuum space. Two turbomolecular pumps continuously evacuated this vacuum space, thereby maintaining an acceptably low pressure for thermal isolation. Nonetheless, helium constantly accumulated on the surfaces at 4~K inside the refrigerator and the icebox. Eventually (about once every month), enough He accumulated to cause a sudden softening of the vacuum (to about $5\times10^{-5}$ torr). This caused a substantial boil-off of liquid helium in a very short time period ($<$1 min.), and temporarily raised the temperature of the detectors as high as 1~K. The leak also increased the average rate of helium boil-off so that two cryogen transfers were required each day.  Detector operation was suspended during these transfers, reducing our detector operation live time fraction. The leak did not expose the detectors to helium, because they were inside a separate, inner vacuum space.

The icebox also contains the cold hardware needed to mount and operate the detectors. Each detector tower has four temperature stages (4~K, 600~mK, 50~mK, base) and contains the wiring and heat-sinking required for six detectors, which are suspended below and are stacked with 2~mm separation between neighboring detectors. The six detectors used in this WIMP-search together with their tower are collectively referred to as ``Tower~I.'' Coaxial wires passing through the tower connect each detector to its own cold electronics card mounted on top of the tower. This card contains two Field Effect Transistors (FETs) for the readout of the two ionization channels. The FETs are weakly heat-sunk to the 4-K stage, and self-heat to 130~K for optimal noise performance. A separate card at 600~mK contains the four arrays of Superconducting Quantum Interference Devices (SQUIDs) required for the readout of each detector's four phonon channels. The signals are carried to room temperature through flexible, 2.5-cm-wide, 3-m-long copper-kapton striplines.  These are heat-sunk at 4~K and~77 K, and pass through a copper radiation shield as they proceed through the electronics stem into a breakout box at room temperature. All these fixtures have been designed to minimize the heat conducted and radiated into the cold stages, and to allow the FETs to operate with suitably low noise.  The cold hardware is constructed from radioactively-screened, low-background materials, in particular copper, kapton, and custom low-activity solder~\cite{cdms_solder}.

\begin{figure*}
\includegraphics[scale=1]{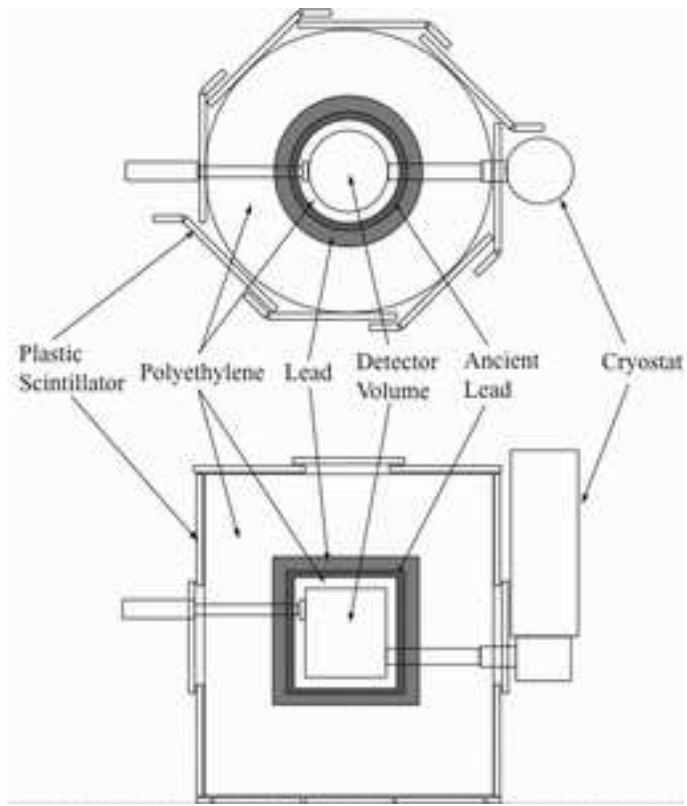}
\caption{Top view and side view of the CDMS-II shielding and veto. The detector volume is referred to as the ``icebox.'' As shown, the stem to the right of the detector volume is the ``cold stem'' and connects the detectors and the copper cans to the cryostat. The stem to the left of the detector volume is the ``electronics stem'' and contains the wiring that connects the cold electronics to the room temperature electronics.}
\label{fig:shieldandveto}
\end{figure*}

\subsection{Shielding and Muon Veto}
\label{sec:shielding}

To minimize the number of interactions that could imitate nuclear recoils and to decrease the overall background rate, the CDMS-II experiment employs several layers of passive shielding as well as an active muon veto. Backgrounds that need to be shielded against include high-energy neutrons produced by cosmic ray muons interacting with the surrounding rock or shield; gamma-rays and neutrons from radioactivity in the surrounding rock; and photons, electrons, and alphas from radioactive impurities on surfaces within the lead shielding.

The CDMS-II experiment at the Soudan Underground Laboratory is situated beneath an overburden of 2090 meters water equivalent (m.w.e.). This overburden reduces the surface muon flux by a factor of $5 \times10^{4}$. Within the RF room, the layers of shielding are arranged concentrically around the icebox (see Fig.~\ref{fig:shieldandveto}). Outermost is an active muon veto which tags the remaining muon flux. 

Forty scintillator panels surrounding the passive shielding and icebox compose the active muon veto system. Each panel consists of a 5-cm-thick slab of Bicron~BC-408 plastic connected to one or two 2-inch Hamamatsu R329-02 photomultiplier tubes. The veto panels are arranged so that adjacent panels overlap. The top panels extend well beyond the side panels, guaranteeing that the small gap between top and side panels does not allow a direct line of sight to the detectors inside the icebox. 

The veto system can distinguish between muons and ambient photons, since minimum ionizing muons typically deposit 10~MeV while the highest energy photon from natural radioactivity present is 2.6~MeV.  The detection efficiency of a veto paddle approaches 100\% for $\geq$ 2~MeV energy deposition. The {\it in situ} measured muon detection efficiency for the entire veto system is $99.4 \pm 0.2$\% for stopped muons and $99.98 \pm 0.02$\% for through-going muons, where the low muon rate limits the precision of this measurement. On average, one muon per minute is incident on the veto, and the combined veto rate (dominated by ambient gammas) is $\sim600$ Hz. Events with detector signals are recorded regardless of whether or not there was veto activity, but a subsequent analysis cut rejects events whose veto activity preceded the detector signal by less than 50~$\mu$s. A veto trigger on events for which multiple scintillator panels have coincident activity is highly efficient at identifying incident muons. 

The veto calibration is checked twice each day. Since there is no significant rate of muons, we calibrate the veto using light from a bank of blue Light Emitting Diodes (LEDs) fed via optical fibers to each veto panel. The light from the LED bank is tuned to match the scintillation light created in each panel from a muon. Additionally, we continually monitor the veto trigger rate and the voltage and current supplied to each veto panel's photomultiplier tube. We found no significant variation in the performance of the veto system over the course of the run.

As depicted in Fig.~\ref{fig:shieldandveto}, within the volume enclosed by the muon veto panels, a 40~cm thick cylindrical outer polyethylene layer first moderates low-energy neutrons from radioactive decays to below-threshold energies. Inside this polyethylene lies a 22.5~cm thick cylindrical lead shield, of which the inner 4.5~cm thickness consists of ancient lead~\cite{nantes}. The lead shielding is constructed in a modular way, with gaps that do not align with the detectors. An inner 10~cm thick cylindrical polyethylene layer provides further neutron moderation. The icebox cans and cold hardware provide an average shielding thickness of about 3~cm of copper directly surrounding the detectors. The shield is 99\% hermetic, with the only penetrations being the cold stem and the electrical stem.

Beginning on November 11, 2003 the air volume between the outermost icebox copper can and the mu-metal shield has been continuously purged to dramatically reduce the concentration of environmental radon in the vicinity of the detectors. The purge gas used is medical grade breathing air that has been stored in metal cylinders for at least two weeks, to allow most of the radon gas to decay. 

\subsection{CDMS ZIP Detectors}
\label{section:zips}

Each CDMS ZIP ($Z$-dependent Ionization- and Phonon-mediated) detector is a cylindrical high-purity Ge or Si crystal that is 1 cm thick and 7.6~cm  in diameter.  A single Ge (Si) ZIP has a mass of 250~g (100~g).  Two concentric ionization electrodes and four independent phonon sensors are photolithographically patterned onto each crystal.

The data described here were obtained with a single tower of six ZIP detectors. Within a tower, the six ZIP detectors are stacked 2~mm apart with no intervening material. This close packing not only shields the detectors from low-energy electron sources on surrounding surfaces but also increases the probability that a background event in one detector would multiply-scatter into another detector. Division of the electrodes into an annular outer ``guard'' electrode and a disk-shaped inner electrode defines an inner fiducial region that is further shielded from low-energy electron sources or x-ray fluorescence.

An external particle scattering in a ZIP detector can interact with an electron (or electrons) in the crystal (\eg\ by Compton scattering, K-capture, etc., any of which we say causes an ``electron recoil''), or with a nucleus (called a ``nuclear recoil''). The interaction deposits energy into the crystal through charge excitations (electron-hole pairs) and lattice vibrations (phonons). A CDMS ZIP detector measures both the ionization and the phonon energy for every event. The simultaneous ionization and phonon measurement not only allows an accurate measurement of the recoil energy independent of recoil type,  but also distinguishes between these two types of recoils.

Depending on the material and the type of recoil,  6\% -- 33\% of the recoil energy is first converted into ionization before subsequent conversion to phonons. On average, one electron-hole pair is produced for every $\epsilon\approx3.0$~eV (3.8~eV) of energy deposited by an electron recoil in Ge (in Si). The ``ionization energy,'' $E_{\mathrm{Q}}$, is defined for convenience as the recoil energy inferred from the detected number of charge pairs, $N_{\mathrm{Q}}$, by assuming that the event is an electron recoil with 100\% charge-collection efficiency:
\begin{equation}
E_{\mathrm{Q}} \equiv N_{\mathrm{Q}} \times \epsilon . 
\label{eqn:ionizationenergy}
\end{equation}
Ionization energy is usually reported in units of ``keVee,'' keV of the equivalent electron recoil determined from electron-recoil calibration measurements. The dimensionless ionization yield parameter, $y$, is the ratio of ionization energy to true recoil energy, $y \equiv E_{\mathrm{Q}} / E_{\mathrm{R}}$, where $E_{\mathrm{R}}$ is the full recoil energy for the event. This definition of yield gives unity for electron-recoil events with complete charge collection measurement.

\begin{figure}
\includegraphics[scale=1]{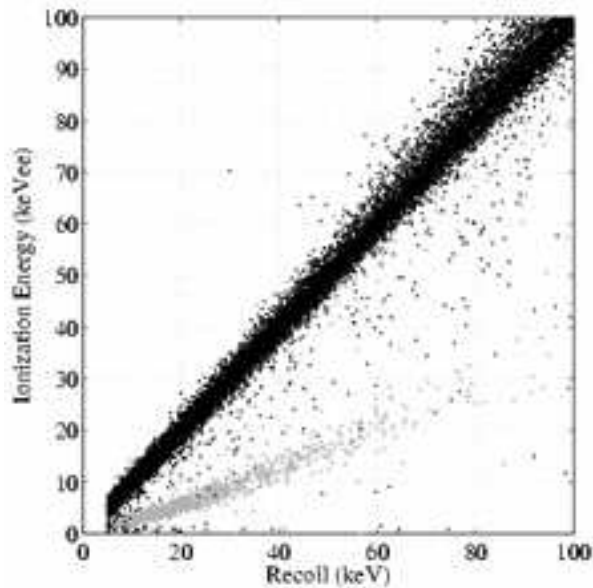}
\caption{Ionization energy versus recoil energy for detector Z5 (Ge). Black dots correspond to calibration events from a $^{133}$Ba source (emits gammas only) and gray dots correspond to calibration events from a $^{252}$Cf source (emits gammas and neutrons).}
\label{fig:2Dplot}
\end{figure}

\begin{figure}
\includegraphics[scale=1]{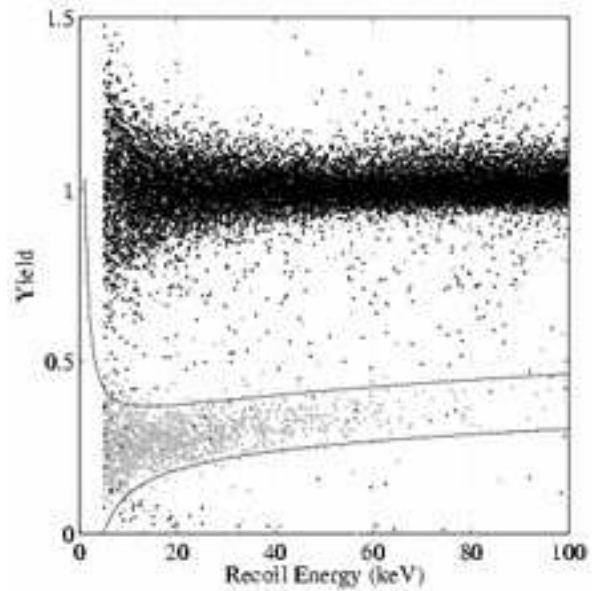}
\caption{Ionization yield versus recoil for detector Z5 (Ge). Black dots correspond to calibration events from a $^{133}$Ba source (emits gammas only) and gray dots correspond to calibration events from a $^{252}$Cf source (emits gammas and neutrons). The upper distribution of events are bulk electron recoils which define  the ``electron-recoil band.'' The lower distribution of events are nuclear recoils which define the ``nuclear-recoil band.''}
\label{fig:yield}
\end{figure}

Nuclear recoils produce fewer charge pairs, and hence less ionization energy, $E_{\mathrm{Q}}$, than do electron recoils of the same recoil energy (see Figs.~\ref{fig:2Dplot} and \ref{fig:yield}). The ionization yield $y$ for nuclear-recoil events depends on both the material and the recoil energy, with $y \sim0.3$ ($y \sim0.25$) in Ge (in Si) for $E_{\mathrm{R}} \agt 20$~keV. The simultaneous measurement of ionization and recoil energy therefore makes it possible to identify and reject most electron-recoil background events.  

\subsubsection{The Ionization Measurement}
\label{section:ionization}

\begin{figure*}
\includegraphics[scale=1]{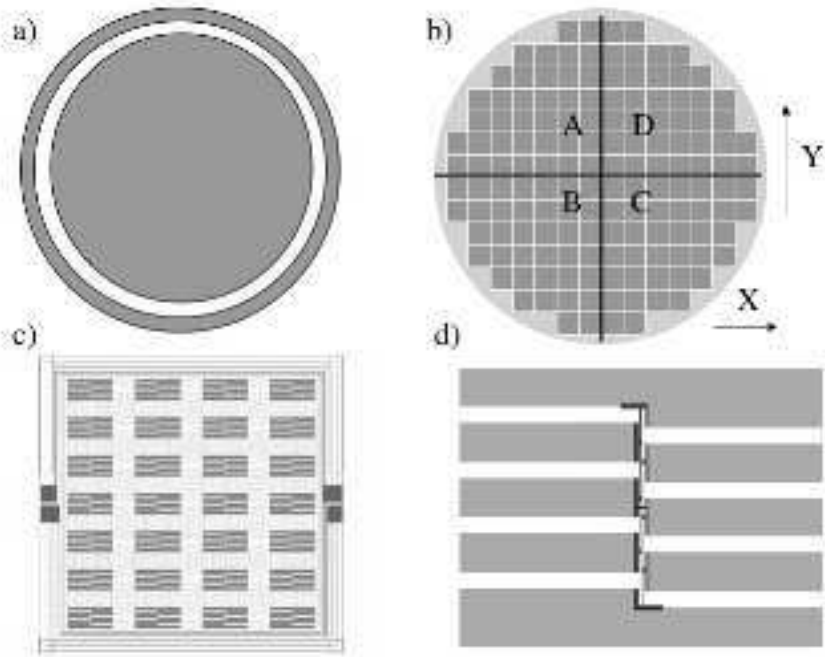}
\caption{Diagram of a CDMS ZIP detector. a) The ``ionization side'' of the detector with a large inner electrode and an outer guard ring electrode. b) The ``phonon side,'' divided into four quadrants labeled A, B, C, and D, each consisting of 37 dies of 28 QETs. The convention for the x-y axes is shown. The area outside the cells consists of a passive Al/W grid that is patterned sparsely (10\,\% area coverage) to minimize athermal-phonon absorption while maintaining field uniformity for the ionization measurement. c) One of the 37 dies constituting a single phonon channel; each die contains 28~QETs. d) One of the QETs consisting of a 1-$\mu$m-wide tungsten strip connected to 8 aluminum fins.}
\label{fig:ZIPcartoon}
\end{figure*}

On each ZIP detector, metal electrodes on the two faces of the crystal substrate serve as the sensors for the ionization measurement.  As shown in Fig.~\ref{fig:ZIPcartoon}, the charge electrodes are divided into an inner fiducial electrode and an outer guard ring. The disk-shaped inner electrode has a diameter of 69~mm and so covers 82--85\% of the ionization side of the detector. The annular outer guard ring is 2.0--2.7~mm wide, and a 1-mm-wide annular gap lies between the two electrodes.  The guard ring is used to reject events occurring near the bare unpolished edges of the crystal. We reject events near these edges for three reasons: the ionization signal may be degraded since the electric field is not uniform in this region, the phonon response is worse since this region is not well covered by the phonon sensors, and background interactions are more likely due to the absence of self-shielding by other detectors in the stack.

The electrodes are used to apply an electric field of a few volts/cm across the crystal, which drifts electrons and holes produced by an interaction to the detector faces. The applied electric field needs to be relatively low in order not to contaminate the phonon signal with too many Neganov-Trofimov-Luke phonons (see Sec.~\ref{section:phonons}), which would compromise the detector's ability to discriminate between electron-recoil and nuclear-recoil events. The number of charges collected on the electrodes, $N_{\mathrm{Q}}$, is proportional to the number of electron-hole pairs produced by the interaction weighted by the drift distance of each of the charges~\cite{R19prd,coldelect}. In the case where all of the electrons and holes drift across the crystal, $N_{\mathrm{Q}}$ (see Eq.~\eqref{eqn:ionizationenergy}) equals the total ionization. The prompt recombination of the drifted electrons and holes in the electrodes releases all of the recoil energy from the electron system into the phonon system.

There are two cases for which the measured charge, $N_{\mathrm{Q}}$, underestimates the total ionization.  In both of these cases, the charges fail to drift across the entire crystal.  The first case is due to poor charge-space neutralization within the crystals. At the low operating temperatures required for the phonon sensors, impurity sites can be left with a net charge. These charged impurity sites can trap the drifting electrons (or holes) from a recoil event. We neutralize the crystals routinely by flashing LEDs, whose photons generate excess electron-hole pairs in the crystal. These excitations can then neutralize ionized impurity sites, reducing their trapping cross-section by several orders of magnitude. The Si crystals contain more impurities than the Ge, and the LED photon energy spectrum is optimized for neutralizing Ge, thus neutralizing the Si detectors takes more flashing cycles than for Ge. For the data run described here, the LED flashing regime was not always sufficient for the Si detector Z6, which suffered incomplete charge collection for a sizable fraction of the data taken. The second case of underestimating the ionization corresponds to events occurring close to one of the charge electrodes. For these events, the relatively low applied electric drift field and self-screening from the initial electron-hole cloud enables some of the electrons or holes to drift into the ``incorrect'' electrode.  By depositing a thin ($\sim$40~nm) layer of lightly doped amorphous silicon between each electrode and the detector surface, this back-diffusion is substantially reduced~\cite{tomltd8, tomltd9}. Interactions within the 10~$\mu$m ``dead layer'' still have deficient charge collection and are referred to as ``surface events.'' As discussed in the following section, information from the athermal phonon measurement makes it possible to identify and reject these surface events.

\subsubsection{The Phonon Measurement}
\label{section:phonons}
A total of 4144 Quasiparticle-assisted Electrothermal-feedback Transition-edge sensors (QETs)~\cite{irwin2} photolithographically patterned onto one of the crystal faces form the phonon sensors for a CDMS ZIP. The QETs are divided into four independent channels, each consisting of 1,036 QETs operated in parallel.  Each QET consists of a 1-$\mu$m-wide strip of tungsten (35~nm thick) connected to eight superconducting aluminum collection fins (300~nm thick), each roughly 380~$\mu$m $\times~55$~$\mu$m, as shown in Fig.~\ref{fig:ZIPcartoon}.

The narrow tungsten strips form the Transition-Edge Sensors (TESs), which are voltage biased, with the current through them monitored by a high-bandwidth SQUID array~\cite{squid}. The tungsten is maintained stably within its superconducting to normal transition by electrothermal feedback based on Joule self-heating: if the sensor were hotter, the resistance would increase, decreasing the current and the Joule heating; an analogous argument applies if the sensor were cooler. To ensure operation in the extreme feedback limit, the substrate is kept much colder ($T\ltwid 50$~mK) than the superconducting-to-normal transition temperature of the tungsten sensors ($T_{\mathrm{c}} \sim80 $~mK). Energy deposited in the tungsten electron system raises the temperature of the film, increasing its resistance and reducing the current.

Most phonons in the crystal that reach the aluminum fins on the surface scatter into the aluminum, creating quasiparticle excitations.  These quasiparticles diffuse through the aluminum and later enter the tungsten TES. Electron-electron interactions between the quasiparticles and the tungsten conduction electrons~\cite{Cabrera:1996mc} cause an increase in the temperature of the tungsten electron system. Strong electrothermal feedback guarantees that the power deposited into the TES is exactly compensated for by a reduction in Joule heating.  Thus, for a voltage bias across the TES, $V_{TES}$, and a TES current decrement $\Delta I$, the total energy deposited, $E_{TES}$,  is just 
\begin{equation}
E_{TES} =V_{TES}\int \Delta I dt \mathrm{.}
\end{equation}

The energy in the phonon system includes not only the full recoil energy, $E_{R}$, but also energy from drifting the electrons and holes across the crystal.  The work done in drifting the electrons and holes manifests itself as Neganov-Trofimov-Luke phonons~\cite{neganov,luke}. The Neganov-Trofimov-Luke phonons contribute to the total phonon energy, $E_{P}$, yielding
\begin{equation}
E_{\mathrm{P}} = E_{\mathrm{R}} + e\sum_{q} Ed_{q}\mathrm{,}
\label{eqn:Ep1}
\end{equation}
where the sum is taken over all charges, $E_{\mathrm{R}}$ is the recoil energy, $E$ is the electric field, and $d_{q}$ is the distance that each charge drifts. Since the electric field is constant throughout most of the detector volume, we can rewrite Eq.~\eqref{eqn:Ep1} as
\begin{equation}
E_{\mathrm{P}} = E_{\mathrm{R}} + e V_{\mathrm{b}} \sum_{q} \frac{d_{q}}{d_{0}}\mathrm{,}
\label{eqn:Ep2}
\end{equation}
where $d_{0}$ is the thickness of the crystal and $V_{b} = E d_{0}$ is the bias voltage between the electrodes. The summation in Eq.~\eqref{eqn:Ep2} is the number of charges weighted by their drift distances and is equal to the charge collected on the electrodes, $N_{\mathrm{Q}}$. The definition of ionization energy given by Eq.~\eqref{eqn:ionizationenergy} gives 
\begin{equation}
E_{\mathrm{P}} = E_{\mathrm{R}} + e V_{\mathrm{b}} N_{\mathrm{Q}} 
               = E_{\mathrm{R}} + \frac{e V_{\mathrm{b}}}{\epsilon}
                 E_{\mathrm{Q}} \textrm{,}
\label{eqn:phononenergy}
\end{equation}
where $V_{\mathrm{b}}$ is the bias voltage across the detector. Equation~\eqref{eqn:phononenergy} is valid even for events with incomplete charge collection (due, for example, to trapping or recombination in the wrong electrode). Since we calibrate electron recoils with full charge collection to have $E_{\mathrm{Q}} = E_{\mathrm{R}}$, $E_{\mathrm{P}} = \left( 1 + e V_{\mathrm{b}}/{\epsilon} \right) E_{\mathrm{R}}$ for these events. Calibration of the detectors at several bias voltages using photon sources confirms that $\epsilon \approx 3$~eV (3.8~eV) in Ge (in Si). For electron recoils with full charge collection in Ge at 3~V bias (the bias voltage for most of the data described here), $E_{\mathrm{P}} = 2E_{\mathrm{R}}$. In practice, the recoil energy $E_{\mathrm{R}}$ of an event is inferred from measurements of the phonon and ionization energies:
\begin{equation}
E_{\mathrm{R}} = E_{\mathrm{P}} - \frac{e V_{\mathrm{b}}}{\epsilon}
                 E_{\mathrm{Q}}  .
\label{eq:recoilenergy}
\end{equation}
Equation~\eqref{eq:recoilenergy} is valid for all events independent of recoil type and the ionization collection efficiency.

One of the advantages of measuring the athermal phonon signal is that it provides sensitivity to phonon physics that is dependent on the nature of the interaction and the event location. This sensitivity makes it worthwhile to discuss the phonon dynamics within the crystal.

An interaction in the crystal produces high-frequency phonons which propagate quasi-diffusively through the crystal~\cite{tamura,paper:tamura1}.  Quasi-diffusive propagation arises from the combination of two scattering processes: elastic scattering, which mixes phonon modes, and anharmonic decay, which increases the total number of phonons while reducing the average phonon frequency.  The net effect of the quasi-diffuse propagation is a phonon ``ball'' expanding at roughly $1/3$ the speed of sound.  When the frequency of a phonon is sufficiently low ($<$ 1~THz), the mean free path becomes comparable to the size of the detector.  Such phonons are termed `ballistic' and travel at the speed of sound (5~mm~$\mu$s$^{-1}$ for Ge and 2.5~mm~$\mu$s$^{-1}$ for Si) through the crystal. They rarely scatter within the bulk, but reflect diffusively at the crystal surfaces, unless they are absorbed by metal films on the surface. Eventually the quasi-diffuse propagation converts all of the initial high-frequency phonons into ballistic phonons, which then eventually thermalize.

In the absence of an electric field through the crystal, no discernible difference in the phonon signal coming from electron recoils versus nuclear recoils has yet been demonstrated~\cite{paper:alee}. However, during normal detector operation an electric field is present. In the first few hundred nanoseconds after an interaction, the freed electron-hole pairs drift across the entire crystal, shedding ballistic Neganov-Trofimov-Luke phonons and producing additional ballistic phonons upon relaxation at the electrodes~\cite{clarkethesis,clarkeapl, ltd5}. Since an electron recoil produces more charges than a nuclear recoil, these processes lead to a larger initial population of ballistic phonons for electron recoils. Ballistic phonons propagate at the speed of sound, while the energy flux of high-frequency phonons moves at approximately one third the speed of sound. This difference in the speed of propagation leads to a faster phonon leading edge for electron recoils when compared to nuclear recoils because of the larger ballistic fraction. 

\begin{figure}
\includegraphics[scale=0.6]{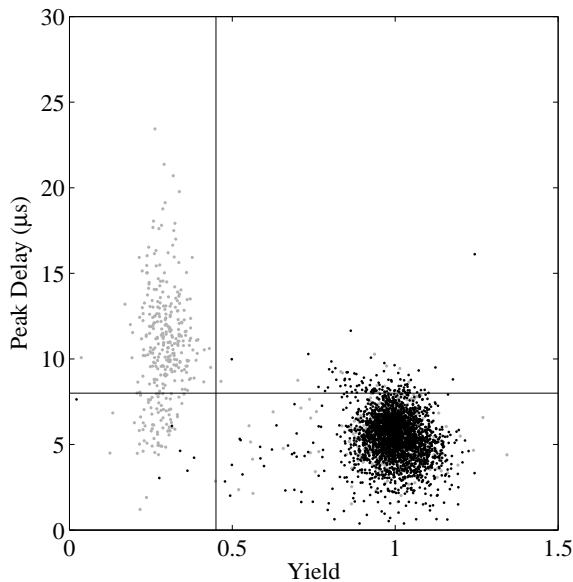}
\caption{Peak phonon delay versus ionization yield for detector Z5. Events from a $^{252}$Cf source (emits both gammas and neutrons) are shown as gray dots. Events from a $^{133}$Ba source (emits gammas only) are shown as black dots. Surface-electron recoils from the $^{133}$Ba calibration appear  as a low-yield tail and having smaller peak phonon delays. Lines for a typical cut excluding events with high yield or low peak phonon delay are shown.}
\label{fig:YvsRT}
\end{figure}

There is an additional increase in the fraction of ballistic phonons arising from the down-conversion of the high-frequency phonons from interactions with metal on the crystal surface~\cite{youngPRL}. When electrons and holes relax to the Fermi surface in the metal electrodes, most of the released energy is rapidly down-converted to a largely ballistic population of phonons. This effect becomes especially significant for the special case where events are very close ($<$ 1 mm) to the detector surface. For these events, the expanding phonon ball is sufficiently close to the metal on the surface that a substantial fraction of the initial high-frequency phonons are down-converted, leading to a third population of prompt ballistic phonons while reducing the population of high-frequency phonons. All of these processes produce ballistic phonons much more rapidly than the down conversion from quasi-diffuse propagation. For such surface events, the phonon signal will rise even faster than for bulk electron recoils. Figure~\ref{fig:YvsRT} illustrates these effects in data.

The measurement of athermal phonons makes it possible to reject events in the dead layer by analysis of the phonon pulse shape.  Additionally, it is possible to reconstruct the interaction location by combining the information from the four independent phonon channels.

\subsection{Data Acquisition Electronics}
\label{sec:daqelectronics}

\begin{figure*}[ht!]
\includegraphics[scale=1]{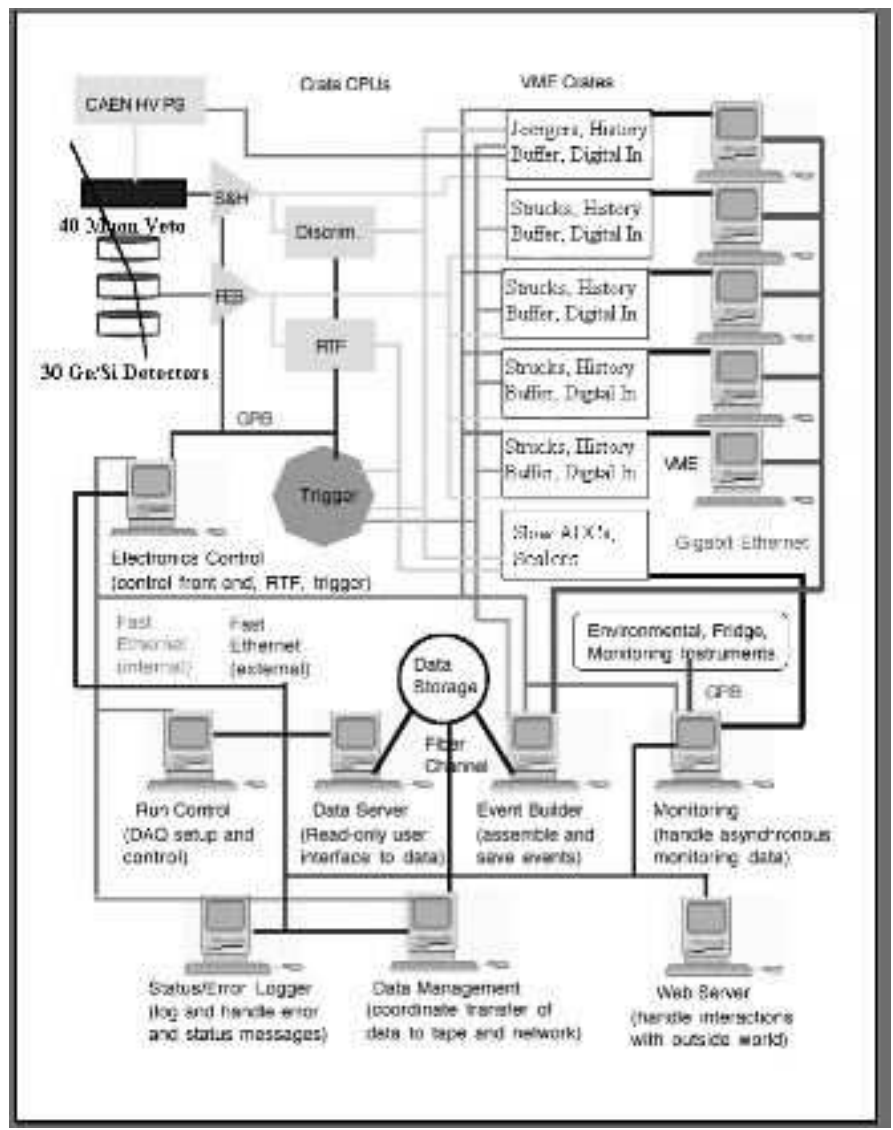}
\caption{CDMS-II data acquisition layout showing the various hardware components in relation to the detector and veto systems. When a detector triggers, various detector and veto elements are read out to form an event. Monitoring processes run continuously to monitor the health of the detector, veto, and cryogenics components.}
\label{fig:daq_layout}
\end{figure*}

Custom-made electronics, the Front-end Electronics Boards (FEBs), are located in 9U crates near the vacuum bulkhead where the striplines from the cold electronics terminate. The back-plane of these crates define the ``star-ground'' for the entire electronics chain and is carefully designed to minimize noise pickup. The FEBs control detector settings, such as bias voltages for both the ionization and phonon measurements. They also contain LED flashing driver electronics, and amplifiers for the readout of the cold electronics' FETs and SQUIDs. The FEBs are programmed by GPIB commands from the Data Acquisition system, as indicated in Fig.~\ref{fig:daq_layout}.

The data acquisition (DAQ) electronics, shown schematically in Fig.~\ref{fig:daq_layout}, records three streams of data. Fast waveform digitizers (ADCs) sample the analog signals and provide time-trace information for both ZIP detectors and veto paddles. In addition, digital time histories are recorded for veto paddle and detector channel triggers. Finally monitoring devices (slow-sampling ADCs and scalers) record trigger thresholds, detector signal offsets, and triggering rates.

Struck SIS~3301 digitizers are used to record the ZIP detector signals. These sample natively at 80~MHz, but we average every 64 samples to achieve a sampling rate of 1.25~MHz, which exceeds the signal bandwidth of interest. This sampling averaging gives a measured accuracy of 14 bits. We take 2048 samples for each of the 6 channels (2 charge and 4 phonon) from each ZIP detector. For the faster scintillation pulses from the muon veto panels, a custom-built circuit stretches the pulses from the PMTs to $\sim\mu$s widths. The signals are then waveform-sampled with Joerger VTR812 digitizers, operating at 5~MHz and taking 1024 samples. 

The trigger information for the detector analog signals is determined by custom-built Receiver-Trigger-Filter (RTF) boards, which filter the detector analog signals, sum the four phonon and sum the two charge channels, and compare the results with user-set thresholds. The digital output from the RTF boards is passed to custom-built Trigger-Conditioner Boards (TCBs), which condition the signal. LeCroy discriminators compare the veto pulses with user-set thresholds (typically below 1~MeV) to generate the digital history for the veto system. Struck SIS~3400 time history buffer modules store the digital trigger information for both the detector and veto systems with 1 microsecond resolution.

A custom-built Trigger Logic Board (TLB), which processes the digital trigger signals for both the detector and veto system, determines whether the signal satisfies the conditions for a global trigger. There are two possible conditions for which the TLB issues a global trigger. The first condition is when there is a signal in more than one veto panel. This condition is designed to issue a trigger for muons passing through the veto. The second condition is when the summed phonon signal of any one detector exceeds a user-set threshold. This condition triggers on events in the ZIP detectors. Since the threshold for this criterion is fixed throughout the experiment, changing phonon pulse-shapes (see Sec.~\ref{sec:lookuptable}) and amplitudes lead to variations in the location of the global trigger in the analog time-stream. For example, events with larger amplitudes (higher energy) have global triggers that are earlier than events with smaller amplitudes (lower energy). Likewise, faster rising events have earlier global triggers than slower ones. This delay between the interaction time and the global trigger can extend up to tens of microseconds.

\subsection{Data Acquisition Software}
The CDMS-II data acquisition software provides control, monitoring, and data recording for the experiment. It consists of a number of independent servers written in JAVA and C++ which communicate via the JAVA Remote Method Invocation (RMI) and CORBA technologies. The use of these technologies provides a scalable system with inherent remote accessibility. There are three important features of the software: high bandwidth, automation, and remote interfacing. Figure~\ref{fig:daq_layout} depicts this software system and its interface with the DAQ electronics.

Though the expected event rate is low at the Soudan mine, the CDMS-II data acquisition software is capable of processing data at a high event rate. Whenever a global trigger is generated, crate-based servers read out each of the VME crates that contain the digitizers and history buffers. This information is combined and recorded to disk, after which the digitizers and Trigger Logic Board are rearmed. The DAQ dead time is about 20~ms, allowing us to take data at 50~Hz. This bandwidth provides the means to acquire calibration data at rates much higher than our WIMP-search data (0.1 Hz).

During standard running most routine operations, such as configuration of the detector and veto electronics, periodic detector neutralization, fast veto calibration with the LED pulser system, data copying, and data archiving, are automatically performed by the data acquisition system. Additionally, if the experimental configuration deviates from the specified parameters, the system automatically pauses and reconfigures the experiment. This automation minimizes user intervention and maintains a stable experimental configuration throughout the entire WIMP search. 

The use of JAVA makes remote control and monitoring an intrinsic part of the data acquisition system. This feature is useful since physical access to the underground laboratory is sometimes limited. By using a thin platform-independent client, any user can monitor the CDMS-II experiment regardless of their location. The system broadcasts information regarding the trigger rates, cryogenics, high voltage, and experimental configuration through data channels so that remote clients can access them. In addition to the remote monitoring of the experiment, a fast analysis of the output data, based on the analog and digital information from the detector and veto systems, is performed to produce various diagnostic plots, which are constantly updated and available on the web. Near real-time diagnostic plots from the online data analysis are available for review of data quality, including some automated features alerting operators to unusual performance of the apparatus.

\section{The First Soudan Data Run}
\label{sec:overview}
\subsection{Run Overview}
\label{runoverview}
\begin{figure}[!thb]
\centering
\includegraphics[scale=0.6]{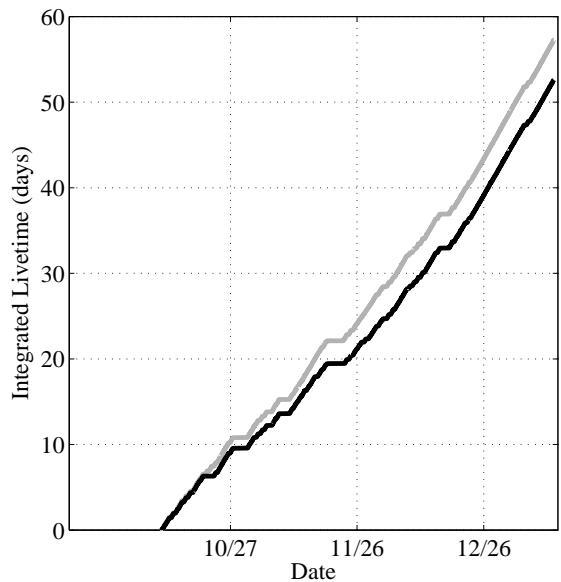}
\caption{Live-time before (gray) and after (black) application of the data-quality cut described in Sec.~\ref{sec:dataquality}. The short regions of down-time correspond to episodes of electronics problems and unintended detector warming above 1~K.}
\label{R118_livetime}
\end{figure}

The first Soudan WIMP-search run started on October 11, 2003 and lasted until January 11, 2004. As shown in Fig.~\ref{R118_livetime}, we acquired 52.6~live-days of WIMP-search data with Tower~1 detectors: Z1 (Ge), Z2 (Ge), Z3 (Ge), Z4 (Si), Z5 (Ge), and Z6 (Si). The same tower, with identical detector configuration, had been used in a WIMP-search run~\cite{R21,saab,driscoll} at the shallow site at Stanford University. The Soudan run was interrupted by brief neutron calibrations with a ${\rm ^{252}Cf}$ source on November 25, 2003, December 19, 2003, and January 5, 2004. In addition, several extensive gamma calibrations with ${\rm ^{133}Ba}$ sources were performed. Most of them were taken in the first half of December 2003, with occasional earlier and later calibration runs for checks on stability. 

Daily maintenance of the experiment, including cryogen transfers into the cryostat, LED flashing periods that ensure detector neutralization, etc., were performed roughly every 12 hours, resulting in approximately 2-3 hours per day of scheduled down-time. Occasional difficulties with cryogenics, electronics, or the noise environment caused additional periods of down-time, as indicated in Fig.~\ref{R118_livetime} and detailed in~\cite{mandicthesis}.

\subsection{Triggering}
\begin{figure}[!thb]
\centering
\includegraphics[scale=.6]{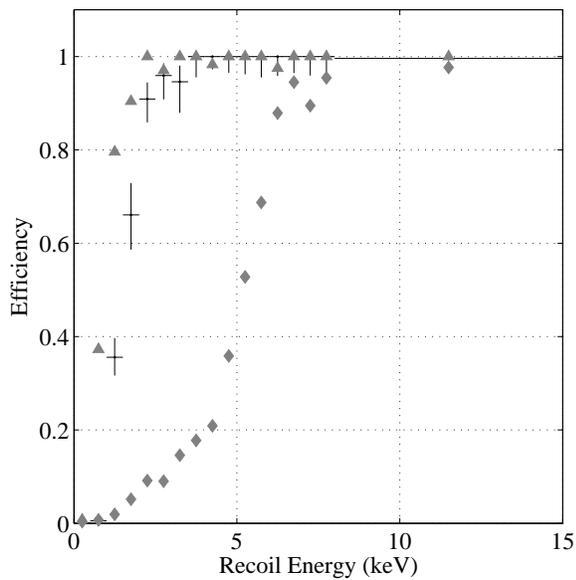}
\caption{Phonon trigger efficiencies for some of the ZIP detectors. Black dots with error bars show the trigger efficiency for the detector Z3 (Ge). The other points are plotted without error bars to avoid confusion. Triangles correspond to the trigger efficiency for Z5 (Ge), and diamonds correspond to the efficiency of Z4 (Si) which has the worst efficiency of all the Tower~I ZIPs.}
\label{R118_trigeff}
\end{figure}

Global triggers for the experiment were issued whenever the sum of the four phonon traces of any detector exceeded a certain threshold. As described in Sec.~\ref{sec:daqelectronics}, the TLB determines whether to issue a global trigger by analyzing the digital output of all of the RTF boards and the veto system. This output is recorded in the digital history buffer. To determine the trigger efficiency for a given detector, we must first preselect events by requiring that the global trigger not be issued by that detector. The detector's trigger efficiency is the fraction of these preselected events for which the history buffer records that the detector's RTF board finds the phonon signal to be above threshold within 100 $\mu$s after the global trigger. We calculate the trigger efficiency using all of the WIMP-search data. For most of the detectors the trigger efficiency reaches 100\% between 5--10~keV true recoil energy, and for some detectors (such as Z3 in Fig.~\ref{R118_trigeff}) even lower, 3--5~keV. To achieve these efficiencies, the trigger thresholds were set low, such that $\sim30$\% of the triggers were caused by noise.

\subsection{Calibrations with External Sources}
\label{sec:calibrations}
We took calibration data on several occasions throughout the WIMP search using external radioactive sources. A $^{133}$Ba source served to characterize the detector response to electron recoils. To avoid the Pb in the passive shield, the source was inserted through a special tube along the electronics stem of the icebox (see Fig.~\ref{fig:shieldandveto}), which penetrates the passive shield. This particular isotope offers several distinct lines: 276~keV, 303~keV, 356~keV, and 384~keV. These lines are sufficiently energetic that the photons can penetrate the copper cans of the cryostat and reach the detectors. However, only the ionization channels have linear response in this high-energy region. The phonon channels typically become significantly non-linear above 200~keV. Hence, we use these lines to calibrate the ionization channels, and then calibrate the phonon channels against the ionization for electron recoils. 

\begin{figure}
\includegraphics[scale=0.6]{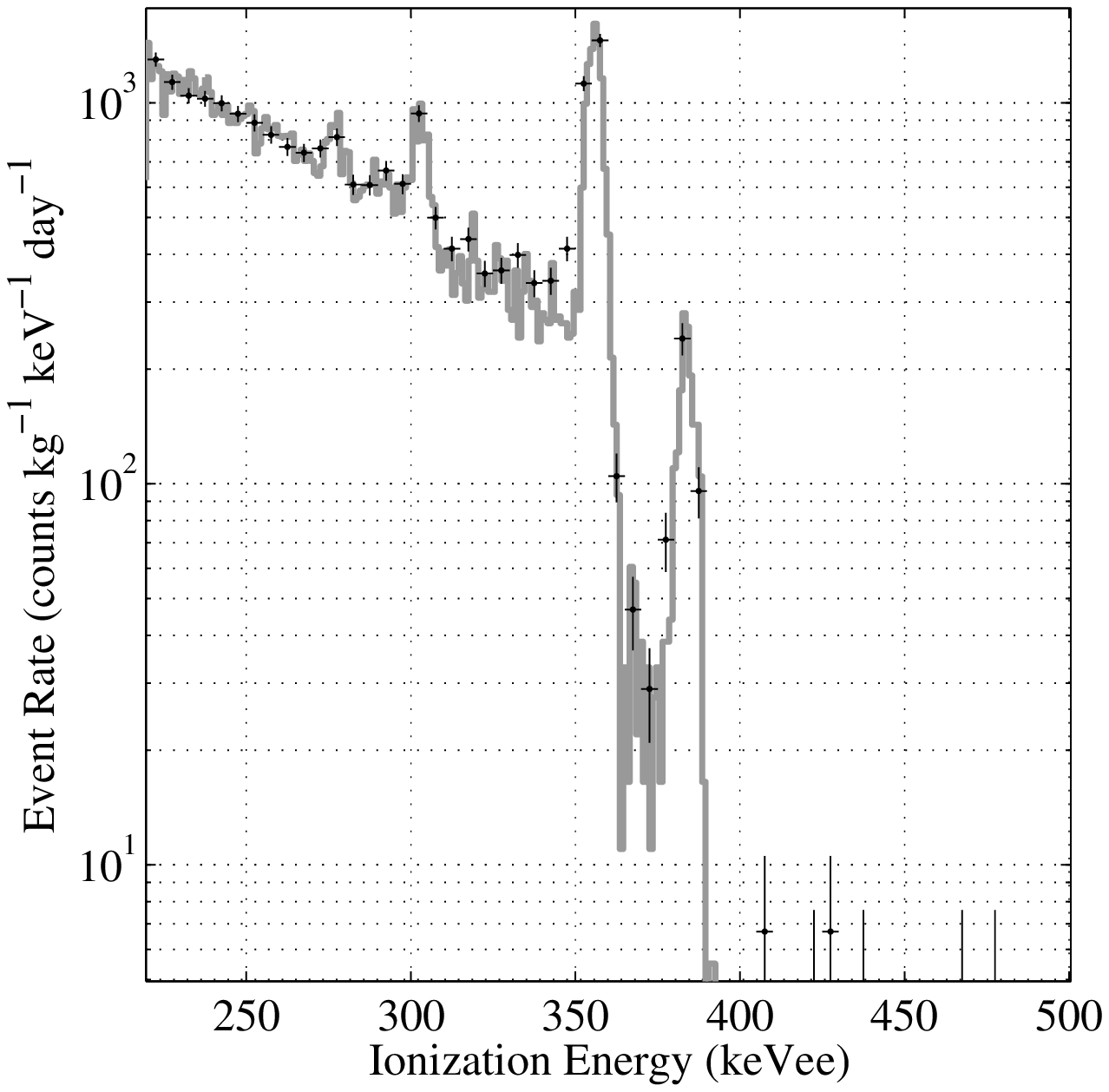}
\includegraphics[scale=0.6]{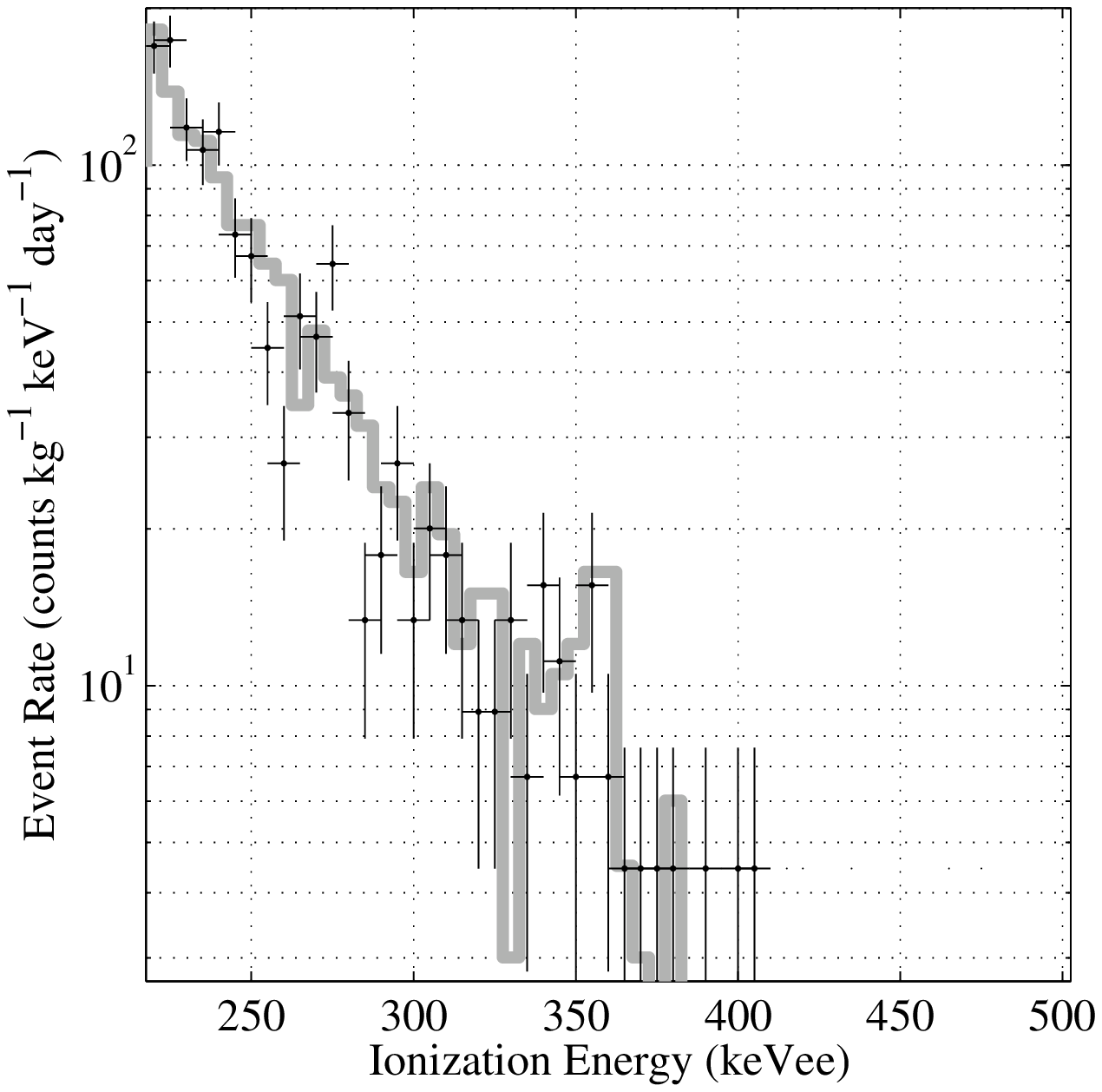}
\caption{Comparison of the measured ${\rm ^{133}Ba}$ charge spectrum (dots with error bars) and the Monte Carlo simulation (gray line) for the Ge detector Z3 (top) and the Si detector Z4 (bottom). The expected lines for this source are at 276~keV, 303~keV, 356~keV, and 384~keV.}
\label{R118_Balines}
\end{figure}

Detailed Monte Carlo simulation of the ${\rm ^{133}Ba}$ calibration was performed using GEANT~3. The results of the simulation were convolved with the energy-dependent resolution of the charge channels. The resulting prediction for Tower~1 is compared to actual calibration data in Fig.~\ref{R118_Balines}. The lines are very clear in all Ge detectors, making the calibration straightforward. For the Si detectors, Compton scattering dominates at these energies, making the lines considerably less defined. Hence, for the case of Si detectors, the calibration is achieved by matching the spectral shapes of the Monte Carlo simulation and the data.

\begin{figure}
\includegraphics[scale=0.6]{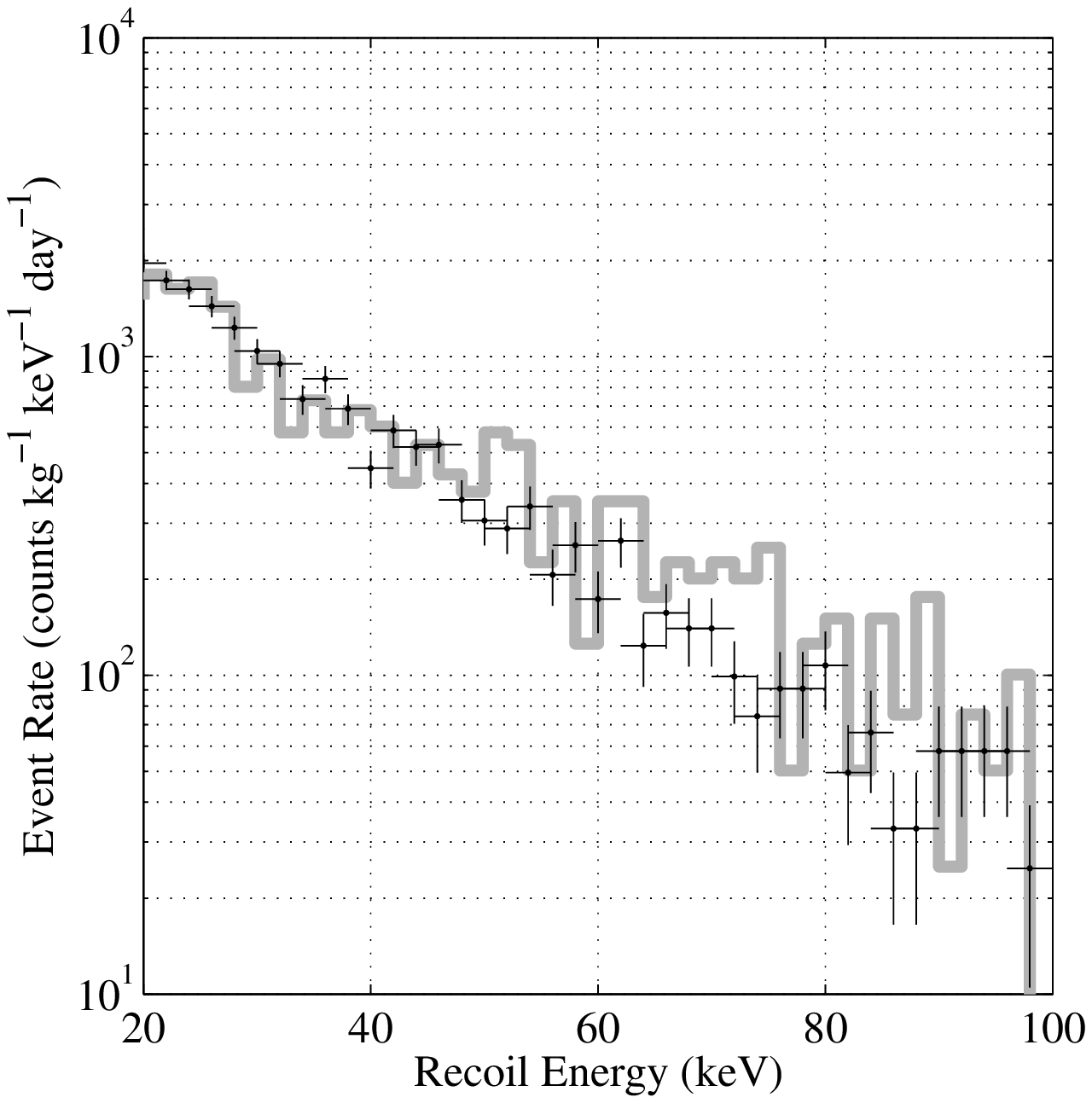}
\includegraphics[scale=0.6]{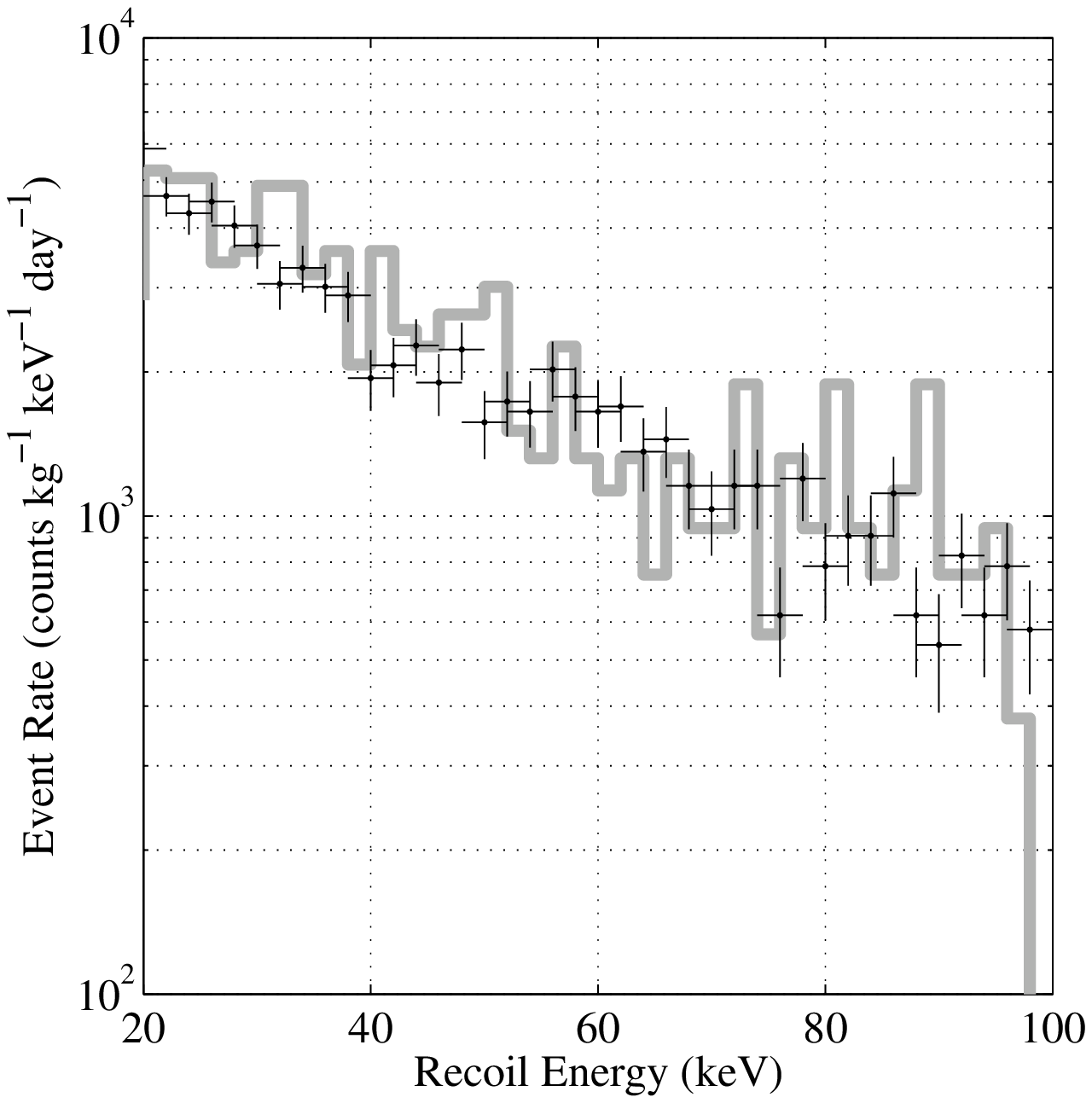}
\caption{Comparison of measured $^{252}$Cf neutron recoil spectrum (dots with error bars) and Monte Carlo simulation (gray line) for coadded Ge detectors (top) and Si detectors (bottom). The excellent agreement suggests that the phonon measurement of the recoil is unquenched.}
\label{fig:cfspectrum}
\end{figure}

We used a $^{252}$Cf source to characterize the detector response to nuclear recoils. Figure~\ref{fig:cfspectrum} shows a comparison of the measured recoil spectrum for neutrons with the spectrum predicted by a simulation of the source. The excellent agreement demonstrates that the calibration of the phonon channels using electron recoils remains valid for nuclear recoils. This suggests that the phonon measurement of the recoil and associated Neganov-Trofimov-Luke phonons is independent of recoil type, or unquenched.

\section{Simulations and Measurements of Expected Background Sources}
\label{sec:Backgrounds}
Four ordinary particles comprise the background for the experiment: neutrons, gammas, betas, and alphas. 

Neutrons produce nuclear recoils and cannot be rejected on an event-by-event basis unless they scatter in more than one detector. We simulated the expected neutron background at Soudan~\cite{kamatthesis} and found that it is insignificant over the course of the run.

Gammas and betas can be misidentified as nuclear recoils when they scatter in the dead layer near the detector surface. The poor charge collection in the dead layer can lead to ionization yields that are low enough to ``leak'' into the nuclear-recoil bands, as shown in Fig.~\ref{fig:yield}. These surface events (also called leakage events) comprise the most significant background for the experiment. The majority of the gamma background interacts in the bulk of the detector where charge collection is complete. Comparison of the WIMP-search bulk-electron recoils to simulation spectra allows us to identify the sources of our background gammas.

Betas arising from contamination on the detector surface are the most difficult background to characterize. There are a number of possible beta emitters such as $^{40}$K, $^{14}$C, and $^{210}$Pb. The detector Z6 has a known $^{14}$C contamination and was included in Tower~1 as an active veto. We have simulated the depth distribution of various sources of surface events. Combining the simulation results with a model of the detector response shows that, for our analysis, betas from surface contamination are the dominant electromagnetic background.

\begin{figure}[!htb]
\centering
\includegraphics[scale=1]{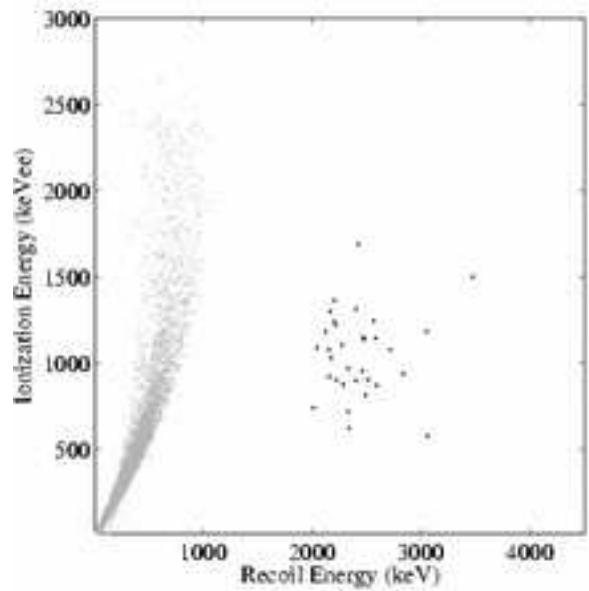}
\caption{Plot of ionization energy versus recoil energy for events during the WIMP search after the initiation of an old-air purge between the outer copper can and the surrounding mu-metal shield. Alphas (black dots) interacting in the detectors are easily identified. Nonlinearity in the phonon energy response at high energies is evident in the gammas (gray dots). This nonlinearity results in the recoil energy of the alphas being underestimated by about a factor of two.}
\label{fig:alphas}
\end{figure}

In addition to emitting betas, contaminants on the detector surface can also emit alphas. Alphas have large recoil energies ($>$1 MeV) and low ionization energies ($<$ 1 MeVee), making them easy to identify (see Fig.~\ref{fig:alphas}). The recoiling nucleus of a typical alpha decay in our detector produces a nuclear recoil with energy of order 20--100~keV. Events produced by the recoiling nuclei of alpha decays can be mistaken for WIMPs if the emitted alpha particles are not detected. Our detection efficiency for alphas is high, so we expect this background to be negligible. Detecting alphas is not only useful for rejecting events produced by the recoiling nuclei from alpha decays, but is also useful for estimating the amount of contamination present that may produce other backgrounds. In particular, we can estimate the amount of $^{210}$Pb on our detectors by detecting alphas produced by the decay of one of the daughters of the $^{210}$Pb decay chain, $^{210}$Po. In the case where the $^{210}$Pb and $^{210}$Po are in equilibrium, a measurement of the alpha rate from the decay of $^{210}$Po yields a beta rate coming from the decay of $^{210}$Pb and $^{210}$Bi. We have used the measured alpha rates during the WIMP search together with simulations of beta contaminants to estimate what fraction of our background is due to betas from the $^{210}$Pb decay chain.

In the following sections, we describe our simulations and measurements of these backgrounds in more detail.

\subsection{Neutron Background}
\label{sec:neutronMC}
Neutrons arise primarily from natural radioactivity, such as ($\alpha,n$) reactions from uranium and thorium nuclei and their decay products, and from muon interactions with nuclei in the apparatus and surrounding materials.  While neutrons produced by radioactivity in the rock are abundant, they have energies of only a few MeV and therefore have a large cross-section on protons.  The flux of such neutrons is attenuated by a factor of $\sim10^6$ by the 50-cm-thick polyethylene shield surrounding the detectors. This attenuation makes the contribution from this background negligible for the data taken thus far, and even for the entire planned CDMS-II exposure. Cosmogenic neutrons produced by muons entering the shield are vetoed with high efficiency on a per-event basis by the scintillator panels surrounding the shield (see Sec.~\ref{sec:shielding}).

The dominant source of any residual neutron-induced background that might be observed during the later course of CDMS-II is expected to arise from muon interactions in the surrounding rock cavern. Such neutrons can have energies well above 50~MeV, which is the energy at which the polyethylene moderator begins to lose effectiveness. These fast neutrons, many of which will not be detected in the present scintillator muon shield, can penetrate the polyethylene and  interact, for example, with a Pb nucleus and produce low-energy secondary neutrons that could recoil from nuclei in the cryogenic detectors. In the past~\cite{R19prd,R21}, we have measured the rate of unvetoed multiple-scatter events due to these ``punch-through'' neutrons to estimate the expected rate due to single scatters. In the data we have taken so far at Soudan, we did not detect any multiple-scatter events in anti-coincidence with the veto, and so we presently rely on Monte Carlo simulations to estimate this background. In addition, we simulate veto-coincident neutrons which are compared with data and serve as a cross-check on the anti-coincident population. We are approaching the expected required exposure to see such events, but to date we have not detected any veto-coincident neutrons.

We carry out simulations of the fast neutrons produced in the rock~\cite{kamatthesis} by using a published spectrum~\cite{khalchukov1983,khalchukov95} in a GEANT~\cite{GEANT} Monte Carlo program which includes the appropriate libraries for hadron cascades. Our parameterization  of this spectrum, which is dominated by hadron cascades caused by primary muon-induced nuclear spallation and by electromagnetic processes, is given by:
\begin{equation}
\frac{dN}{dE} = 
    \begin{cases}
    6.05~e^{(-E/77)} & \textrm{for $50<E<200$ MeV}\\
                e^{(-E/250)} & \textrm{for $E>200$ MeV}
    \end{cases}
    \label{neutronMChighE}
    \end{equation}
where E is the neutron energy.

The normalization is based on the neutron yield in rock at Soudan, which has been measured as $(3.3 \pm 1.0) \times 10^{-4}~\rm{n}/(\mu\,g\,{\rm cm}^{-2})$~\cite{ruddick}. This yield is consistent with FLUKA~\cite{FLUKA} simulations of the neutron yield of $3.56 \times 10^{-4}~\rm{n}/(\mu\,g\,{\rm cm}^{-2})$ at 270~GeV~\cite{wulandari} scaled to $2.96 \times 10^{-4}~\rm{n}/(\mu\,g\,{\rm cm}^{-2})$ at 210~GeV, the mean energy of downward going muons at Soudan~\cite{ruddick}. The scaling is based on the mean muon-energy dependence of the fast neutron yield, first pointed out by~\cite{bezrukov} and since confirmed by a variety of measurements and simulations (see~\cite{YFWang} and references therein). The results of this simulation, which model the full geometry of the CDMS-II apparatus in the configuration of the data set and WIMP-search exposure reported in this paper, predict $0.051 \pm 0.024$ events in the Ge detectors and $0.024\pm0.011$  events in the Si detectors from neutrons produced in the rock. GEANT~4 and FLUKA simulations of muon interactions in the rock cavern predict that, of these events, at least 60\% will be coincident with activity in the veto.

We also carry out simulations of the neutrons produced by muons in the lead shield and copper cryostat~\cite{kamatthesis}, which represent the majority of material internal to the scintillator veto. In addition to the high-energy component described by Eq.~\eqref{neutronMChighE} , we simulate the low-energy component which proceeds through the giant dipole resonance, yielding neutrons with the evaporation spectrum used to describe muon capture processes at low depths. The lower-energy component of the primary spallation spectrum can thus be described by the functional form~\cite{dasilvathesis}:
\begin{equation}
\frac{dN}{dE} = 
    \begin{cases}
    0.812\, E^{5/11}~e^{(-E/1.22)} & \textrm{for $E <4.5$ MeV}\cr
           0.018~e^{(-E/9)} & \textrm{for $4.5<E<50$ MeV}.
           \end{cases}
\end{equation}
This low-energy component is important for neutrons produced in the shielding because there is only 10\,cm of polyethylene internal to the lead and none inside the copper. The normalization of the neutron yield is inferred from measurements in lead for a mean muon energy of 110~GeV~\cite{gorshkov1968, gorshkov1971, gorshkov1974} and scaled with mean muon energy, as before. The results of these simulations predict that neutrons produced in the shield will yield $1.94 \pm 0.44$ events in the Ge detectors and $0.89\pm 0.18$ in the Si detectors for the exposure reported in this paper. Since the muons producing these neutrons must pass through the scintillator, all of these events would be vetoed. We find no such nuclear recoils, which is consistent at 6\% confidence with the predictions of our simulations, indicating that the neutron background may be less than our conservative estimate.

\subsection{Gamma Backgrounds}
\label{sec:gammabackground}

\begin{figure}[!htb]
\includegraphics[scale=0.6]{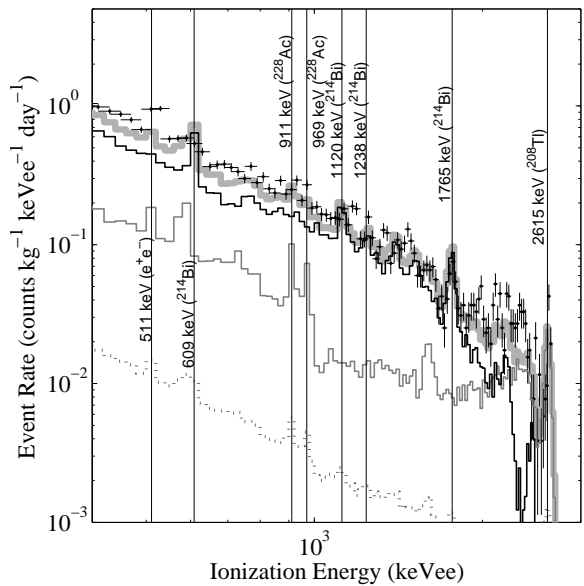}
\caption{Measured gamma spectra in the coadded Ge detectors along with Monte Carlo simulations of the most likely contributions: U/Th/K on the inner polyethylene (dotted), U/Th/K on the Cu cans (thin gray), Rn outside the mu-metal shield (thick black), and the sum of all these contributions (thick gray).}
\label{fig:gamma2}
\end{figure}

Simulations of the detector response (see Sec.~\ref{sec:betaMC}) show that the vast majority ($>$99.99\%) of gammas in the detector are easily discriminated from WIMPs. We can identify the different components of our gamma background by comparing our WIMP-search gamma spectrum with that predicted by GEANT~\cite{GEANT} Monte Carlo simulations of our expected gamma background. The gamma background in the 5\,keV to 3\,MeV region for 39\,days of livetime is shown in Fig.~\ref{fig:gamma2}. For energies greater than 2.6\,MeV, the rates at Soudan are considerably lower than those of SUF~\cite{saab,driscoll}. This is due to the absence of high-energy gamma cascades caused by neutron capture in the materials surrounding and within the Ge and Si ZIP detectors. From Z1 to Z6 we observe a total of 0, 1, 4, 7, 3 and 3 events, respectively, in the energy region 2.7-8.0\,MeV. All these events are multiple scatters depositing energy in the outer charge electrode. In SUF, after the muon veto cut, we observed about 240 events per detector for a similar exposure of 39\,days and the same energy interval. These were due to muon-induced neutron activation of the detector materials and are significantly reduced at Soudan. Below 2.6\,MeV, which is the endpoint of natural radioactivity originating from the $^{238}$U and $^{232}$Th chains, the gamma background is higher by about a factor of 2 compared to SUF. This factor is, {\it a priori}, consistent with the lack of lead shielding inside the Soudan icebox which was present for runs at SUF. 

We have simulated the contributions to the background from the U/Th/K content of the Cu cans, of the inner and outer polyethylene shields, and from $^{210}$Pb in the inner lead shield. These materials had been previously screened with HPGe spectrometers, and the results were used to normalize the Monte Carlo simulations. In the Cu used to build the icebox cans, U/Th/K contaminations of less than 0.5\,ppb were measured. For the inner and outer polyethylene shields, the measured U/Th contents were 30\,ppt and 1\,ppb, respectively. No $^{40}$K was detected, the upper limit being 0.25\,ppm. The upper limit on $^{210}$Pb in the inner lead shield was 0.2\,ppb. The sum of these normalized contributions yields a gamma background smaller by about a factor of four than the observed one. 

A likely additional source of gammas, especially the observed $^{214}$Bi line at 609\,keV in the Ge detectors, is the decay of radon outside the currently purged volume. To keep radon gas outside the space directly surrounding the icebox, we purge this volume with air which has been stored long enough in metal cylinders to allow the radon to decay (T$_{1/2}$ = 3.2 days) and the daughters to plate out. Radon can still penetrate and diffuse into the inner polyethylene shield, however, and Monte Carlo simulations show that about 35\,Bq/m$^3$ of radon gas outside the purged volume is sufficient to explain most of the remaining gamma background (see Fig.~\ref{fig:gamma2}). The measured radon levels in the Soudan laboratory in the CDMS experimental room show a mean of about 500\,Bq/m$^3$, with large seasonal variations (up to 700\,Bq/m$^3$ in summer and down to 200\,Bq/m$^3$  in winter).
 
\subsection{Beta Backgrounds}
\label{sec:betaMC}

The dominant background for this analysis comes from electron recoils within 10~$\mu$m of the detector surface. This region is referred to as the dead layer. As described in Sec.~\ref{section:ionization}, these events have suppressed ionization collection and could be mistaken for nuclear recoils. Surface events come from three sources: gammas interacting within the first 10~$\mu$m of the surface, low-energy electrons ejected from nearby material by high-energy x-rays, and electrons produced by radioactive beta decays from surface contamination. The first two sources are related to the ambient gamma background and are negligible for the background at Soudan. The third source, however, can limit the sensitivity of the experiment. 

\subsubsection{Detector Response to Betas}
\label{sec:BetaResponse}

We characterize the detector response to contamination betas by using the surface events produced in the very large gamma calibration runs. To simulate betas produced by calibration gammas, we used a simplified geometry of CDMS-II Tower~1 where six ZIPs were vertically stacked and surrounded by a thin copper cylinder. Simulated source photons originated isotropically from a spherical surface outside of the copper and with energies evenly distributed between 0-1~MeV. For each of these events, a weight was assigned based upon the initial energy of the source particle for that event, such that the detector recoil spectrum calculated by the Monte Carlo matched that observed in the gamma calibration. We simulated $^{133}$Ba and $^{60}$Co gamma calibration data. In addition to the gamma calibrations, simulations of possible beta-emitter ZIP contaminants were conducted with sources of $^{40}$K, $^{210}$Pb, and $^{14}$C, where each source is distributed uniformly between the pair of Ge ZIPs Z2 and Z3. Each event, as described by the Monte Carlo simulation, is reported as a list of positions and energy depositions for individual scatters or ``hits.'' An effective single position for each event is formed by taking the energy-weighted mean position of the hits. 

\begin{figure}[!htb]
\includegraphics[scale=0.6]{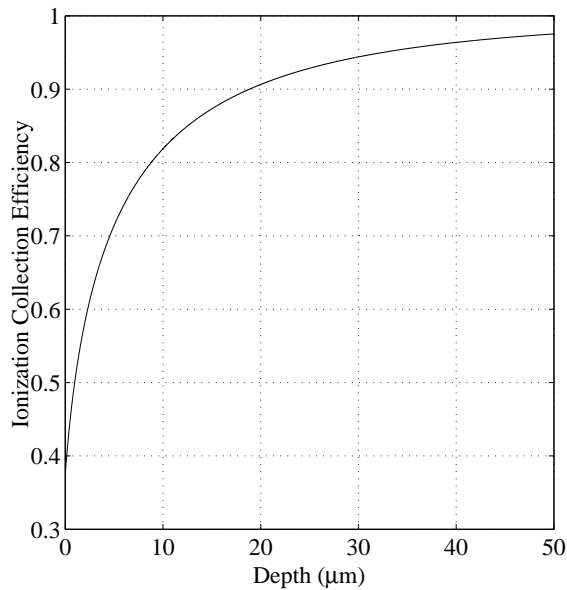}
\caption{Charge collection efficiency as a function of depth from the surface of a Germanium ZIP at a 3.0~V ionization bias, as derived from Monte Carlo simulations and analysis of $^{109}$Cd data.}
\label{fig:yvsdepth}
\end{figure}

In order to compare the results from simulations of beta emitters with actual data, we need a method for estimating the detector ionization yield of a Monte Carlo event. It has been shown~\cite{shuttthesis} that the dead layer effect, which leads to incomplete charge collection near the surface of the ZIP detectors, can be described by a function relating depth to charge collection efficiency (see Fig.~\ref{fig:yvsdepth}). Combining that function with the Monte Carlo event depth information allows us to model the yield response of the ZIP detectors to simulated surface events. This method of simulating betas has been shown to accurately reproduce the detector response to a $^{109}$Cd source~\cite{G31}.

Those leakage events with a signal in more than one detector can be excluded as WIMP candidates because WIMPs are not expected to scatter in multiple ZIPs.  However, leakage events with a signal in just one detector do remain a problem.

\begin{figure}[!htb]
\includegraphics[scale=0.6]{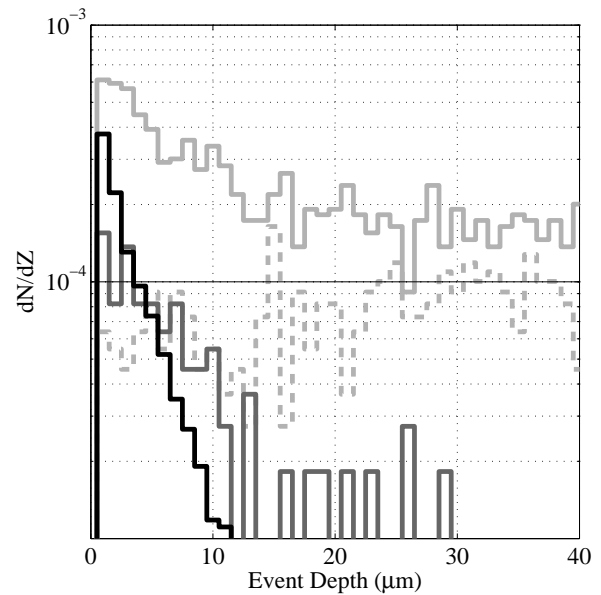}
\caption{Depth distributions for simulated surface events corresponding to $^{133}$Ba single scatters (dashed light gray), $^{133}$Ba nearest neighbor doubles (solid light gray), $^{133}$Ba beta-beta doubles (dark gray) and $^{40}$K singles (black). The $^{133}$Ba simulation results are normalized to all events. The $^{40}$K simulation results have been scaled down by a factor of $\sim$100 for easier comparison with the results from the $^{133}$Ba simulations. The horizontal thin black line indicates the expected distribution for events interacting uniformly throughout the entire detector.}
\label{fig:MCBetadepth}
\end{figure}

The results from the depth distribution analysis of the $^{133}$Ba calibration simulation (see Fig.~\ref{fig:MCBetadepth}) show that single-scatters have a flat depth profile extending into the 10 $\mu$m dead layer. We therefore expect very large $^{133}$ Ba calibrations to have a low-yield tail coming from single-scatters within this region of suppressed charge collection. These events can be used to characterize the detector response to interactions in the dead layer. In the region 15~$\mu$m from the surface, there is a factor of 3 enhancement in nearest-neighbor double events. Beta-Beta Double events (BBD), nearest-neighbor double events that have low-yield events in both neighboring detectors, show high preference for the surface with an exponential decrease and 1/e folding scale of $\sim10~\mu$m.

Simulations of a $^{40}$K beta-emitter (see Fig.~\ref{fig:MCBetadepth}) show that single-scatters from contamination on the detector surface have a depth distribution very similar to that of $^{133}$Ba BBD events, suggesting that we could improve our characterization of the detector response to beta-emitter singles through BBD events from gamma calibrations. 

We can use the function relating the charge collection efficiency to depth to estimate what fraction of contamination betas we expect to have measured ionization that is comparable to that of a nuclear recoil. The simulations indicate that $\sim$85\% of the betas will have yields below the electron-recoil band, shown in Fig.~\ref{fig:yield}, of which $\sim25$\% may be misidentified as nuclear recoils.

After initiating the old-air purge at Soudan, the beta rates in the ZIP detectors decreased by a factor of 1.5-2.5.  By relating the decrements in both gamma and  beta rates, we have estimated that surface-electron recoils produced by external photons contribute about 40-50\% of our beta background. However, simulations of $^{133}$Ba and $^{60}$Co sources and comparison with calibration data indicate that only about 2\% of these surface-electron recoils will have their ionization sufficiently suppressed that they would appear as nuclear recoils (prior to any timing cuts), implying that this background is negligible for the CDMS-II experiment.

\subsubsection{Detector Surface Contamination}
\label{sec:contamination}

The remainder of the beta rate is most probably caused by long-lived beta emitters contaminating the detector surfaces. A likely candidate is $^{210}$Pb, which is a long-lived (T$_{1/2}$ = 22.3\,years) radioisotope in the decay chain of airborne $^{222}$Rn.  Our detectors are exposed to radon during fabrication, mounting, and testing. We have measured the ambient radon levels at all the locations where detectors travel and have taken precautions to store them in nitrogen-gas purged cabinets  or evacuated vessels for most of the time.

The $^{210}$Pb $\beta^-$ contamination decays to $^{210}$Bi, with an endpoint of 63\,keV. $^{210}$Bi decays to $^{210}$Po, which then decays with a half-life  of 138\,days to the stable  $^{206}$Pb, emitting an $\alpha$ particle with an energy of 5.3\,MeV. As shown in Fig.~\ref{fig:alphas}, the alpha decays from the decay of $^{210}$Po produce a cluster of events in the detectors above 1~MeV in phonon recoil energy and with a significantly lower ionization yield than electron recoils in the bulk. Prior to this WIMP-search run the detectors had not been exposed to ambient air for about two years, well exceeding the 138-day half-life of $^{210}$Po, and leaving the $^{210}$Pb in equilibrium with $^{210}$Po. Measuring the alpha rate thus translates directly into a total beta rate due to radon contamination.

The first column of Table~\ref{alphas} lists the total number of alphas detected over the course of the entire WIMP search. Our analysis rejects events for which there is a significant amount of charge collected by the outer electrode (see Sec.~\ref{sec:qinner}), so only alphas that produced signals in just the inner electrode are considered. We will assume that we have 100\% detection efficiency for alphas and that all of the alphas come from $^{210}$Po decays which are in equilibrium with $^{210}$Pb. Consideration of the two relevant beta-emitters suggests that we will detect roughly the same number of low-energy betas in our detector as alphas. Simulations indicate that $\sim$10\% of these low energy betas will be single scatters having ionization yields comparable to nuclear recoils.  The second column of Table~\ref{alphas} lists the expected number of single-scatter betas that will have yields comparable to nuclear recoils under these assumptions. This analysis indicates that $\sim$35\% of the single-scatters in the WIMP-search nuclear-recoil band may come from beta-decay of $^{210}$Pb and $^{210}$Bi. 

\begin{table}[!htb]
\centering
\begin{tabular}{|c|c|c|}
\hline
Detector & Number of & Expected Number  \\
& alphas under the & of betas in the \\
& inner electrode & nuclear-recoil band \\
\hline
Z1 & 26 & 2.6 \\
\hline
Z2 & 20 & 2.0 \\
\hline
Z3 & 14 & 1.4 \\
\hline
Z4 & 17 & 1.7 \\
\hline
Z5 & 31 & 3.1 \\
\hline
Z6 & 39 & 3.9 \\
\hline
\end{tabular}
\caption{Alpha rates and associated beta-rate predictions assuming the $^{222}$Rn decay chain is in equilibrium after the $^{210}$Pb plate-out. The first column lists the number of alphas under the inner electrode found in the WIMP-search data. The second column lists the expected number of betas that are misidentified, assuming that we have a 100\% detection efficiency for alphas and that all of the alphas are from $^{210}$Po.}
\label{alphas}
\end{table}

Another possible source of beta contamination in our experiment is $^{14}$C, produced in the atmosphere by cosmic-ray spallation. Natural carbon could be introduced during processing, and traces of carbon are easily measured by various materials-science surface screening methods (e.g., Auger depth profiling, Rutherford Backscattering, and Particle Induced X-ray Emission). Results of these analyses indicate that about 2.5 monolayers of C are present on the ZIP surfaces, consistent with exposure to air, but there is no evidence of buried carbon. This yields a predicted $^{14}$C beta-event rate of 0.3 betas/day per ZIP, with a 156~keV endpoint, or less than 10\% of the total observed 15--45~keV beta rate.

\section{Analysis}
\label{sec:analysis}
With the WIMP-search data in-hand, along-with calibration and simulations of the expected backgrounds, we will now consider in detail the pulse-shape analysis, cut-setting, efficiency estimation,  leakages, and systematics. Note that during this whole analysis process the WIMP-search data itself is blinded, with a generous nuclear-recoil band defined such that any possible WIMP-candidate events are not present until ``the box is opened'' so that no human bias enters the construction of the data cuts.

\subsection{Event Reconstruction}
\subsubsection{Waveform Analysis}
\label{sec:fitting}
For the analysis of the ionization waveforms, we first construct detector-specific templates, both for the primary pulse and for crosstalk between the inner and outer electrodes. We restrict the template construction to using pulses in the 10--100~keV energy range in order to minimize non-linear effects in this energy range of interest for a candidate WIMP-recoil event. The templates are constructed by averaging a number of ionization pulses from {\it in situ} calibrations with external $^{133}$Ba photon sources. Each of these selected ionization pulses is for an event that produced ionization that was collected in only one of the two electrodes. We use two algorithms to estimate the amplitude and time offset of an ionization pulse. In the optimal-filter algorithm~\cite{golwalathesis,numrecipes} we construct a filter in the frequency domain from the pulse template and the noise spectrum for a given charge channel. This filter weights frequency bins based on their signal-to-noise. Convolving the filter with each ionization pulse yields an estimate of the amplitude and time offset relative to trigger for each pulse. In the second algorithm, we use a time-domain $\chi^2$ minimization of the template for each pulse to estimate the amplitude. The quantity we minimize is 
\begin{equation}
\chi^2 = \sum_{i = 1}^{N} \frac{| V_i - V_0\, s_i |^2}{\sigma^2}
\end{equation}
where $V_i$ are the ($N = 2048$) digitized data samples, $s_i$ is the pulse-shape template, $V_0$ is the fitted pulse amplitude, and $\sigma$ is the rms noise per sample. The time-domain algorithm is ``non-optimal,'' as it assumes that the frequency dependence of the noise can be discarded in the amplitude and time offset estimation process. This algorithm is of interest, however, for high-energy (1 MeVee) events with pulses that saturate (exceed the range of) the digitizers. The time-domain fit is applied only to the unsaturated digitizer samples (the rising and falling edges of the pulse), making it a better estimator than the optimal-filter algorithm for these high-energy events. For both algorithms, we analyze the inner and outer electrode signals simultaneously to account for crosstalk between the two channels. The time-domain fit was inadvertently applied to some of the low-energy events in our initial analysis. At lower energies, this fit is more sensitive to non-white noise in the charge channels, resulting in a slightly worse energy resolution than was obtained using the optimal-filter algorithm.
 
For the analysis of the phonon waveforms, we use a single phonon-pulse template for each detector type (Ge or Si). The templates have the functional form of a double exponential with risetime $\sim$30~$\mu s$ (15~$\mu s$ for Si) and falltime $\sim$300~$\mu s$ (150~$\mu s$ for Si).  For each phonon channel, we determine the area under the phonon pulse, and hence the phonon energy, using two different algorithms for different energy regimes. At relatively low energies ($<$ 100~keV) we use optimal filtering, where, as with the ionization signal, a filter generated from the pulse template and the measured noise is used to estimate the pulse area. At higher energies, where physical saturation of the phonon sensors leads to large variations of the pulse shape, integration of the pulse, after subtraction of the average pre-pulse baseline and high-pass filtering, provides a more accurate estimate for the area of the pulse.

For each phonon pulse, we also estimate the times at which the pulse reaches 10\%, 20\%, and 40\% of the peak along the rising edge. From these phonon pulse parameters, we construct additional quantities which are useful for event position reconstruction. The phonon ``delay'' time $t_{i}$ ($i=$A, B, C, or D for each channel) is the difference between the 20\% time of the phonon pulse and the similarly calculated 20\% time of the ionization pulse. We define the phonon ``risetime'' as the difference between the 10\% and 40\% phonon times.  For each detector, the  ``peak delay'' $t$ and the ``peak risetime'' $\tau$ are defined as the delay and risetime of the channel that has the most energy, which we call the ``peak'' sensor.  As shown in Sec.~\ref{sec:phonontiming}, both the peak delay and the peak risetime provide excellent rejection of surface events.

\subsubsection{Position Reconstruction}
\label{sec:position}
\begin{figure}[!htb]
\includegraphics[scale=.6]{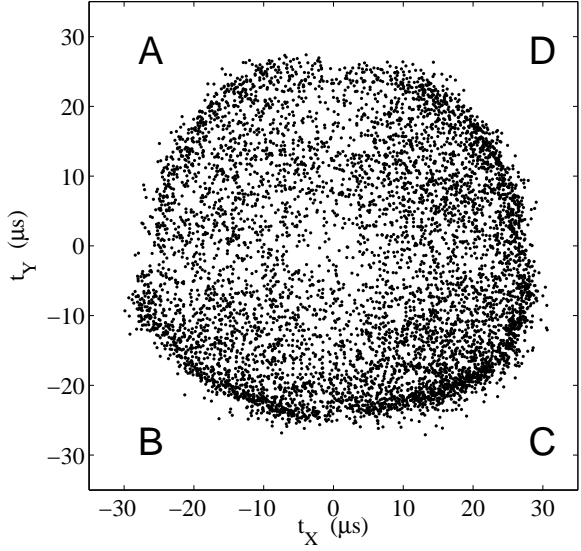}
\caption{Plot of the time delay parameters $t_{Y}$ versus $t_{X}$ for detector Z5 with $^{133}$Ba calibration data. The barrel shape of the plot indicates that this event reconstruction is non-linear.}
\label{fig:delay}
\end{figure}

\begin{figure}[!htb]
\includegraphics[scale=0.6]{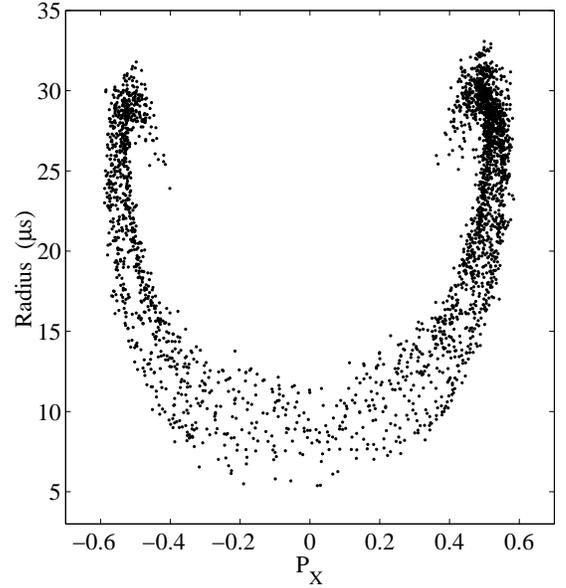}
\caption{Plot of delay ``radius,'' $R = \sqrt{t_X^2 + t_Y^2}$, versus phonon partitioning parameter $P_{X}$ for $-0.3 < P_{Y} < -0.2$. Events shown are for detector Z5 when exposed to a $^{133}$Ba source.}
\label{fig:radius}
\end{figure}

It is possible to combine the phonon parameters to extract information regarding the interaction location for any given event.  We define the x and y axes as shown in Fig.~\ref{fig:ZIPcartoon}.  The z-axis corresponds to the orthogonal coordinate, i.e. depth into the crystal. We calculate x-y delay coordinates $t_{X},t_{Y}$ from the relative phonon delays in the peak sensor and its two neighbors according to Table~\ref{table:delaydef}. Figure~\ref{fig:delay} shows that these coordinates are related to the physical location of each event. Under the approximation that each sensor is a point far from the interaction region, these formulas would yield accurate physical x-y coordinates as defined in Fig.~\ref{fig:ZIPcartoon}. However, the finite size of the sensors results in a non-linear mapping between the parameters $t_{X},t_{Y}$ and the physical position. As indicated in Fig.~\ref{fig:delay}, the expected azimuthal symmetry is broken by the sharp increase in sensitivity near the boundaries between the sensors. The ``barrel'' shape of the event distribution suggests that the radial dimension may not be monotonic (see Fig.~\ref{fig:radius}). Modeling of the dynamics of the phonon system to obtain a more linear reconstruction from the timing parameters~\cite{wangthesis} has been achieved, but for simplicity, we use the parameters described above for our current analysis.

\begin{table}[!tb]
\centering
\begin{tabular}{|c|c|c|}
\hline
Peak Quadrant & $t_{X}$ & $t_{Y}$ \\
\hline
A&$t_{A}-t_{D}$&$t_{B}-t_{A}$\\
\hline
B&$t_{B}-t_{C}$&$t_{B}-t_{A}$\\
\hline
C&$t_{B}-t_{C}$&$t_{C}-t_{D}$\\
\hline
D&$t_{A}-t_{D}$&$t_{C}-t_{D}$\\
\hline
\end{tabular}
\caption{Definition of $t_{X}$ and $t_{Y}$ parameters where $t_{i}~$(i=A,B,C,D) is the time the pulse in the $i^{\rm th}$ channel reaches 20\% of its peak.}
\label{table:delaydef}
\end{table}

\begin{figure}[!htb]
\includegraphics[scale=0.6]{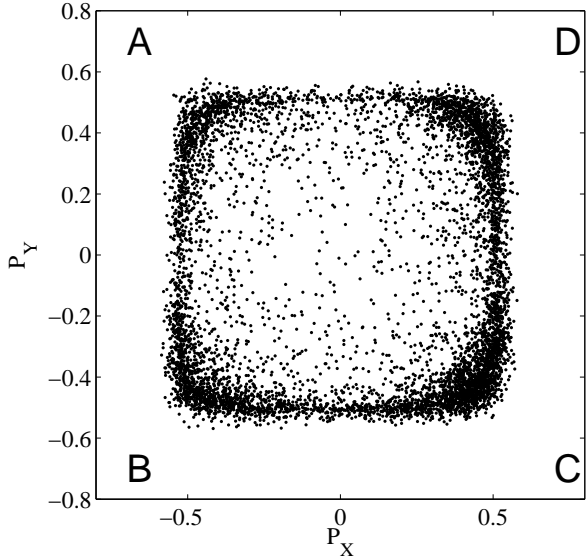}
\caption{Plot of the phonon partitioning parameter $P_{Y}$ versus $P_{X}$ for detector Z5 with $^{133}$Ba calibration data. $P_{Y}$ and $P_{X}$ denote relative partitioning of the phonon energy as defined by Eq.~\eqref{eq:ppart}.}
\label{fig:boxplot}
\end{figure}

A second nonlinear measure of an event's x-y position is determined from the relative partitioning of energy among the four phonon channels. These ``phonon partition'' position variables are defined by 
\begin{equation}
\begin{split}
&P_{X} = \frac{(P_{C}+P_{D})-(P_{A}+P_{B})}{\sum P_{i}}\\  \textrm{and} &\\
&P_{Y} = \frac{(P_{A}+P_{D})-(P_{B}+P_{C})}{\sum P_{i}}
\end{split}
\label{eq:ppart}
\end{equation}
where $P_{i}~(i=A,B,C,D)$ is the phonon energy measured in the $i$th channel. The layout of the QETs produces a square (or ``box'') distribution of the events in the detector, as shown in Fig.~\ref{fig:boxplot}. There are two important features of this distribution. First, the corners of the box correspond to the centers of each detector quadrant, where we expect to find the maximum partitioning of the phonon energy. We note that for any event, enough phonon energy is distributed to all of the channels such that there is an upper bound for $P_{X,Y}$ of $\sim0.5$. Second, the plot folds over (see Fig.~\ref{fig:radius}) with events near the outer edge of the detector appearing closer to the center of the plot.
 
The non-linearity inherent in both measures of event position is evident in the sparsity of events in the central regions of the two plots, and the over-crowding towards the perimeter (where both measures lose sensitivity). The ``button-hook'' feature shown in Fig.~\ref{fig:radius} illustrates that both the phonon-timing and phonon-energy-partitioning positions are not single-valued. However, the combination of both the phonon-timing and phonon-energy-partitioning variables does allow accurate parameterization of an event according to its physical x-y position. This method does not reconstruct an event's depth (z position), but instead allows us to account for position variation of the peak risetime and peak delay parameters so that they can be used to reject surface events (see Sec.~\ref{sec:lookuptable}).

For every event, this method defines a three-dimensional vector
\begin{equation}
\Theta= (P_{X}, P_{Y},R/R_{0}),
\label{eqn:pxpyrad}
\end{equation}
where $R=\sqrt{t_{X}^{2}+t_{Y}^2}$ is the delay ``radius,'' and $R_{0}=$22.5 $\mu$s in Ge (12.25 $\mu$s in Si). We chose $R_{0}$ so that the normalized radius, $R/R_{0}$, would have a maximum value that is comparable with the maxima of $P_{X,Y}$. This vector provides a unique mapping for the x-y position of an event as shown in Fig.~\ref{fig:radius}.

\subsubsection{Position Dependence of the Detector Response}
\label{sec:lookuptable}

The phonon pulse shapes depend on the event location within the detector.  There are two reasons for this variation. The first is that the superconducting transition temperature of the tungsten TESs is not perfectly uniform across the face of a detector. Since we bias the TESs in parallel, TESs with different transition temperatures are biased at slightly different points within their transitions, and this varies their response. Detector Z1 has a very large variation in its TES transition temperatures which dominates the position dependence of the phonon response. The other detectors in this tower had their transition temperatures tuned~\cite{ionimplant} to be sufficiently uniform so that this contribution to the position dependence of the phonon response is negligible.The second reason why the phonon-pulse shapes vary is that the physical arrival time of phonons at the QETs depends on position. For events near the center of a detector, a large number of the phonons detected have their last scattering point close to the initial interaction location. For events closer to the edge of the QET array, more of the phonons detected have scattered from the crystal surface.

\begin{figure}[!htb]
\centering
\includegraphics[scale=0.6]{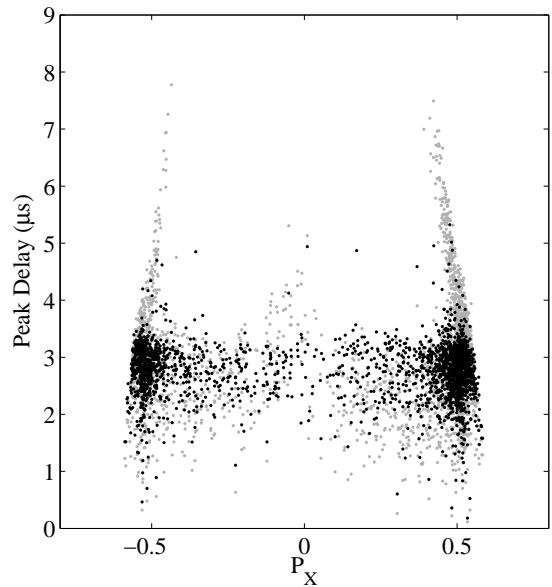}
\caption{Plot of Peak Delay versus phonon partitioning parameter $P_{X}$ for the slice $-0.3<P_{Y}<-0.2$. Gray dots are uncorrected distributions and black dots are corrected distributions. The corrected Peak Delay has been rescaled to 3 $\mu$s for easier comparison.}
\label{fig:positionDelay}
\end{figure}

\begin{figure}[!htb]
\centering
\includegraphics[scale=0.6]{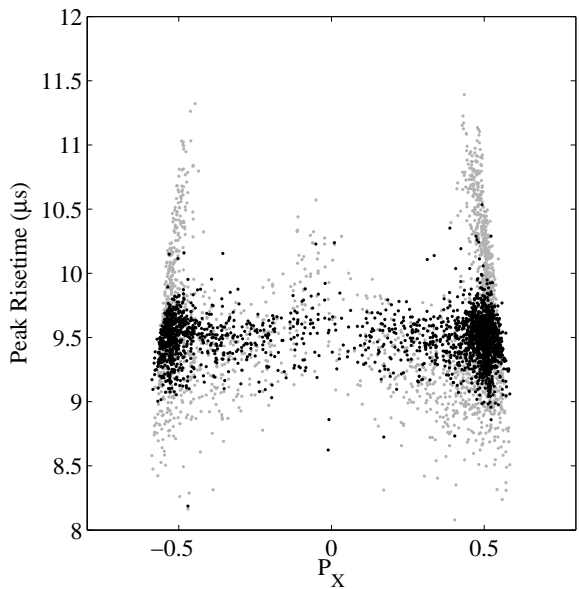}
\caption{Plot of Peak Risetime versus phonon partitioning parameter $P_{X}$ for the slice $-0.3<P_{Y}<-0.2$. Gray dots are uncorrected distributions and black dots are corrected distributions. The corrected Peak Risetime has been rescaled to 9.5 $\mu$s for easier comparison.}
\label{fig:positionRT}
\end{figure}

\begin{figure}[!htb]
\centering
\includegraphics[scale=0.6]{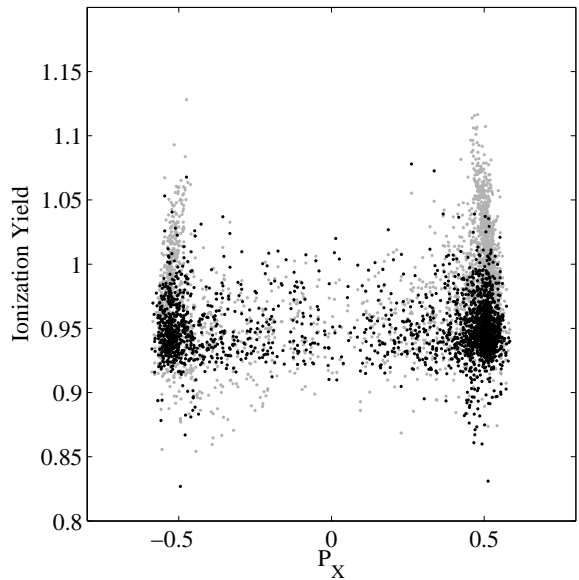}
\caption{Plot of Ionization Yield versus $P_{X}$ for the slice $-0.3<P_{Y}<-0.2$. Gray dots are uncorrected distributions and black dots are corrected distributions. The corrected yield has been rescaled to 0.95 for easier comparison.}
\label{fig:positionPamplitude}
\end{figure}

The varying phonon pulse shape leads to a position variation of our extracted event parameters. Figures~\ref{fig:positionDelay}, \ref{fig:positionRT}, and \ref{fig:positionPamplitude} show that the peak delay, peak risetime, and the phonon amplitude parameters vary with position. Such variation, if left uncorrected, would degrade the energy resolution and event-type discrimination based upon the ionization yield. In addition, variation in the timing parameters would prevent the use of a single cut to discriminate surface events from events in the detector bulk. To remedy this, we apply position-dependent corrections~\cite{changthesis} to make each parameter independent of the x-y position of the event.

First, we correct for a non-linearity of the phonon energy response arising from the saturation of the tungsten TESs. The correction has the form 
\begin{equation}
\widetilde P_{T} = \lambda(e^{P_{T}/\lambda}-1),
\end{equation}
where $P_{T}$ is the total phonon energy and $\lambda$ is determined empirically based on the fact that the ionization energy approaches linearity at low energies (see Section~\ref{sec:calibrations}). We correct for a weak energy dependence of the phonon timing parameters by using 
\begin{equation}
\widetilde\tau = \frac {\tau}{a+bP_{T}^{c}},
\label{eq:time energy correction}
\end{equation}
where $\widetilde\tau$ and $\tau$ correspond to the energy-corrected and original timing parameters and $P_{T}$ is the total (unlinearized) phonon energy. We empirically determine the constants, $a$, $b$, and $c$ such that the corrected timing parameter, $\widetilde\tau$, is linear and has the following normalization:  6~$\mu s$ (3~$\mu s$ for Si) for the peak delay, and 12~$\mu s$ (6~$\mu s$ in Si) for the peak risetime. Finally, we use a lookup-table algorithm (described below) to remove the position dependence of these energy-corrected quantities.

A position correction for each event is obtained using a lookup table specific to each detector. Each  lookup table consists of approximately 12,000 single-scatter inner-electrode electron recoils with energies between 10--200~keV. These electron recoils come from calibrations with external photon sources. For each event, $i$, in the lookup table, we determine a ``distance,'' $\rho_{ij}$, to any other event, $j$, in the table according to the following metric:
\begin{equation}
\rho^{ij} = \sqrt{\sum_{k=1}^{3}(\Theta^{i}_{k}-\Theta^{j}_{k})^{2}},
\end{equation}
where $\Theta$ is as defined by Eq.~(\ref{eqn:pxpyrad}). For each event in the lookup table, we determine the $N$ nearest events in the table. The value of $N$ is the same for every event in a given detector. $N$ must be sufficiently small to remove the variation due to the x-y position dependence while being sufficiently large to prevent statistical variations in the correction from contributing significantly to the position-independent resolution. For these lookup tables, we chose $N$ to be around 80.

The correction value for the phonon energy of a particular event in the lookup table is the average of the ionization yield for the event's $N$ nearest neighbors. The correction values for the timing parameters are the average of that timing parameter's values for the event's $N$ nearest neighbors divided by that parameter's normalization. Events that are not part of the lookup table use the correction values for the nearest event in the lookup table.

\subsection{Cut Definitions}
\label{sec:blind}
In this Section we describe all the cuts of our analysis and estimate their efficiencies. Nearly all efficiencies are estimated as a function of the event's recoil energy in the following energy bins: 5--8, 8--12, 12--16, 16--20, 20--30, 30--40, 40--50, 50--60, 60--70, and 70--100~keV. As mentioned earlier, all cuts were finalized prior to inspecting the WIMP-search data set. For the initial analysis~\cite{R18prl}, we inadvertently used the time-based fit to determine the amplitude of some of the ionization pulses. To account for this, we calculated two sets of efficiencies: one for the initial implementation of the cuts, the other for the current analysis which applies the cuts as intended. Only the fiducial-volume cut for detector Z5 was found to have a significantly different efficiency. We will show the reduced efficiency in Figs.~\ref{R118_qineff} and \ref{R118_finaleff} and describe both estimates in Section \ref{sec:qinner}. Tables~\ref{R118OFGe_breakdown}~and~\ref{R118OFSi_breakdown} summarize the effect of the ten individual data analysis cuts described in detail below.

\subsubsection{Data-Quality Cut}
\label{sec:dataquality}
A data-quality cut removes all non-optimal data sets (or parts of data-sets).  This cut excludes data-sets with known problems (such as non-operational channels), data sets that failed in the off-line diagnostics (usually due to increased noise), and events that were triggered by noise bursts. Figure~\ref{R118_livetime} shows that, over the course of the entire run, 52.6 live-days of WIMP-search data pass the data-quality cut. Less than 5\% of the data fail the data-quality cut.

\subsubsection{Phonon Pre-trigger Cut}
\label{sec:pstd}
The phonon pre-trigger cut rejects events for which the pre-trigger part of a phonon trace (about 400 ${\rm \mu s}$ long) is unusually ``noisy,'' more specifically, for which the standard deviation of the ADC values for the first 400 0.8~$\mu$s time bins is large. This cut complements the ionization $\chi^2$ cut described below for purposes of eliminating pile-up events and noise-triggered events. For the duration of this run, the phonon pre-trigger cut was set at $5 \sigma$, where $\sigma$ is the standard deviation of a Gaussian fit performed separately for each phonon channel of each detector using data from ${\rm ^{133}Ba}$ gamma calibrations.

The efficiency of the phonon pre-trigger cut is calculated as the fraction of WIMP-search events passing the cut.  The efficiency is better than 99.999\%.

\subsubsection{Ionization $\chi^2$ Cut}
\label{sec:chi2}
\begin{figure}[!thb]
\centering
\includegraphics[scale=1]{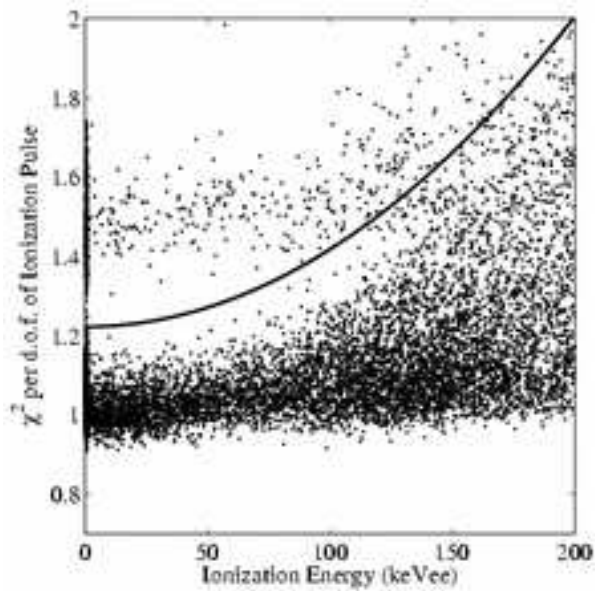}
\caption{Reduced $\chi^{2}$ (for 4092 degrees of freedom) of the ionization pulse as given by the optimal-filter algorithm plotted against ionization energy (electron equivalent) from a $^{133}$Ba calibration for detector Z3. Only events beneath the solid line pass the $\chi^{2}$ cut. The faint upper band consists of events taken during the $\sim5$\% of the time that a noisy heater was on; see Fig.~\ref{fig:slidingseal}.}
\label{R118_QSOFchisq}
\end{figure}

Events with anomalous ionization pulse shapes are rejected by a cut on $\chi^2$ of the optimum-filter fit.  Such events can be from pile-up (in which more than one pulse takes place in the same 1.6 ms trace) or from noise glitches. Since the start times of the optimal filter's templates can vary only by integer numbers of time bins, the fits are slightly inaccurate for events that start between two bins. For high energy pulses, this inaccuracy is large compared to the noise, so this sub-digitizer-bin time-jitter causes some pulses to have worse fits. To keep these events, the cut is implemented as a quadratic function of the ionization energy, as illustrated in Fig.~\ref{R118_QSOFchisq} for detector Z3.  The cut is defined using the ${\rm ^{133}Ba}$ calibration data.

\begin{figure}[!htb]
\centering
\includegraphics[scale=1]{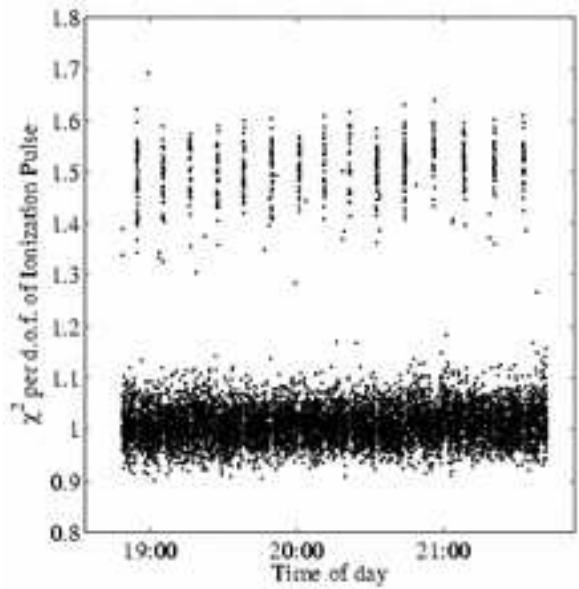}
\caption{Plot of the reduced $\chi^{2}$ (for 4092 degrees of freedom) of the ionization pulse from the optimal-filter algorithm versus time of day for a $^{133}$Ba calibration of detector Z3. Events are between 10--60~keVee. The events in the upper band of Fig.~\ref{R118_QSOFchisq} are clearly localized in time. These excursions are correlated with the time a noisy heater was active.}
\label{fig:slidingseal}
\end{figure}

Figure~\ref{R118_QSOFchisq} shows the existence of two bands.  As indicated by Fig.~\ref{fig:slidingseal}, the upper band corresponds to events during times of increased noise produced by a heater on the cryostat.  This heater intermittently introduces additional noise at a few hundred kHz in the ionization channels and causes worse fits for the events occurring while it is on (approximately 5\% of the total). Though this additional noise did not impact the energy resolution of the optimal-filter algorithm, we decided to reject these events in order to simplify the cut.

\begin{figure}[!htb]
\centering
\includegraphics[scale=0.6]{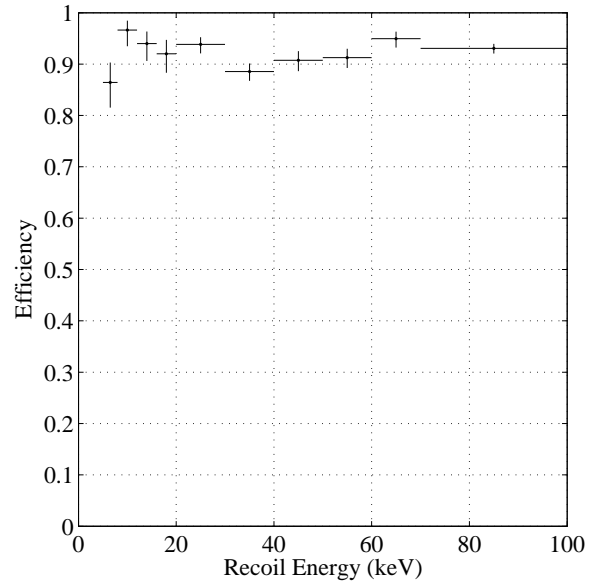}
\caption{Efficiency of the ionization $\chi^2$ cut for nuclear-recoil candidate events (see Sec.~\ref{sec:chi2}) in the Ge detector Z3. Note that the plot is versus recoil energy and not ionization.} 
\label{R118_chisqeff}
\end{figure}

The efficiency of the ionization $\chi^2$ cut is estimated using the WIMP-search electron-recoil events. We use WIMP-search data rather than the calibration data because we have observed small variations in the ionization $\chi^2$ parameter among the WIMP-search data sets. The efficiency is estimated as a function of ionization energy. We calculate the efficiency as a function of recoil energy for nuclear-recoil events using the observed relationship between ionization and recoil energy for neutrons from calibration with external sources. As an example, Fig.~\ref{R118_chisqeff} shows the efficiency of these two cuts on the nuclear-recoil events for the detector Z3. 

\subsubsection{Fiducial-Volume Cut}
\label{sec:qinner}
\begin{figure}[!thb]
\centering
\includegraphics[scale=0.6]{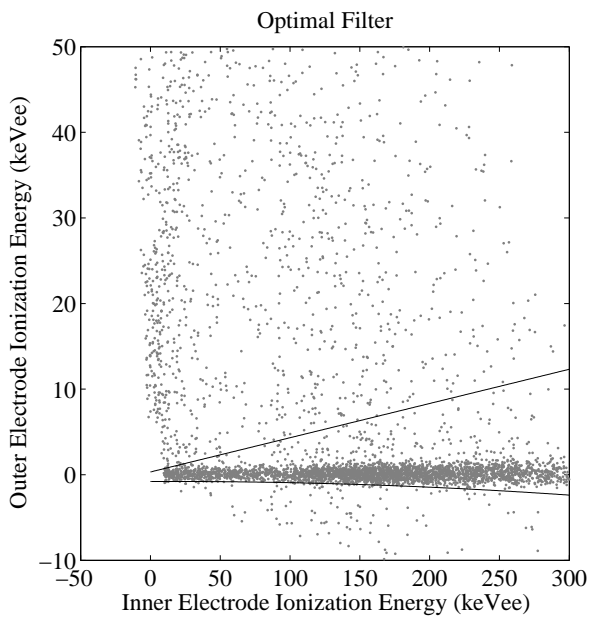}
\caption{Ionization collected in the Outer electrode versus ionization collected in the Inner electrode using the optimal-filter (see Sec.~\ref{sec:fitting}) algorithm for Ge detector Z5. Calibration data taken with an external $^{133}$Ba source are shown. Events between the two lines pass the fiducial volume cut.}
\label{R118_qinner}
\end{figure}

As described in Sec.~\ref{section:ionization}, we use the amplitude of the pulse in the outer ionization electrode, or ``guard ring,'' to reject events that take place close to the edge of the crystal. The cut is defined in the $Q_o$-$Q_i$ plane, where $Q_o$ and $Q_i$ stand for ionization energy in the outer and inner ionization channels, respectively. ${\rm ^{133}Ba}$ calibration data are used to fit the $Q_o$ distribution to a Gaussian in several bins of $Q_i$ energy. The events passing the cut lie in a band centered around $Q_o \approx 0$, defined by
\begin{equation}
    \mu - 3\sigma < Q_{o} < \mu + \sigma + Q_i/25 ,
\end{equation}
where $\mu$ and $\sigma$ are the ($Q_{i}$-dependent) mean and standard deviation determined by the Gaussian fits. The last term is introduced in order to reduce the impact of the outer-electrode noise on the high-energy inner-electrode events. Figure~\ref{R118_qinner} illustrates this cut for the Ge detector Z5.

\begin{figure}[!htb]
\centering
\includegraphics[scale=0.6]{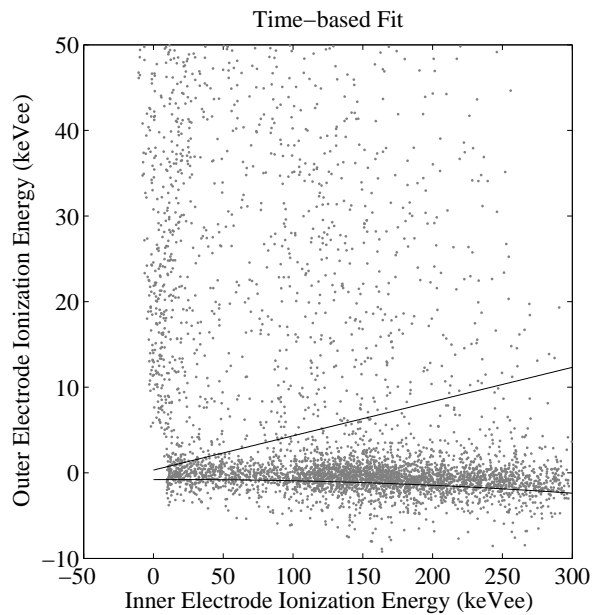}
\caption{Ionization collected in the Outer electrode versus ionization collected in the Inner electrode using the time-based fitting (see Sec.~\ref{sec:fitting}) algorithm for Ge detector Z5. Calibration data taken with an external $^{133}$Ba source are shown. Events between the two lines pass the fiducial volume cut.}
\label{R118F5qinner}
\end{figure}

The efficiency of this cut for WIMPs should be nearly 82\%, which is the volume fraction covered by the inner electrode.  The efficiency should be slightly less than 82\% (especially at high energies) because events near the gap between the inner and outer electrodes may have some ionization collected in the outer electrode. Additionally, noise in the outer electrode will cause some events in the inner electrode to fail the fiducial volume cut. To be conservative, we used the expected efficiency for neutrons (which could be checked more directly) as the efficiency for WIMPs as well. The efficiency of the fiducial-volume cut for neutrons should be lower than the efficiency for WIMPs due to two effects, both of which are related to WIMPs' small interaction probability.  First, some neutrons multiple-scatter within a detector and thereby deposit energy in both the inner and outer electrodes.  Second, due to the shielding of the inner region of the detectors by their outer regions and by other detectors in the stack, neutrons that single scatter are more likely to deposit energy in the outer electrode than would be calculated by the volume fraction alone.

\begin{figure}
\centering
\includegraphics[scale=0.6]{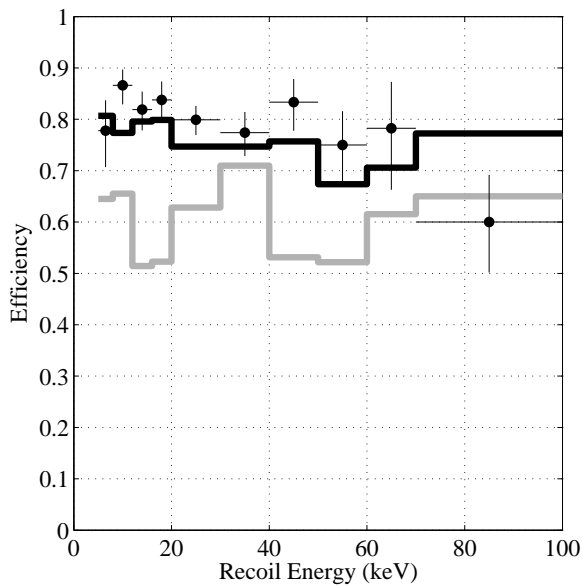}
\caption{Efficiency of the fiducial-volume cut for detector Z5 (Ge). The black line corresponds to the Monte-Carlo simulation of the $^{252}$Cf neutron calibration estimate when the intended optimal-filter algorithm is used to estimate the ionization pulse amplitude. The gray line corresponds to the estimate when using the time-based fit. The black dots with error bars are estimates using the actual nuclear-recoil events due to neutrons from the $^{252}$Cf calibration runs using the cuts for the current analysis. The roll-off above 60~keV is an artifact due to surface-electron contamination of the data set. Only the fiducial-volume cut for this detector (Z5) varied substantially ($\sim$25\% reduction) with the the time-based fit.}
\label{R118_qineff}
\end{figure}

We estimate the efficiency of the outer-ionization-electrode cut using a Monte Carlo simulation of the ${\rm ^{252}Cf}$ neutron calibration. The simulation determines the recoil energy and ionization energy of each detector hit and sums the inner and outer ionization energy of all hits in a detector based on the simulated positions of the hits. For this simulation, we do not account for any possible sharing of ionization between the two electrodes. Any hit had its full ionization collected by the electrode it was closest to. Then we convolve the ionization energy with the observed noise performance of the ionization channels. We characterize the noise of each channel by a Gaussian fit of the distribution of ionization-energies for random-trigger events. We then estimate directly the fraction of the neutrons that fail the outer-ionization-electrode cut. To estimate the efficiency for the time-based peak-height analysis, we simply use the noise performance given by the time-based algorithm. Only Z5 had significantly worse noise with this algorithm (see Fig.~\ref{R118F5qinner}). The increase in noise results in the initial implementation of this cut being more severe than intended, reducing its efficiency as shown in Fig.~\ref{R118_qineff}. 

Figure~\ref{R118_qineff} also shows a check of our calculated fiducial-volume cut efficiency by comparison with a measurement using $^{252}$Cf neutron calibration events.  These events were preselected to pass the ionization $\chi^2$, phonon pre-trigger, and ionization threshold cuts and to have ionization yields within the $2\sigma$ nuclear-recoil band (see Sec.~\ref{sec:nrband}). Our efficiency estimate is the fraction of the preselected events that pass the outer-ionization-electrode cut. The efficiency estimated in this way suffers from electron-recoil misidentification in the outer electrode. Electron-recoil events that take place close to the edge of the crystal may not have full charge collection and can leak into the nuclear-recoil band, contaminating the sample of true neutron events. The efficiency is consequently underestimated, especially at high recoil energies ($>70$~keV) where the number of neutrons in the neutron calibration is relatively low. 

\subsubsection{Electron-recoil and Nuclear-recoil Bands}
\label{sec:nrband}
To calculate the electron-recoil bands and the nuclear-recoil bands, we first fit Gaussians to the distributions of ionization yield (= ionization / recoil energy) for both electron-recoil and nuclear-recoil events (from the ${\rm ^{133}Ba}$ and ${\rm ^{252}Cf}$ calibrations, respectively) in several recoil-energy bins. The estimated means and standard deviations are then fitted versus recoil energy. Unless specified otherwise, the bands are always taken to be $\pm 2 \sigma$. 

\begin{figure}[!thb]
\centering
\includegraphics[scale=0.6]{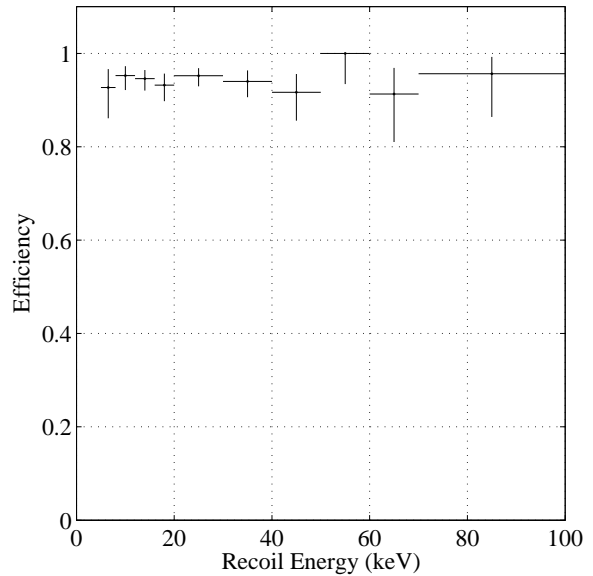}
\caption{Efficiency of the nuclear-recoil-band cut for the Ge detector Z3 (see Sec.~\ref{sec:nrband}).}
\label{R118_cnreff}
\end{figure}

We determine the efficiency of the nuclear-recoil-band cut using the nuclear-recoil events in the neutron calibration. We preselect events by imposing the charge $\chi^2$, phonon pre-trigger, charge threshold, fiducial-volume, and phonon-timing cuts. Additionally, we only consider events in the $4\sigma$ nuclear-recoil band, which are expected to include almost all of the true nuclear-recoil events. The efficiency estimate is the fraction of the preselected events that fall in the $2\sigma$ nuclear-recoil band. Figure~\ref{R118_cnreff} shows the efficiency of this cut for Z3. Neutrons which multiply scatter in the same detector tend to have lower yield than do single scatters with the same total recoil energy. This effect should cause our measured nuclear-recoil-band cut efficiency to be slightly lower than the efficiency for WIMPs.  We make no correction at this time for this small effect.

\subsubsection{Ionization Threshold Cut}
\label{subsec:QThresEff}
The goal of the ionization threshold cut is to select events that have measurable pulses in the charge channels. This cut is particularly important at low energies, for which an event with only noise in the inner ionization channel could otherwise be mistaken for a low-energy nuclear-recoil event. 

This cut is defined using the distribution of ionization energy for noise events (obtained by random triggering) in the ${\rm ^{133}Ba}$ calibration data. In particular, we examine the distribution of the optimal-filter amplitudes of pulse-less traces in the inner charge electrode. We fit a Gaussian to this distribution and set the cut at $\mu+5\sigma$, where $\mu$ and $\sigma$ are the mean and the standard deviation estimated in the Gaussian fit, respectively. 

\begin{figure}[!thb]
\centering
\includegraphics[scale=0.6]{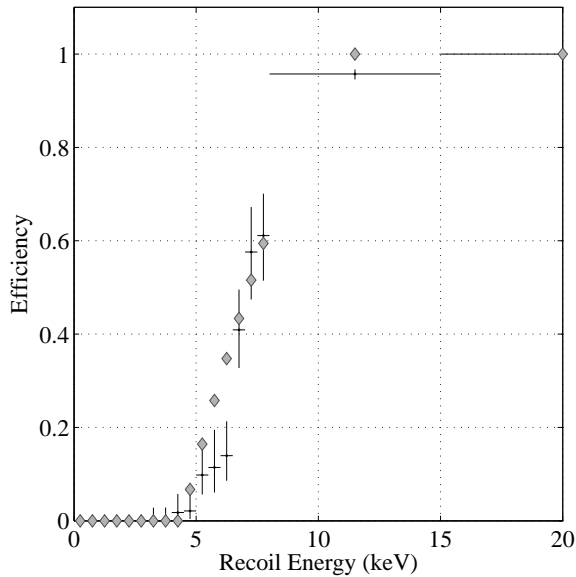}
\caption{Efficiency of the ionization-threshold cut (see Sec.~\ref{subsec:QThresEff}) on nuclear-recoil events. Black points with error bars indicate the efficiency as determined from data using nuclear recoils from the $^{252}$Cf calibration runs. Diamonds correspond to the calculated efficiencies assuming that the yield is a Gaussian.}
\label{R118_qthreff}
\end{figure}

The efficiency of the ionization threshold cut is measured with inner-electrode, $^{252}$Cf calibration events which trigger on the phonon signal and fall in the nuclear recoil band.  The efficiency is the fraction of such events which pass the ionization threshold cut.  The efficiency for Z3 is shown in Fig.~\ref{R118_qthreff}. Figure~\ref{R118_qthreff} also checks this result by comparison with a calculation of the efficiency of a nearly equivalent cut in the yield vs recoil-energy plane, assuming the yield distribution is Gaussian.  For events with zero outer electrode signal, the ionization cut is equivalent to a hyperbolic cut in the yield vs recoil-energy plane.  This alternative efficiency estimation suffers from a systematic error because the yield actually depends on the sum of the inner and outer charge channels. But its results are nonetheless very similar to those obtained using real events.

\subsubsection{Muon-Veto Cut}
\label{sec:vetocut}
\begin{figure}
\centering
\includegraphics[scale=0.6]{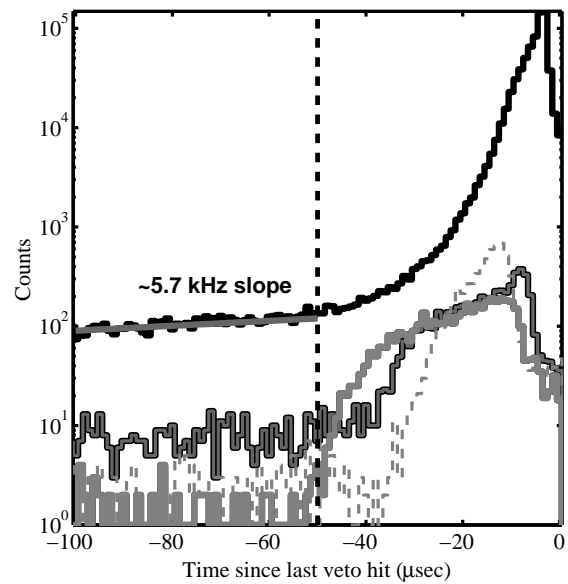}
\caption{Histogram of time from last veto hit for Tower 1 at the Stanford Underground Facility. An excess above the 5.7-kHz rate expected for uncorrelated events is seen for times near zero.  The cut at $-50$~$\mu$s is set in order to remove the excess for single scatters with recoil energies between 5--10~keV in Ge detectors Z2, Z3, and Z5 (solid dark gray curve); in Z1 (solid light gray line); and in the two Si detectors Z4 and Z6 (dashed gray line).}
\label{fig:vetotimewalk}
\end{figure}

We define the muon-veto cut using the time between the global trigger and the last hit in the muon veto. Events for which this time is long are unlikely to be caused by a muon interaction inside the veto. As described in Section~\ref{sec:shielding}, events occurring within 50~$\mu$s after veto activity are rejected, resulting in removal of $>99.4$\% of events caused by muons entering the shielding. The size of this window is set primarily to ensure that low-energy muon-induced events which have delayed global triggers (see Sec.~\ref{sec:daqelectronics}) are removed. Figure~\ref{fig:vetotimewalk} shows a histogram of the delays for these detectors when operated at the shallow Stanford Underground Facility~\cite{R21}, indicating that setting the cut at 50~${\rm \mu s}$ should be sufficient even for energies as low as 5~keV.

The efficiency of the muon-veto cut for WIMP events is determined by considering the effective deadtime induced by each veto hit. Given the 600~Hz rate of the muon veto (dominated by ambient photons) and the $50~{\rm \; \mu s}$ rejected window after each hit, about 3\% of the data-acquisition time is lost.  The efficiency of the muon-veto cut is therefore 97\%, independent of the energy deposited in a detector.

\subsubsection{Singles and Multiples Cuts}
\label{sec:singles}
To define single-scatter events (events in which only one detector was hit), we use the distribution of phonon energy for noise events obtained by random triggering. In particular, single-scatter events are events in which only one detector had a phonon signal larger than $6 \sigma$ of this distribution.

The multiple scatter events have to satisfy the following criteria:
\begin{itemize}
\item ALL detectors have to pass the data-quality cut, the ionization $\chi^2$ cut, and the phonon pre-trigger cut.
\item At least two detectors have to pass their fiducial-volume cut and have recoil energies above 5~keV. 
\end{itemize}

Note that these two definitions do not account for all events. For example, an event with 6~keV and 4~keV recoil energies in two detectors would not qualify as a multiple scatter ($<5$~keV in the second detector), but it may also not qualify as a single scatter since both detectors may have phonon signals larger than $6 \sigma$ of the phonon noise.

We can place a lower bound on the efficiency of our singles cut on WIMPs by considering the event rate and the duration of the post-trigger time window. The average event rate for all the detectors is 0.1 Hz, and the post-trigger digitization window is 1.2 ms, giving a lower bound of 99.99\% for the singles-cut efficiency.

\subsubsection{Phonon Timing Cut}
\label{sec:phonontiming}
As discussed earlier in Sec.~\ref{section:phonons}, the sensitivity of a ZIP detector to athermal phonons produced by an interaction provides for background rejection using phonon pulse-shape information. In particular, electron recoils near the detector surface result in a larger ballistic fraction than nuclear recoils in the bulk, both from the larger ionization yield and the more rapid down-conversion of phonons generated in a near-surface event. Figure~\ref{fig:YvsRT} illustrates these effects in calibration data. Using the phonon pulse shape to reject surface-electron recoils is especially significant, since these events can have reduced ionization collection, leading to misidentification if using the ionization yield alone.

A detailed beta calibration of a Ge ZIP detector performed at the UC Berkeley test facility~\cite{G31} indicated that using two phonon timing parameters, the peak delay and the peak risetime (see Sec.~\ref{sec:fitting}), provides good rejection of surface-electron recoils while retaining reasonable acceptance of nuclear recoils.

\begin{figure}
\centering
\includegraphics[scale=0.6]{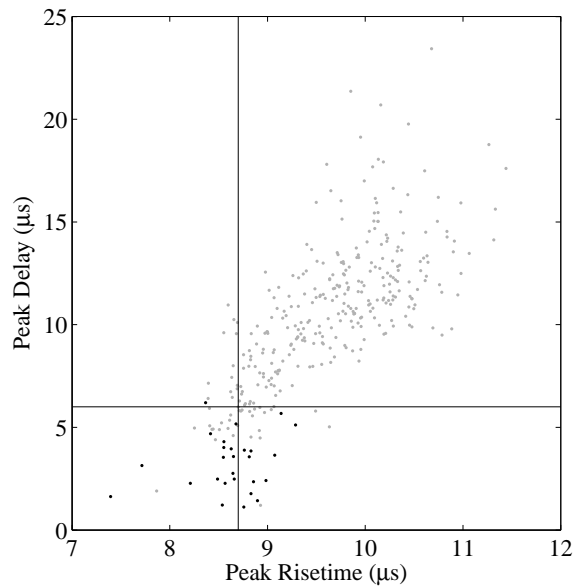}
\caption{Plot of phonon peak delay versus phonon peak risetime (defined in Sec.~\ref{sec:fitting}) for Ge detector Z5 in the 20--40~keV recoil bin. $^{252}$Cf neutrons in the $2\sigma$ nuclear-recoil band (gray) and $^{133}$Ba surface-electron recoils in the $4\sigma$ nuclear-recoil band (black) are shown. The first step in determining the timing cuts is to set energy-independent cuts within the bin for each of these two parameters so that all of the surface-electron recoils are rejected. Events below or to the left of the lines illustrating hypothetical cut values would be rejected.}
\label{R118_timingparams}
\end{figure}

\begin{figure}
\centering
\includegraphics[scale=0.6]{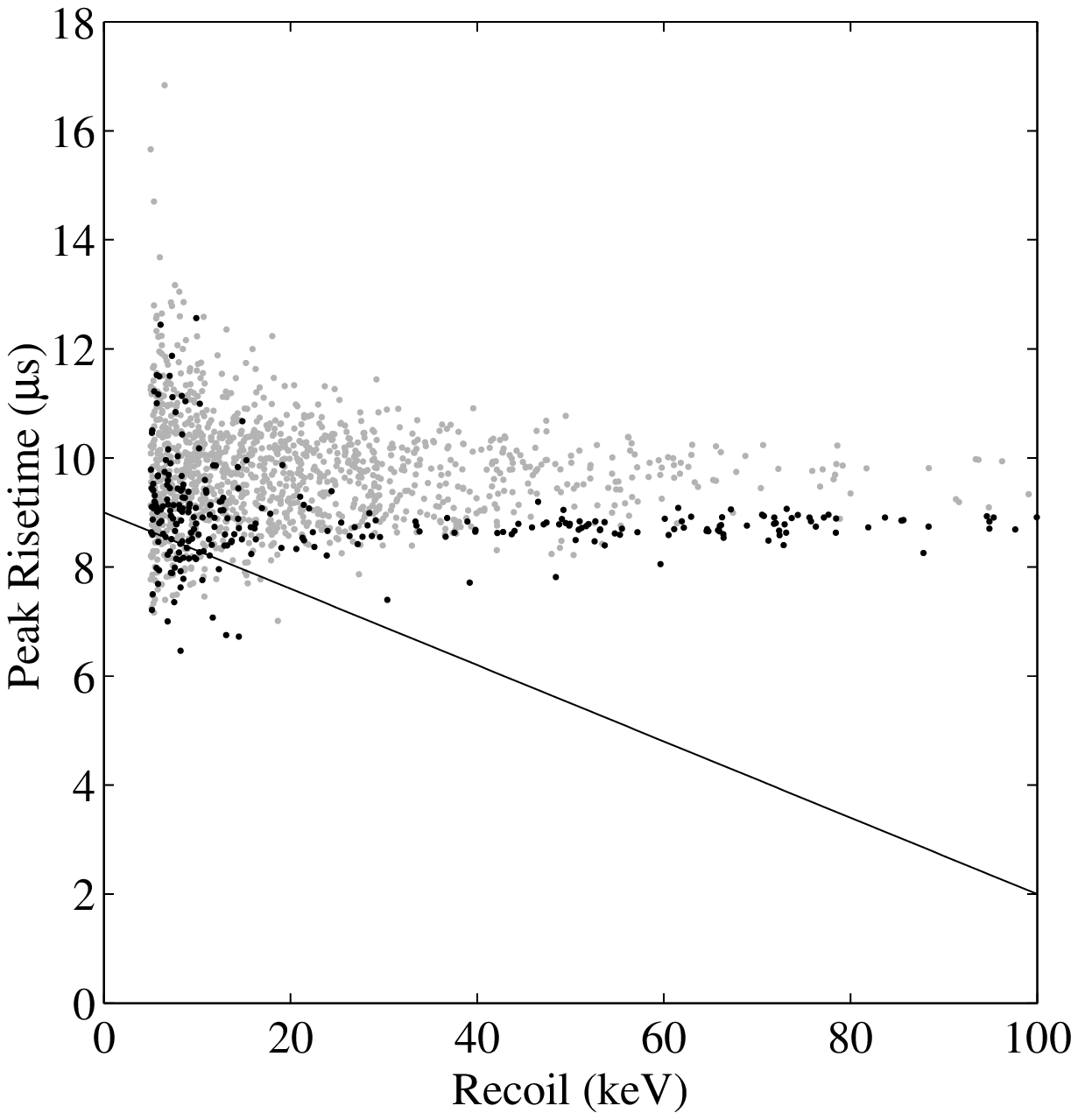}
\includegraphics[scale=0.6]{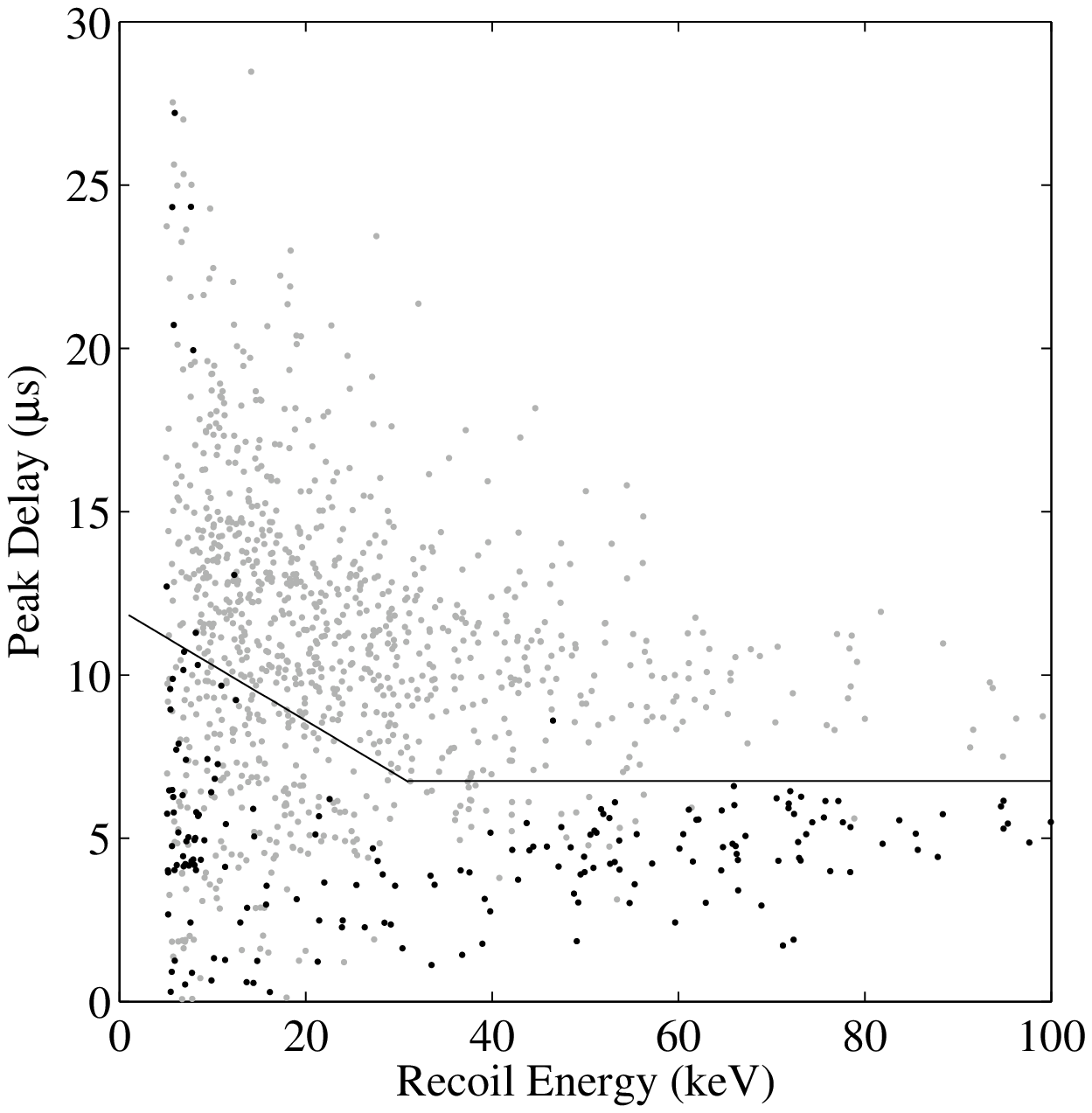}
\caption{Plots of peak risetime (top) and peak delay (bottom) versus recoil energy for detector Z5 (Ge). Gray dots correspond to $^{252}$Cf neutrons in the $2\sigma$ nuclear-recoil band. Black dots correspond to $^{133}$Ba surface-electron recoils in the $4\sigma$ nuclear-recoil band. The piecewise linear functions corresponding to the cut in each parameter are also shown. Events below these lines are cut. For this detector, above $\sim$20~keV all of the discrimination comes from the peak delay.}
\label{fig:RTvsPRC}
\end{figure}

To define the phonon timing cut, we first determine cuts (see Fig.~\ref{R118_timingparams}) in each of the two parameters for the following recoil energy bins: 5--10~keV, 10--20~keV, 20--40~keV and 40--100~keV. In each bin, we determine a pair of cut values that would reject every event from a subset of $^{133}$Ba calibrations that was in the $4\sigma$ nuclear-recoil band. Using only a portion of the $^{133}$Ba calibrations allows us to cross check our misidentification estimates with an independent set of events (see Sec.~\ref{sec:leakage}). For each parameter, we then perform a piecewise linear fit to the cut values to define a continuous, energy-dependent cut (see Fig.~\ref{fig:RTvsPRC}). By defining the cut in this way, we expect to misidentify a fraction of an event in each of the energy bins (see Sec.~\ref{sec:leakage}).

\begin{figure}[!thb]
\centering
\includegraphics[scale=0.6]{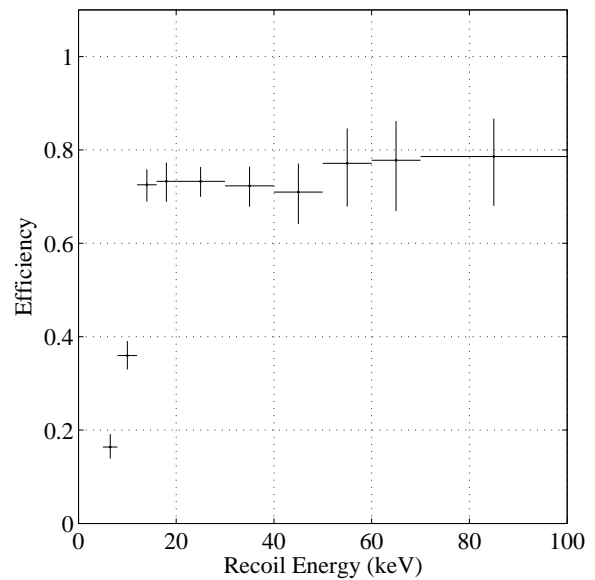}
\caption{Efficiency of the phonon-timing cut for the Ge detector Z3.}
\label{R118_crteff}
\end{figure}

We estimate the efficiency of the phonon-timing cut on WIMPs by using neutrons from the $^{252}$Cf calibrations. We preselect events in the $2\sigma$ nuclear-recoil band passing data-quality cuts (charge $\chi^2$, phonon pre-trigger, charge threshold) and the outer charge-electrode cut. The efficiency estimate is the fraction of these events that also pass the phonon-timing cut (see Fig.~\ref{R118_crteff}). 

\subsubsection{Analysis Thresholds}
\label{subsec:PThres}
Effective discrimination of electromagnetic backgrounds is achieved for energies as low as 5~keV in most detectors. However, to be conservative, we set the analysis thresholds for most detectors at 10~keV. The exceptions are Z1 and Z4, whose surface-event rejection was inadequate for recoil energies $<$~20~keV. For these two detectors, we set the analysis threshold at 20~keV. All detectors have an upper analysis bound on the recoil energy of 100~keV as this is an appropriate upper limit on the WIMP-search candidate-event recoil energy of interest.

\subsection{Overall Cut Efficiency}
\label{sec:totaleff}
\begin{figure}[!htb]
\centering
\includegraphics[scale=0.6]{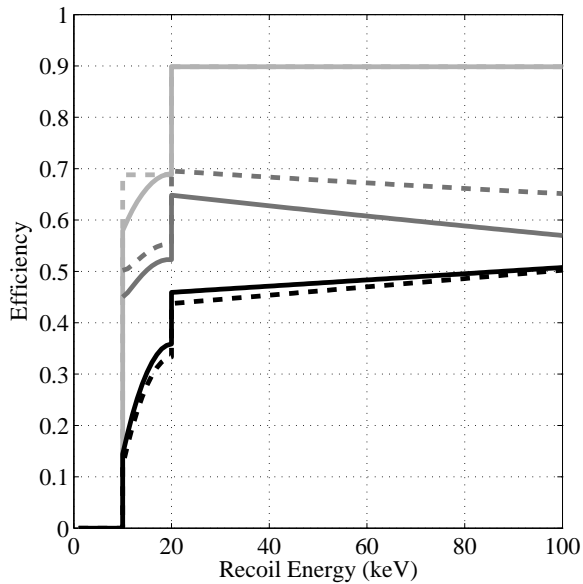}
\caption{Overall WIMP efficiency when cuts are applied. Light gray dashed line corresponds to application of data-quality, pile-up, and muon-veto cuts and, above 20~keV, overlaps with the light gray solid line which corresponds to imposing the ionization analysis threshold. Dark gray dashed line corresponds to application of the fiducial-volume cut. Dark gray solid line corresponds to the $2\sigma$ nuclear-recoil-band cut. The black solid line corresponds to application of the phonon timing cuts and is the overall efficiency for our current analysis. The black dashed line is the overall efficiency for the initial analysis.}
\label{R118_finaleff}
\end{figure}

The cut efficiencies described above are multiplied together to obtain the overall cut efficiency. Note that the efficiency is forced to zero below 10~keV for all detectors, and below 20~keV for Z1 and Z4, to reflect their analysis thresholds. Furthermore, Z6 is not used in this analysis because of the $^{14}$C contamination mentioned in Sec.~\ref{sec:Backgrounds} and neutralization problems mentioned in Sec.~\ref{section:ionization}. We average over the Ge detectors Z1, Z2, Z3 and Z5 and perform a fit to the efficiency estimate versus recoil energy: a straight line above 20~keV and a parabola below 20~keV, with the slopes matched at 20~keV. Figure~\ref{R118_finaleff} shows the effect of imposing each individual cut as well as the overall efficiency estimate for this analysis. The discontinuity at 20~keV comes from the higher analysis threshold of Z1. Integrated over the 10--100~keV bin, the overall efficiency yields a net Ge exposure of 19.4~kg-days for a 60~GeV~c$^{-2}$ WIMP.
 
\begin{figure}[!htb]
\centering
\includegraphics[scale=0.6]{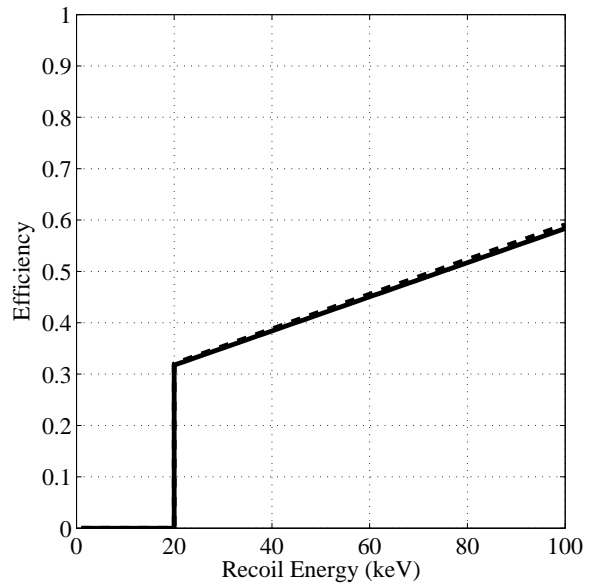}
\caption{Overall WIMP efficiency for the Si detector Z4. Dashed line corresponds to the efficiency for the initial analysis and the solid line is the efficiency for the current analysis. The two estimates differ by less than 1\% since, for this detector, there was little difference between using the time-based pulse height algorithm and the optimal filter algorithm.}
\label{fig:SiEff}
\end{figure}

For the initial analysis, where we inadvertently used the time-based algorithm to determine the ionization energy for some of the data, we had to calculate the efficiencies for both of the algorithms and make a weighted average of the two, where the weights were determined from the fractions of the WIMP-search data that were analyzed by each algorithm (see Table~\ref{R118_off5weights}). As shown in Fig.~\ref{R118_finaleff}, the efficiency for this analysis is slightly lower than the efficiency for the analysis using just the optimal filter algorithm. As discussed in Sec.~\ref{sec:qinner}, the reason for this loss in efficiency is that the time-based fit leads  to worse resolution for the outer electrode in Z5. This resolution loss made the guard ring cut for this detector more severe, reducing the efficiency of the cut. The efficiency for the Si detector (Z4) was unaffected (see Fig.~\ref{fig:SiEff}).

\begin{table}[!th]
\centering
\begin{tabular}{|c|c|}
\hline
Detector & Fraction analyzed by time-based fit\\
\hline
Z1 & 66.1\%\\
\hline
Z2 & 57.2\%\\
\hline
Z3 & 50.1\%\\
\hline
Z4 & 41.0\%\\
\hline
Z5 & 51.2\%\\
\hline
Z6 & 37.0\%\\
\hline
\end{tabular}
\caption{Fractions (in live-time) of the WIMP-search data analyzed by the time-based fitting algorithm, for all six detectors. The remainder of the data was analyzed by the frequency-based optimal-filter algorithm which has better resolution at low energies.}
\label{R118_off5weights}
\end{table}

\subsection{Background Rates and Leakages}
\label{sec:leakage}
The backgrounds for the WIMP-search results presented in this paper consist of two classes of interactions that could be misidentified as WIMPs: nuclear recoils from ordinary particles and electron recoils in the dead layer. As discussed in Sec.~\ref{sec:neutronMC}, Monte Carlo estimates of the punch-through neutron background at Soudan predict 0.051 +/- 0.024 events in the coadded Ge detectors and 0.024 +/- 0.011 events in the Si detector over the course of the entire run. Of these events, more than 60\% will have activity in the veto and will be rejected by the muon-veto cut. This gives an expected neutron background of $0.018\pm0.008$ in Ge and $0.004\pm0.002$ in Z4. Nuclei recoiling from alpha decays are also expected to produce negligible background because the efficiency for detecting an emitted alpha particle is high. As discussed in Sec.~\ref{sec:Backgrounds}, gammas and betas constitute a background through interactions in the dead layer where the ionization collection can be incomplete. In this section, we describe our estimated misidentification of dead-layer events. Table~\ref{R118_summarybkgd} summarizes our expected background misidentification.

We defined the timing cuts to reject all of the events in the $4\sigma$ nuclear-recoil band coming from a subset of the $^{133}$Ba calibrations (see Sec.~\ref{sec:phonontiming}). Since the cut is set to reject all of the events, we expect, on average, to misidentify one event for every $N^{\gamma}_{4\sigma}+1$ events, where $N^{\gamma}_{4\sigma}$ is the number of events in the $4\sigma$ nuclear-recoil band that were used to determine the cut. It is important that we separately estimate our leakage for each of the energy bins defined in Sec.~\ref{sec:phonontiming}. If the surface events produced during $^{133}$Ba calibrations yield distributions of the phonon timing parameters that are reasonable approximations to the distributions of these timing parameters for the sources of beta contamination (see Sec.~\ref{sec:betaMC}), we can estimate the number of misidentified surface-electron recoils by 
\begin{equation}
\frac{n_{2\sigma}}{N^{\gamma}_{4\sigma}+1},
\label{eq:betaleakage}
\end{equation}
where $n_{2\sigma}$ is the number of events in the WIMP-search data passing all but the timing cuts. This leakage estimate gives an expected surface-electron-recoil misidentification of $0.7\pm0.3$ events for Ge and $0.15\pm0.12$ events in Si (Z4 only) for our entire WIMP-search data.

As a crosscheck of our algorithm for estimating misidentification, we use Eq.~(\ref{eq:betaleakage}) to calculate the expected leakage for a set of $^{133}$Ba calibrations that were not used to define cuts. For these calibrations, there are a total of 141 events in Ge and 51 in Si (Z4 only) that are in the nuclear-recoil band. Substituting these numbers for $n_{2\sigma}$ (in the appropriate energy bins) in Eq.~(\ref{eq:betaleakage}) we expect $4.55\pm1.44$ events in Ge and $0.42\pm0.31$ events in Si to pass the timing cuts. In the calibration data, three events pass the timing cuts in Ge and one passes the timing cuts in Si, consistent with our calculated leakage. This agreement gives us confidence in our method of estimating our surface-electron-recoil misidentification.

We note that there could be an important systematic associated with this calculation. Misidentified events from the $^{133}$Ba calibrations consist entirely of surface events produced by ambient gammas. As discussed in Sec.~\ref{sec:Backgrounds}, dead-layer events produced during the WIMP search are dominated by beta decay from radioactive contamination on the detector surface. Differences in the spatial distribution between dead-layer events coming from contamination and events produced by ambient gammas can lead to a systematic error in our leakage estimate. We utilize the contamination on the silicon detector Z6 to estimate the impact of this systematic for both Ge and Si detectors.

Over the course of the entire WIMP search, there were 150 single-scatter events in the Z6 nuclear-recoil band that were anticoincident with the veto. Of those events, we expect fewer than 15 to be produced by the ambient gammas, indicating that the majority of the events came from contamination on the detector itself. The estimate of our surface-event misidentification predicts that, out of these 150 events,  $3.06\pm1.23$ events should pass the timing cuts. There were five events that pass the timing cuts, which is consistent with our expectation. This agreement suggests that the differences between surface events produced by  contamination and surface events coming from ambient gammas do not lead to significant differences in discrimination by the timing cuts.

\begin{table}[!th]
\centering
\begin{tabular}{|l|c|c|}
\hline
& Ge & Si \\
\hline
surface-electron recoils & $0.7\pm0.3$ & $0.15\pm0.12$\\
\hline
Neutrons & $0.018\pm0.008$ & $0.004\pm0.002$ \\
\hline
\end{tabular}
\caption{Total number of background events appearing in the nuclear-recoil band after cuts expected for the 52.6 live-days of the Tower~1 WIMP-search run.}
\label{R118_summarybkgd}
\end{table}

\subsection{Systematics}
Small variations in the detector performance throughout the course of the WIMP search lead to a systematic uncertainty in our calculation of the cut efficiencies, since these efficiency estimates are determined from the brief neutron calibrations. We estimate this uncertainty by comparing the cut efficiencies on the electron recoils over the course of the WIMP-search data with the efficiencies on  electron recoils during the neutron calibrations. The systematic uncertainty can be divided into two parts: one component comes from variation in the calibration of the ionization and phonon measurements, and one component comes from variation in the detector phonon timing response. Despite the use of the lookup-table correction described in Sec.~\ref{sec:lookuptable}, there may still be a residual position dependence which would correlate the two effects. In order to properly account for any such correlations, we compare the efficiencies of the two cuts combined. 

Estimating the uncertainty in the timing-cut efficiencies is especially difficult since, as shown in Fig.~\ref{fig:YvsRT}, the timing distribution of the electron recoils is intrinsically different from that of the nuclear recoils. To compute the uncertainty due to variation in the phonon timing response, we first parameterize the electron-recoil timing distributions as Gaussians with means and standard deviations given by:
\begin{equation}
\mu^{ER}_{WIMP} = \mu^{ER}_{Cf} + \Delta\mu, 
\end{equation}
and
\begin{equation}
\sigma^{ER}_{WIMP} = \sigma^{ER}_{Cf}\times\Lambda,
\end{equation}
where $\mu^{ER}_{Cf~(WIMP)}$ and $\sigma^{ER}_{Cf~(WIMP)}$ are the mean and standard deviation of the distribution in the neutron calibrations (WIMP-search data). Variations between the electron recoils of the calibration and those of the WIMP search are described by a shift in the means of the distributions, $\Delta\mu$, and a scaling of the standard deviations, $\Lambda$. We then adjust the timing distributions of the neutron calibration according to the equation:
\begin{equation}
t' = \Lambda\times(t-\mu^{NR}_{Cf}) + \mu^{NR}_{Cf} + \Delta\mu,
\label{rtshiftstretch}
\end{equation}
where $\mu^{NR}_{Cf}$ and $\sigma^{NR}_{Cf}$ are the mean and standard deviation of the nuclear-recoil timing distributions in the $^{252}$Cf data. Equation~\eqref{rtshiftstretch} simply shifts the mean of the timing distributions by $\Delta\mu$ and scales the standard deviation by $\Lambda$. We estimate the variation of our timing-cut efficiency to be the fraction of events that pass the timing cut using these adjusted timing parameters.

The results of this computation of our systematic uncertainty indicate that our efficiency estimates are good to within 10\% in the energy range of 10--100~keV. These estimates are consistent with other checks of experimental stability including comparison of the ionization peaks in three separate $^{133}$Ba calibrations, observation of the 10~keV line from Ga activation throughout the WIMP search, analysis of the timing distribution of surface events over the course of the WIMP search, monitoring of the bandwith of the TES-readout electronics, and comparison of the nuclear-recoil timing distributions for the three separate neutron calibrations.

\section{Results and Discussion}
\label{sec:results}
\begin{figure*}[!htb]
\centering
\includegraphics[scale=1]{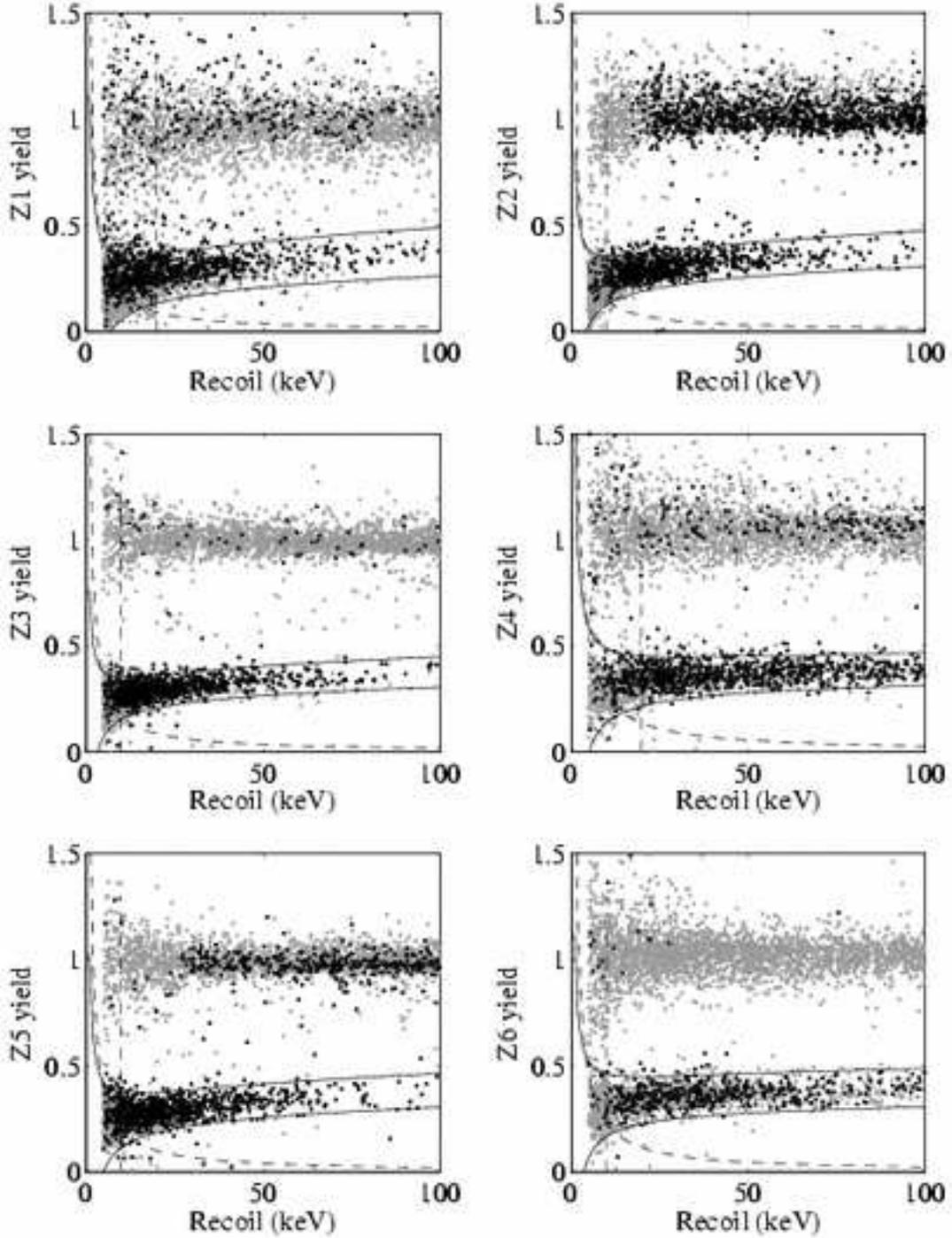}
\caption{Events from the $^{252}$Cf calibrations with recoils between 5--100~keV. Solid lines indicate the $2\sigma$ nuclear-recoil bands. Dashed lines indicate the recoil (vertical) and ionization (curved) analysis thresholds. All the events shown satisfy all of the WIMP-search cuts (e.g.single-scatter and fiducial volume), excluding the ionization-yield and phonon-timing cuts. Gray points fail the timing cuts and black points pass the timing cuts. Compare with Fig.~\ref{fig:OFyield} and Fig.~\ref{fig:blindyield}. Note that the timing cut rejects $\sim$20\% of the events in the nuclear-recoil band while rejecting $\gtrsim$75\% of the events in the electron-recoil band.}
\label{fig:cfyield}
\end{figure*}

\begin{figure*}[!htb]
\centering
\includegraphics[scale=.8]{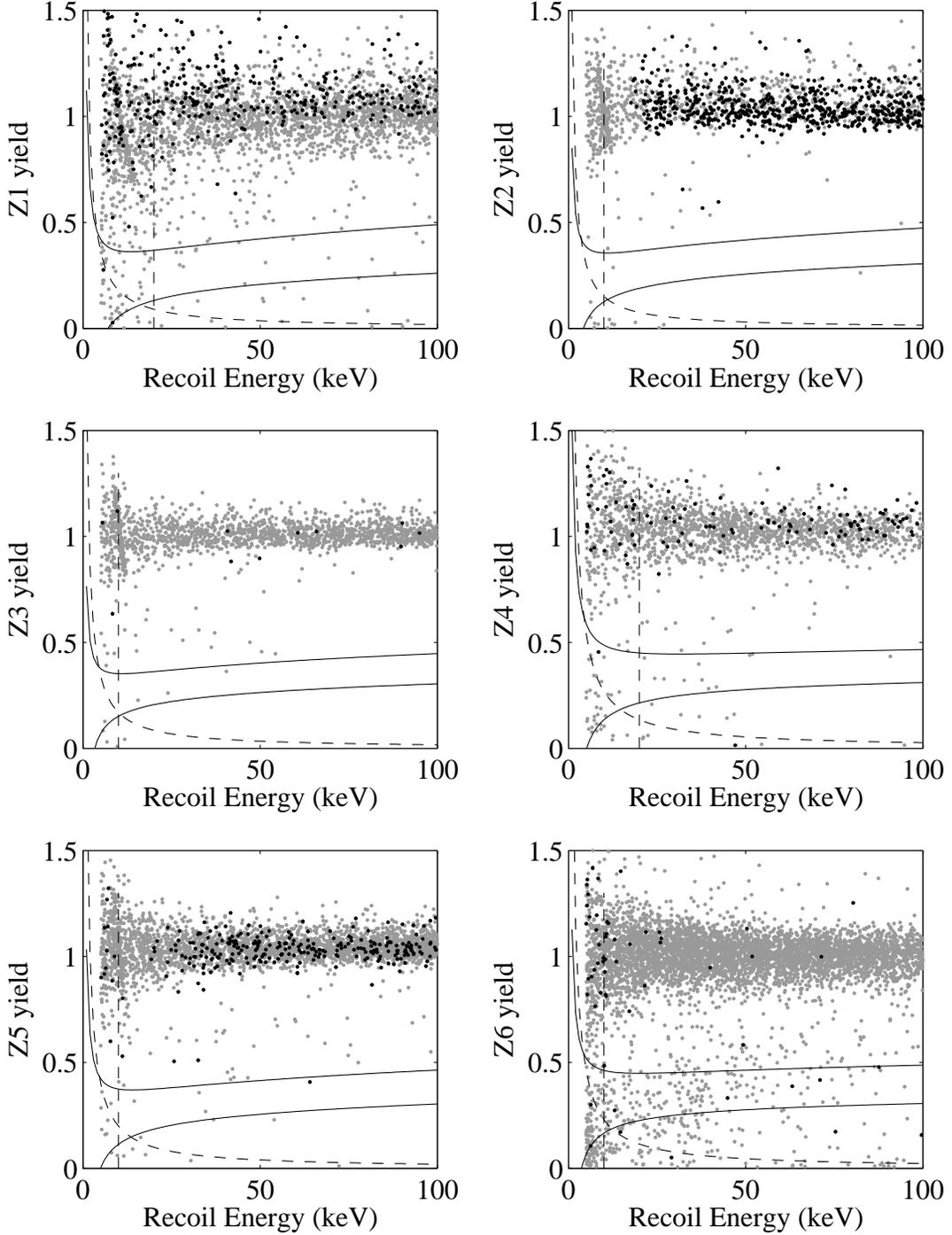}
\caption{Unvetoed single-scatters in the fiducial volume for each of the six ZIP detectors of Tower~1. These data correspond to the current analysis of 52 livedays of the Run 118 WIMP search at Soudan. Solid lines indicate the $2\sigma$ nuclear-recoil bands. Dashed lines indicate the recoil (vertical) and ionization (curved) analysis thresholds. Gray points fail the timing cuts and black points pass the timing cuts. A single event in Z5 with a 64~keV recoil satisfies all of the criteria for a WIMP candidate but is consistent with our expected misidentification of electron recoils near the surface. Compare with Fig.~\ref{fig:cfyield}. Detector Z6 was not included in this analysis due to a known $^{14}$C contamination.}
\label{fig:OFyield}
\end{figure*}

\begin{figure*}[!htb]
\centering
\includegraphics[scale=.8]{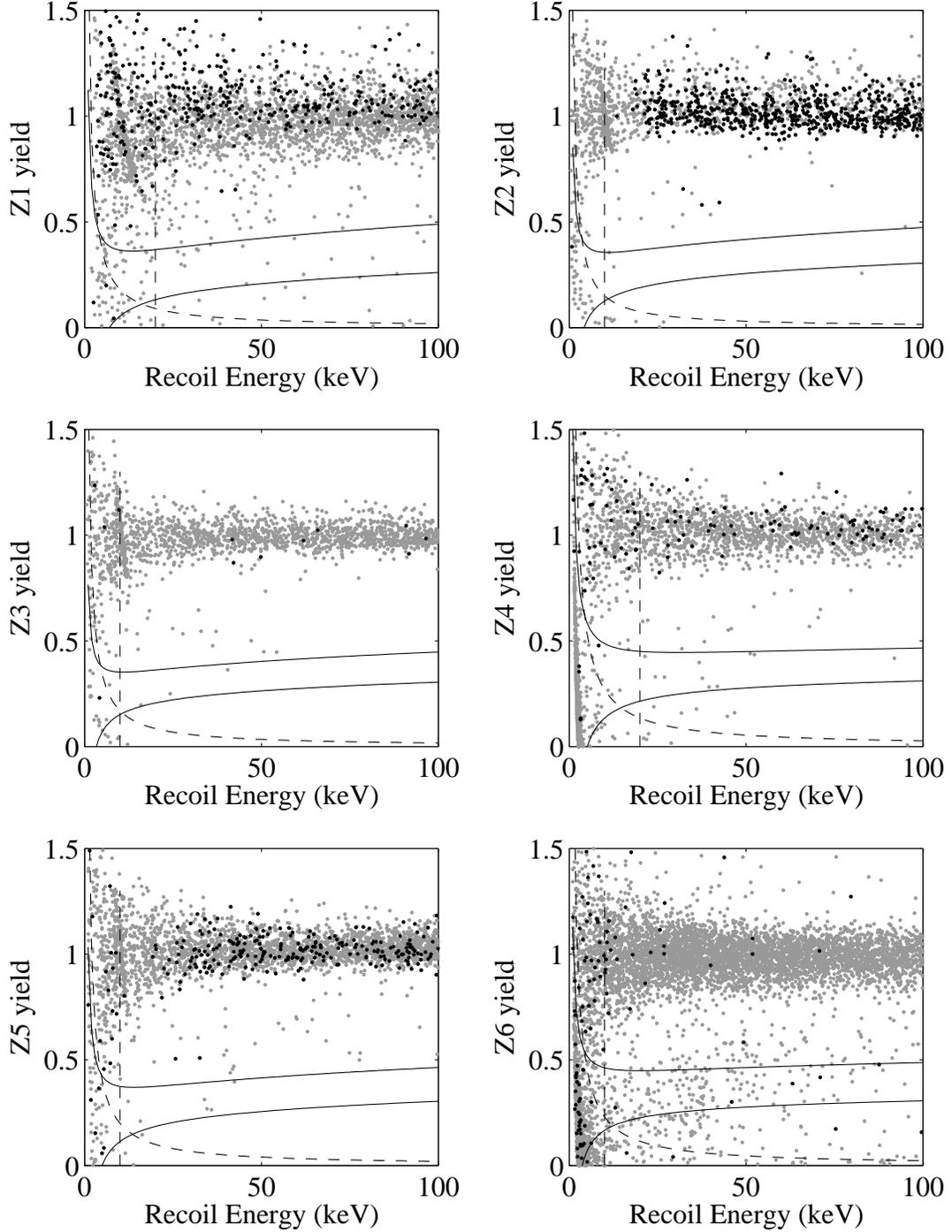}
\caption{Unvetoed single-scatters in the fiducial volume for each of the six ZIP detectors of Tower I. These data correspond to the initial analysis of 52 livedays of the Run 118 WIMP search at Soudan for which we inadvertently analyzed some of the ionization pulses using the time-based fit. Solid lines indicate the $2\sigma$ nuclear-recoil-acceptance bands. Dashed lines indicate the recoil (vertical) and ionization (curved) analysis thresholds. Gray points fail the timing cuts and black points pass the timing cuts. This  algorithm led to an overly severe fiducial volume cut which rejected the event in Z5 that passed all of the other cuts. Compare with Fig.~\ref{fig:cfyield}. Detector Z6 was not included in this analysis due to a known $^{14}$C contamination.}
\label{fig:blindyield}
\end{figure*}

In the 52.6 live days for this first WIMP search at Soudan, we find one event with a recoil energy of 64~keV passing all of the cuts in our current analysis (see Fig.~\ref{fig:OFyield}). We note that the timing cuts are much more severe in rejecting gammas compared to nuclear recoils (see Fig.~\ref{fig:YvsRT}), leading to a significant reduction of events in the electron-recoil band. Figure~\ref{fig:cfyield} illustrates that our analysis still retains sensitivity to nuclear recoils while rejecting a significant fraction of the electron recoils. For the initial analysis where we used the time-based algorithm to determine the ionization peak height, we found no events passing all of our cuts (see Fig.~\ref{fig:blindyield}). These two results are consistent with each other, given the differences in exposure and statistical uncertainty. The observation of one event at 64~keV is consistent with our expected (surface) electron recoil misidentification. Tables~\ref{R118OFGe_breakdown}~and~\ref{R118OFSi_breakdown} show the number of events passing the cuts for the Ge and Si detectors, respectively. We note that several new advanced analysis techniques (e.g. \cite{wangthesis}), which are beyond the scope of this paper, are substantially improving the rejection of surface electrons, so that we do not expect to be limited by this background for the remaining CDMS-II runs.

\begin{table}[!th]
\centering
\begin{tabular}{|l|r|r|}
\hline
& Initial & Current\\
\hline
All events & 968,680 & 968,680\\
\hline
Not random trigger & 940,619 & 940,619\\
\hline
Analysis thresholds (Sec.~\ref{subsec:PThres}) & 79,655 & 79,460\\
\hline
Singles (Sec.~\ref{sec:singles}) & 20,715 & 20,907\\
\hline
Data quality (Sec.~\ref{sec:dataquality}) & 18,852 & 19,027\\
\hline
Pile up (Sects.~\ref{sec:pstd} and~\ref{sec:chi2}) & 17,622 & 17,793\\
\hline
Muon veto (Sec.~\ref{sec:vetocut}) & 17,171 & 17,339\\
\hline
Ionization threshold (Sec.~\ref{subsec:QThresEff})& 14,697 & 14,835\\
\hline
Fiducial volume (Sec.~\ref{sec:qinner})& 7,187 & 7,615\\
\hline
Nuclear-recoil band (Sec.~\ref{sec:nrband}) & 29  & 23\\
\hline
Phonon timing (Sec.~\ref{sec:phonontiming})& 0 & 1\\
\hline
\end{tabular}
\caption{Breakdown of the events in the Ge WIMP-search data as we apply each cut. One event with a recoil of 64~keV passes all of the cuts in the current analysis.}
\label{R118OFGe_breakdown}
\end{table}

\begin{table}[!th]
\centering
\begin{tabular}{|l|r|r|}
\hline
& Initial & Current \\
\hline
All events & 968,680 & 968,680\\
\hline
Not random trigger & 940,619 & 940,619\\
\hline
Analysis thresholds (Sec.~\ref{subsec:PThres})  & 79,655 & 79,460\\
\hline
Singles (Sec.~\ref{sec:singles}) & 3,718 & 3,734\\
\hline
Data quality (Sec.~\ref{sec:dataquality})& 3,389 & 3,406\\
\hline
Pile up (Sects.~\ref{sec:pstd} and~\ref{sec:chi2}) & 3,144 & 3,163\\
\hline
Muon veto (Sec.~\ref{sec:vetocut})& 3,074 & 3,093\\
\hline
Ionization threshold (Sec.~\ref{subsec:QThresEff})& 2,590 & 2,608\\
\hline
Fiducial volume (Sec.~\ref{sec:qinner})& 1,575 & 1,595\\
\hline
Nuclear-recoil band (Sec.~\ref{sec:nrband})& 14  & 15\\
\hline
Phonon timing (Sec.~\ref{sec:phonontiming})& 0 & 0\\
\hline
\end{tabular}
\caption{Breakdown of the events in the Si WIMP-search data as we apply each cut.}
\label{R118OFSi_breakdown}
\end{table}

\begin{figure}[!htb]
\centering
\includegraphics[scale=0.6]{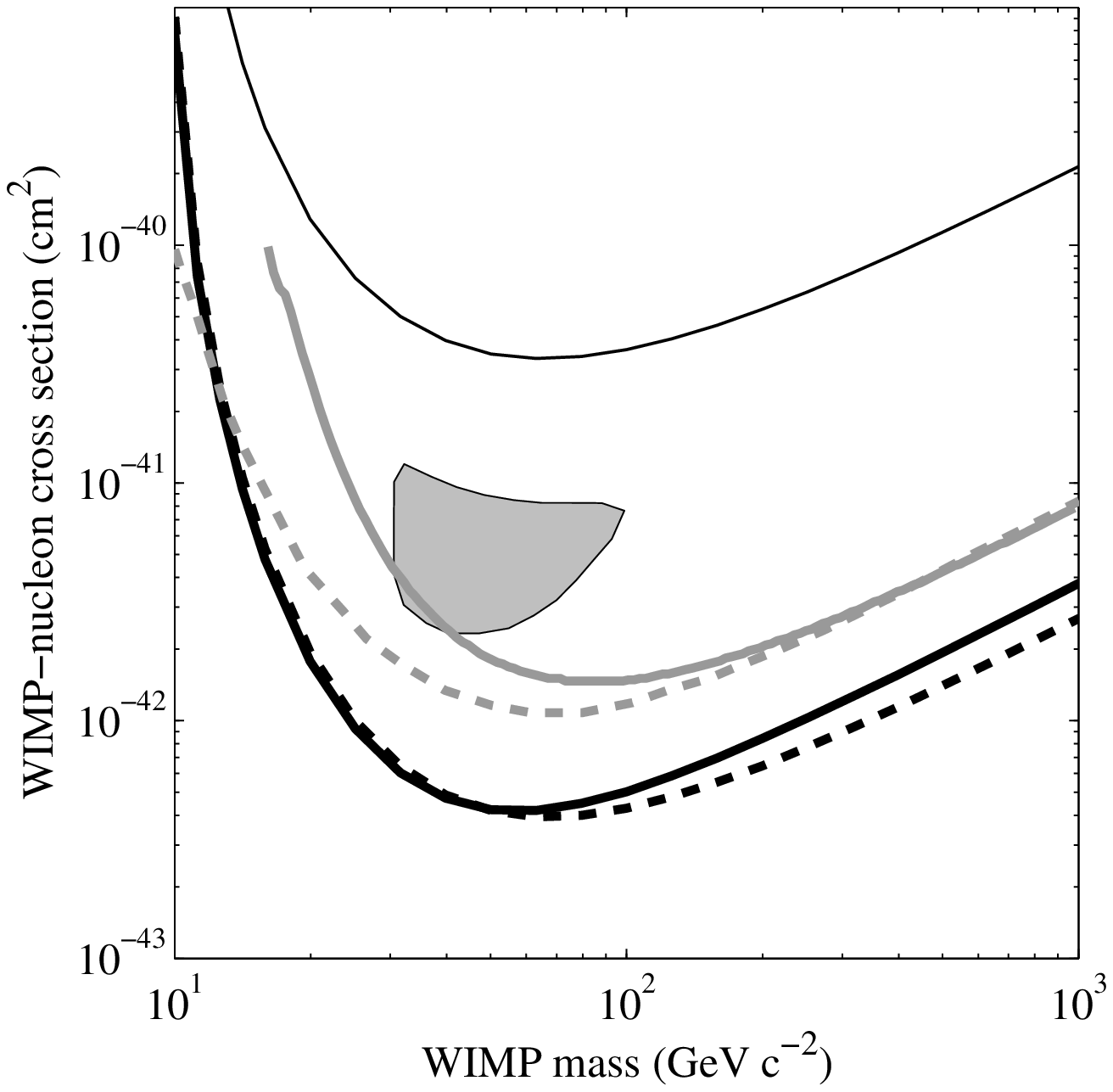}
\caption{Experimental limits in the WIMP-nucleon cross-section versus WIMP mass parameter space for spin-independent interactions. The cross-section is normalized to a single nucleon. The region above the solid (dashed) black curves are excluded at 90\% confidence by the current (initial) analysis of the CDMS-II Soudan WIMP search (Ge detectors). The upper thin black curve is the CDMS-II Soudan 90\% C.L. exclusion limit for the Si detector. The solid gray curve is the EDELWEISS~\cite{edelweiss} exclusion limit. The dashed gray curve is the ZEPLIN~I~\cite{zeplin1} exclusion limit. The $3\sigma$ DAMA~\cite{dama} detection region is shown in light gray. Plot courtesy of~\cite{dmplotter}.}
\label{fig:SIlimitexp}
\end{figure}

\begin{figure}
\centering
\includegraphics[scale=0.6]{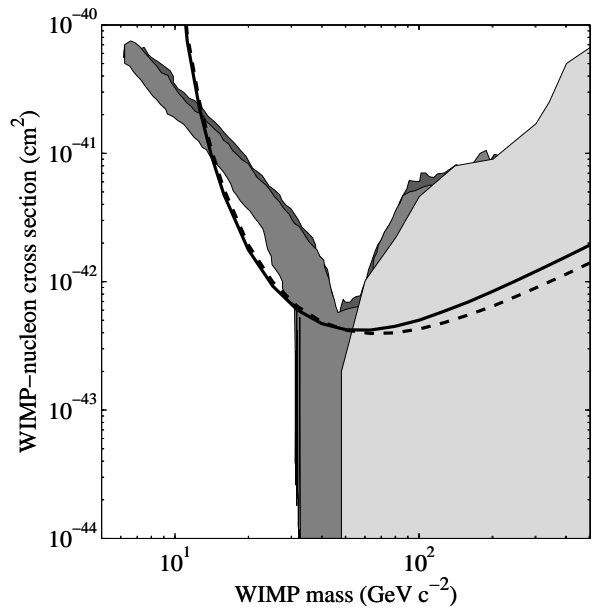}
\caption{Comparison of the current (solid curve) and initial (dashed curve) CDMS-II exclusion limits with predicted regions in the WIMP-nucleon cross-section versus WIMP mass parameter space for spin-independent interactions. The cross-section is normalized to a single nucleon. The light gray region corresponds to parameter space that is consistent with the standard constraints of the MSSM framework and measurements of the relic abundance of dark matter~\cite{kim}. The two dark gray regions correspond to additional allowed parameter space under the relaxation of some of the grand unification constraints~\cite{Bottino:noGUT}. Plot courtesy of~\cite{dmplotter}.}
\label{fig:SIlimittheory1}
\end{figure}

Under the assumptions of a standard galactic halo (as described in~\cite{lewin}), these data yield the lowest limit, calculated using the Optimal Interval Method~\cite{Yellin},  on the spin-independent WIMP-nucleon elastic-scattering cross-section for WIMP masses larger than 10~GeV~c$^{-2}$ (see Fig.~\ref{fig:SIlimitexp}). These limits begin to constrain some of the cosmologically-interesting supersymmetric theories under the MSSM framework~\cite{kim} (see Fig.~\ref{fig:SIlimittheory1}). Of particular importance is the exclusion of most of the funnel-shaped region in Fig.~\ref{fig:SIlimittheory1}, which corresponds to parameter space allowed by certain MSSM models where grand unification constraints are relaxed~\cite{Bottino:noGUT}. Other supersymmetric frameworks such as mSUGRA~\cite{baltz,chattopadhyay} and split-SUSY~\cite{Masiero,Pierce} remain beyond the sensitivity of this WIMP search. These data confirm that events detected by CDMS at SUF~\cite{R21} and those detected by EDELWEISS~\cite{edelweiss} were not a WIMP signal.Ê In addition, these data show that the claimed DAMA signal~\cite{dama} does not come from WIMPs of masses greater than 10~GeV~c$^{-2}$ distributed in a standard galactic halo and interacting through spin-independent interactions.  

\begin{figure}
\centering
\includegraphics[scale=0.6]{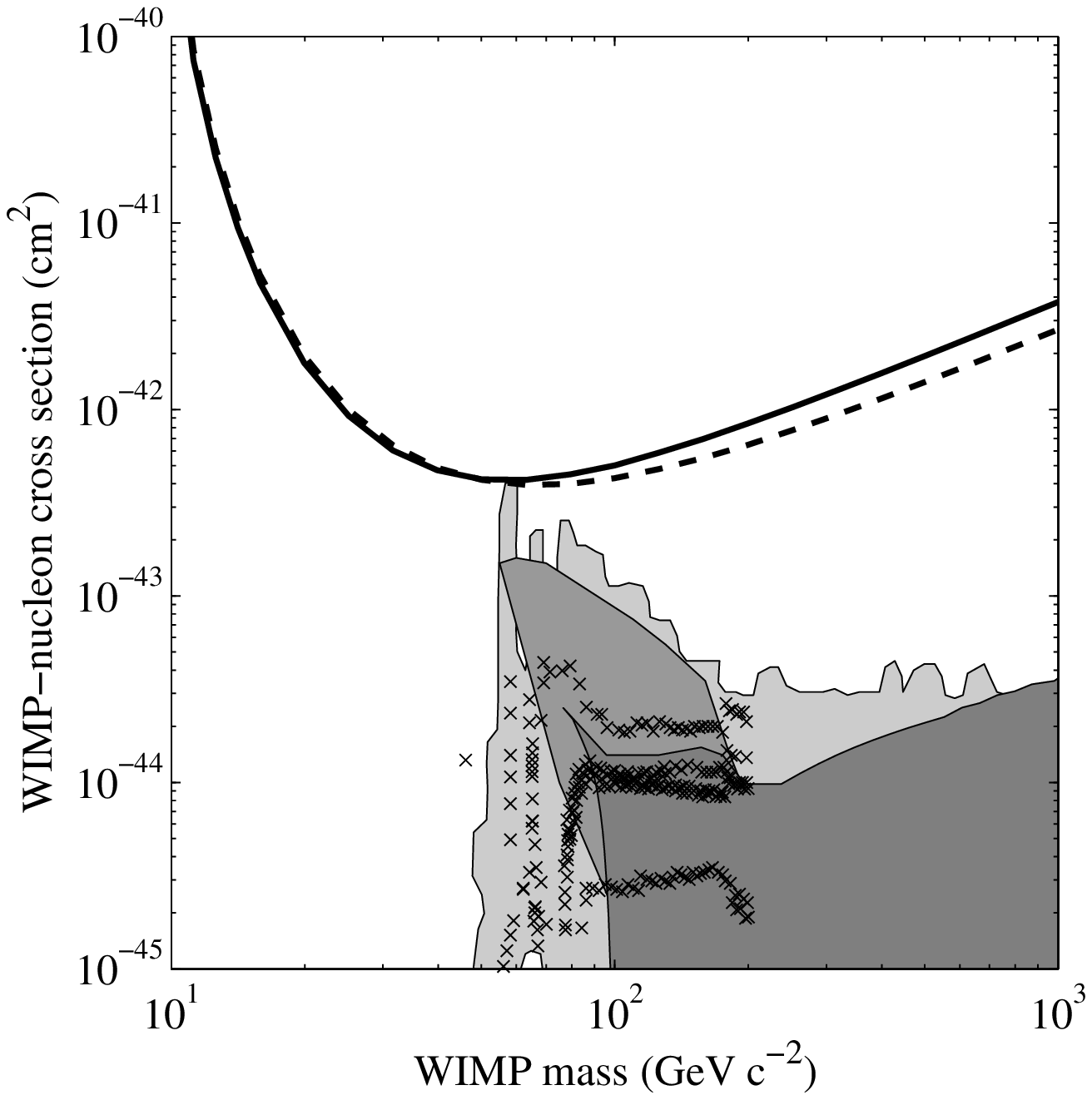}
\caption{Comparison of the current (solid curve) and initial (dashed curve) CDMS-II exclusion limits with predicted regions in the WIMP-nucleon cross-section versus WIMP mass parameter space for spin-independent interactions. The cross-section is normalized to a single nucleon. The two light gray regions correspond to parameter space allowed in the mSUGRA framework~\cite{baltz,chattopadhyay}. The dark gray region and black $\times$'s correspond to allowed regions under the assumptions of split-SUSY~\cite{Masiero,Pierce}. Plot courtesy of~\cite{dmplotter}.}
\label{fig:SIlimittheory2}
\end{figure}

\begin{figure}
	\centering
		\includegraphics [width = 8cm]{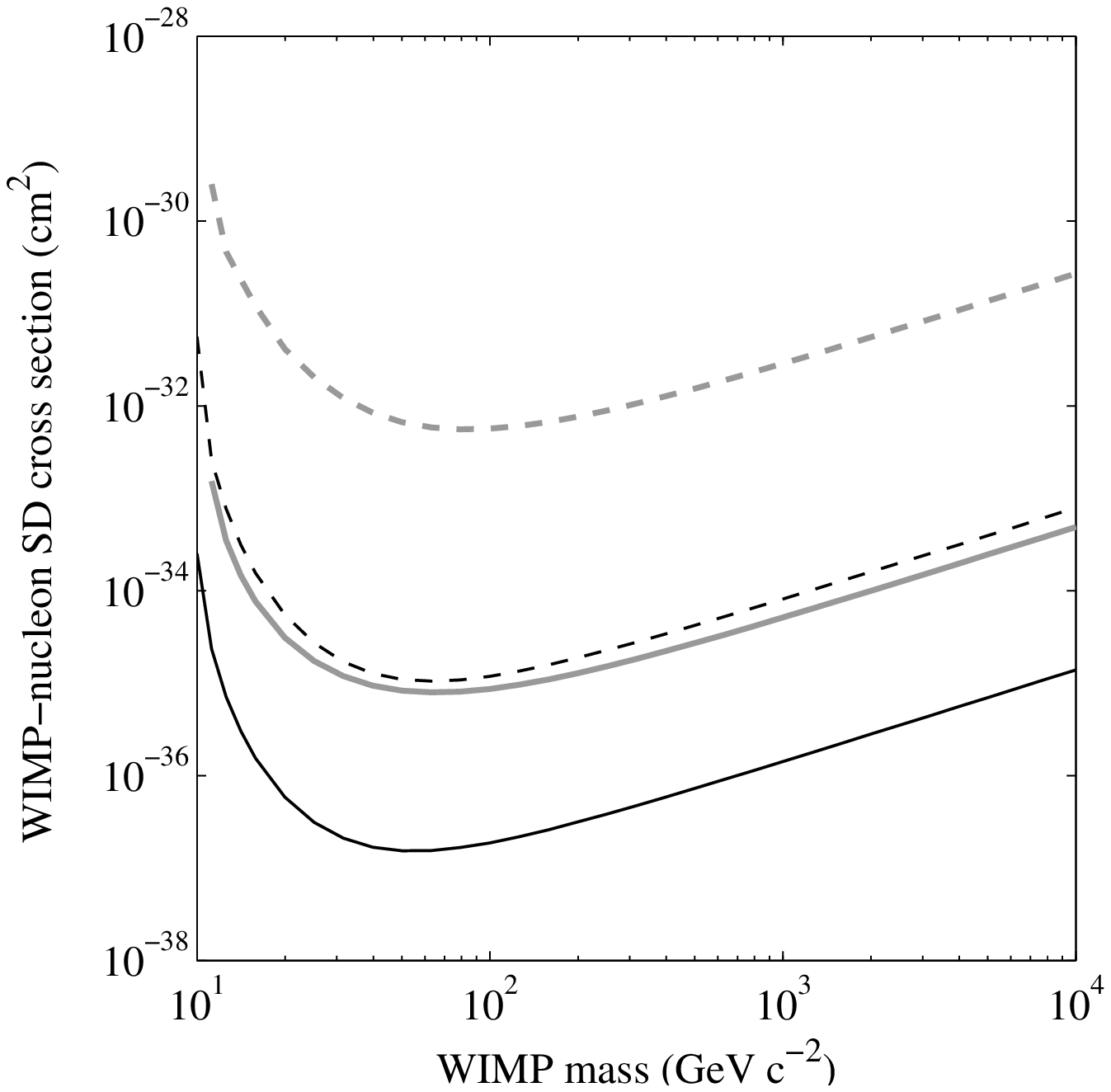}
	\caption{CDMS-II (Tower~1) spin-dependent limits.  Black lines are Ge limits, gray lines are Si limits.  Dotted lines are proton limits and solid lines are neutron limits.}
	\label{fig:SDLimPlot}
\end{figure}

\begin{figure}
	\centering
		\includegraphics[width=8cm]{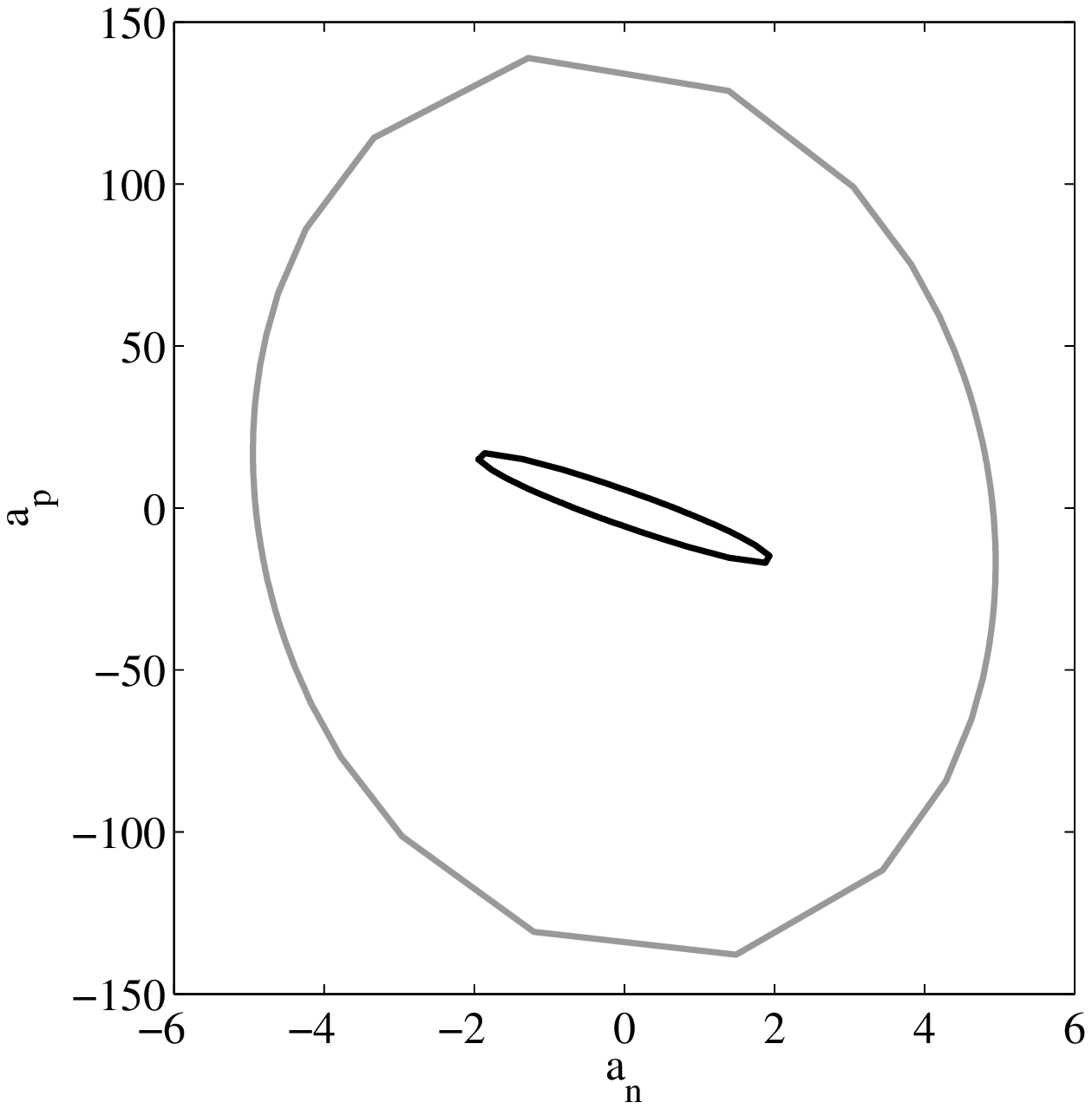}
	\caption{Limits on WIMP-nucleon couplings ($a_p$ and $a_n$) with a WIMP mass of $50$~GeV c$^{-2}$.  The black curve is the Ge limit and the gray curve is the Si limit.}
	\label{fig:apan50GeV}
\end{figure}

In addition to spin-independent (SI) WIMP interactions, it is also possible that WIMPs may possess spin-dependent (SD) interactions with atomic nuclei~\cite{jkg}. Such interactions provide additional avenues by which direct detection experiments may constrain WIMP models, as well as alternative interpretations for reports of positive WIMP detection signals~\cite{Bottino:noGUT}. CDMS obtains its SD sensitivity through the presence of two odd-neutron isotopes in its target materials: $^{29}$Si (4.68\% of natural Si) and $^{73}$Ge (7.73\% of natural Ge). These nuclides permit us to interpret our limits on nuclear-recoil rates in terms of limits on WIMP-nucleon spin-dependent cross-sections. We proceed under the same assumptions of a standard galactic halo, following the method described in~\cite{Tovey} and using the form factors given in~\cite{Ressell93} for Si and~\cite{Engel95} for Ge. Figure~\ref{fig:SDLimPlot} shows the resulting limits, calculated using the Optimal Interval Method~\cite{Yellin}. These data yield the lowest WIMP-neutron limit over a substantial range of WIMP masses~\cite{giuliani,savage} and are comparable to those quoted in~\cite{savage}, but are $\sim$20\% stronger at the minimum due to our full knowledge of the detector efficiencies. We may also express our limits in terms of allowed regions in the $a_p-a_n$ plane of WIMP-nucleon SD couplings, taking proper account of the finite-momentum effects described in~\cite{Ressell93,Engel95}. The allowed region for WIMPs of mass 50~GeV~c$^{-2}$ is shown in Fig.~\ref{fig:apan50GeV}.  Further details of this analysis may be found in~\cite{CDMSSpinDep}.

\section{Summary}
From October 11, 2003 until January 11, 2004, the CDMS collaboration searched for WIMP dark matter at the Soudan Underground Laboratory using four Ge and two Si ZIP detectors. The expected background consists of $0.7 \pm 0.3$ betas from contamination on the detector surfaces. Out of nearly one million events, only one survives all of our analysis cuts. Under the assumptions of a standard galactic halo, these data set the world's lowest limits on the WIMP-nucleon cross-section in the case of spin-independent interactions or spin-dependent interactions with neutrons.

\section{Acknowledgements}

We wish to thank the technical and engineering support at our home institutions for their contributions over the years to bring the CDMS-II Soudan apparatus and first data run reported here to fruition. At the Soudan Underground Laboratory we wish to thank the Department of Natural Resources (DNR) for access and operation of the Soudan site. The University of Minnesota Soudan laboratory staff for their assistance in construction and operation of the experiment. In particular, Jim Beaty is owed a special thanks by many of us for his dedication to the success of this experiment's operation. At Fermilab, engineering support for the cryogenic systems and overall infrastructure from Richard Schmitt, Lou Kula and Stan Orr is noted, along-with the technical support of Bruce Lambin, Bryan Johnson, Rodney Choate, and James Williams. 

At Santa Barbara, the muon veto shield construction and installation was supported by Dave Hale, Susanne Kyre, Sam Burke,  and Dan Callahan. ZIP detector fabrication and testing support were provided by Robert Abusaidi, Pat Castle, Larry Novak, Astrid Tomada, Mike Hennessy, and Jim Perales at Stanford, with further invaluable support from John Emes at Lawrence Berkeley National Laboratory. Support for the cold electronics and associated infrastructure were provided by Judith Alvaro-Dean and Garth Smith at Berkeley. Detector testing at Case Western Reserve University was supported by Aaron Manalysay and Adam Sirois. Undergraduates at University of Colorado at Denver and Health Sciences Center, Denver, were instrumental in characterizing the SQUID amplifier chips now in use at Soudan. We also thank the undergraduates at the collaboration's other home institutions for their various contributions to CDMS-II. 

This work is supported by the National Science Foundation under Grant No.\ AST-9978911 and No.\ PHY-9722414, by the Department of Energy under contracts DE-AC03-76SF00098, DE-FG03-90ER40569, DE-FG03-91ER40618, and by Fermilab, operated by the Universities Research Association, Inc., under Contract No.\ DE-AC02-76CH03000 with the Department of Energy. The ZIP detectors were fabricated in the Stanford Nanofabrication Facility operated under NSF. We dedicate this paper to Ron Ross, who passed away before these results came to fruition.

\end{document}